\documentclass[aps, twocolumn, superscriptaddress, 10pt]{revtex4-2}

\usepackage{amsmath}
\usepackage{amsmath, amssymb}
\usepackage{braket}
\usepackage[normalem]{ulem}
\usepackage{physics}
\usepackage{quantikz}
\usepackage{color}
\usepackage{subfigure}
\usepackage{orcidlink}

\usepackage{todonotes}

\newcommand{\difft}[1]{\frac{\mathrm{d} #1}{\mathrm{d} t}}
\newcommand{\diff}[2]{\frac{\partial #1}{\partial #2}}
\newcommand{\ddiff}[2]{\frac{\partial^2 #1}{\partial #2^2}}

\begin{document}
\title{Fully quantum algorithmic framework for partial differential equation-constrained optimization}

\author{Yuki Sato}
\email[]{yuki-sato@mosk.tytlabs.co.jp}
\affiliation{Toyota Central R\&D Labs., Inc., 1-4-14 Koraku, Bunkyo-ku, Tokyo 1120004, Japan.}
\affiliation{Quantum Computing Center, Keio University, 3-14-1 Hiyoshi, Kohoku, Yokohama, Kanagawa 223-8522, Japan}

\author{Jumpei Kato}
\affiliation{Mitsubishi UFJ Financial Group, Inc.~and MUFG Bank, Ltd., 4-10-2 Nakano, Nakano-ku, Tokyo 164-0001, Japan}
\affiliation{Quantum Computing Center, Keio University, 3-14-1 Hiyoshi, Kohoku, Yokohama, Kanagawa 223-8522, Japan}
\affiliation{Graduate School of Science and Technology, Keio University, 3-14-1 Hiyoshi, Kohoku, Yokohama, Kanagawa 223-8522, Japan}

\author{Hiroshi Yano}
\affiliation{Toyota Central R\&D Labs., Inc., 1-4-14 Koraku, Bunkyo-ku, Tokyo 1120004, Japan.}
\affiliation{Quantum Computing Center, Keio University, 3-14-1 Hiyoshi, Kohoku, Yokohama, Kanagawa 223-8522, Japan}

\author{Kosuke Ito}
\affiliation{Advanced Material Engineering Division, Toyota Motor Corporation,
1200 Mishuku, Susono, Shizuoka 410-1193, Japan}
\affiliation{Quantum Computing Center, Keio University, 3-14-1 Hiyoshi, Kohoku, Yokohama, Kanagawa 223-8522, Japan}

\author{Naoki Yamamoto}
\affiliation{Quantum Computing Center, Keio University, 3-14-1 Hiyoshi, Kohoku, Yokohama, Kanagawa 223-8522, Japan}
\affiliation{Department of Applied Physics and Physico-Informatics, Keio University, 3-14-1 Hiyoshi, Kohoku, Yokohama, Kanagawa 223-8522, Japan}

\begin{abstract}
    Partial differential equation (PDE)–constrained optimization, where an optimization problem is subject to PDE constraints, arises in various applications such as design, control, and inference. 
    Solving such problems is computationally demanding because it requires repeatedly solving a PDE and using its solution within an optimization process.
    In this paper, we first propose a fully quantum algorithmic framework for solving PDE-constrained optimization problems. 
    The proposed method combines a quantum PDE solver that prepares the solution vector as a quantum state, and a quantum optimizer that assumes oracle access to a quantized objective function. 
    The central idea is the explicit construction of the oracle in a form of block-encoding for the objective function, which coherently uses the output of a quantum PDE solver.
    This enables us to avoid classical access to the full solution that requires quantum state tomography canceling out the potential quantum speedups.
    We also derive the overall computational complexity of the proposed method with respect to parameters for optimization and PDE simulation, where quantum speedup is inherited from the underlying quantum PDE solver.
    We numerically demonstrate the validity of the proposed method by applications, including a parameter calibration problem in the Black-Scholes equation and a material parameter design problem in the wave equation. 
    This work presents the concept of composing quantum subroutines so that the weakness of one (i.e., prohibitive readout overhead) is neutralized by the strength of another (i.e., coherent oracle access), toward a bottleneck-free quantum algorithm.
\end{abstract}

\maketitle

\section{Introduction}
Partial differential equation (PDE)–constrained optimization~\cite{hinze2008optimization, de2015numerical} arises in various applications, including design optimization~\cite{deb2012optimization, martins2021engineering} such as shape~\cite{sokolowski1992introduction, haslinger2003introduction} and topology optimization~\cite{bendsoe1988generating}, optimal control~\cite{troltzsch2024optimal}, and parameter estimation of a physical system.
PDE-constrained optimization is a kind of inverse problems where one seeks design or control parameters $\xi$ so that the state $u$ of a physical system, governed by a PDE, minimizes an objective function.
In an abstract setting, one solves the following problem:
\begin{align}
    \min_{\xi \in \Xi_\mathrm{ad}} ~ \mathcal{F}(u, \xi) \text{ subject to } ~ \mathcal{G}(u, \xi) = 0, \label{eq:PDECO}
\end{align}
where $\Xi_\mathrm{ad}$ is an admissible space of design parameters, $\mathcal{F}$ is an objective function, and $\mathcal{G}$ is a governing equation of a physical system of interest given by PDEs with appropriate boundary conditions.
If we consider $\Xi_\mathrm{ad} \subset \mathbb{R}^M$ with $M$ the number of parameters, we have continuous optimization, while if $\Xi_\mathrm{ad} \subset \mathbb{Z}^M$, we have discrete optimization.
When focusing on continuous optimization, one typically relies on gradient-based approaches to solve the problem in Eq.~\eqref{eq:PDECO}, where each optimization step requires at least one forward and adjoint numerical analyses of PDE $\mathcal{G}$ to compute the objective function $\mathcal{F}$ and its derivative $\nabla_\xi \mathcal{F}$.
For discrete optimization, one usually uses heuristics in which many forward numerical analyses are required to compute the objective function $\mathcal{F}$.
As system sizes of PDEs grow, the computational cost of these numerical analyses can be computationally prohibitive, which makes it intractable to solve PDE-constrained optimization problems.

So, could quantum computing provide an efficient solution to those problems? 
In fact, first, quantum computing offers asymptotic speedups for linear-algebra primitives underlying the numerical analysis of PDEs.
For stationary (elliptic) problems, spatial discretization yields a sparse linear system, which can be solved by quantum linear system algorithms (QLSAs) such as the HHL~\cite{harrow2009quantum}, linear combination of unitaries (LCU)-based~\cite{childs2017quantum} and quantum singular value transformation (QSVT)-based algorithms~\cite{martyn2021grand}.
Under standard access models (sparse-access oracles for HHL and LCU; block-encoding for QSVT), these algorithms can offer quantum speedups with respect to the system size $N$ of discretized PDEs.
Several works proposed methods for solving elliptic PDEs by quantum algorithms~\cite{cao2013quantum, montanaro2016quantum}.
For evolution (parabolic or hyperbolic) problems, space and time discretization yield a linear system, which can be solved by QLSAs~\cite{berry2014high}.
Also, spatial discretization results in an ordinary differential equation (ODE), which can be simulated by quantum ODE solvers such as linear combination of Hamiltonian simulation (LCHS)~\cite{an2023linear, an2023quantum, pocrnic2025constant, low2025optimal}, Schr\"{o}dingerization~\cite{jin2023aquantum, jin2024quantum, jin2026schrodingerization}, and Lindbladian simulation-based approach~\cite{shang2025designing, fang2025qubit}.
These approaches can provide quantum speedups in terms of the system size $N$ of discretized PDEs under Hamiltonian simulation oracles, serving as an access model.
There are several works that applied these algorithms or simply Hamiltonian simulation for solving PDEs~\cite{costa2019quantum, miyamoto2024quantum, brearley2024quantum, sato2024hamiltonian, wright2024noisy, sato2025quantum, hu2024quantum}.
In particular, some of these works provided an explicit quantum circuit with gate counts~\cite{sato2024hamiltonian, wright2024noisy, hu2024quantum, sato2025quantum}.
These results suggest that quantum algorithms have potential for accelerating the simulation of PDEs on future fault-tolerant quantum computers. 
However, we cannot retrieve the entire solution vector, because this needs the full state tomography; what we can do is to extract only partial information of the solution by measuring some appropriate observables~\cite{aaronson2015read, williams2025vortex, williams2024addressing}.

Quantum algorithms are also promising for optimization problems.
For discrete optimization, the variational quantum algorithms (VQAs)~\cite{cerezo2021variational, tilly2022variational} and the quantum approximate optimization algorithm (QAOA)~\cite{farhi2014quantum, zhou2020quantum, willsch2020benchmarking} are among the most widely studied quantum algorithms, which are hybrid quantum–classical algorithms for combinatorial optimization. 
A possible drawback of these variational methods is that estimation of the gradient needs multiple running of the circuit, and the gradient can become exponentially small as the system size increases, which is known as the barren plateau phenomenon~\cite{cerezo2021cost, arrasmith2022equivalence}.
For continuous optimization, on the other hand, quantum dynamics-based optimization~\cite{leng2023quantum, leng2025quantum, chen2025quantum, catli2026efficient} is a promising method in which an optimization problem is converted into a coherent quantum dynamics that does not involve any measurement process. 
In particular, quantum Hamiltonian descent (QHD)~\cite{leng2023quantum, leng2025quantum}, which maps gradient-based optimization into time-dependent Hamiltonian simulation, has the potential for offering exponential speedups under access to an objective function via bit or phase oracles~\cite{catli2026efficient}. 
However, this means that the practical use of these coherent optimization methods could be limited because such oracle construction is in general quite non-trivial and possibly hard. 
Moreover, even if such oracles are available, it is also uncertain whether employing QHD offers a practical advantage in terms of end-to-end gate complexity including the cost of constructing the oracle rather than query complexity.

\begin{figure*}[t]
    \centering
    \includegraphics[width=\textwidth]{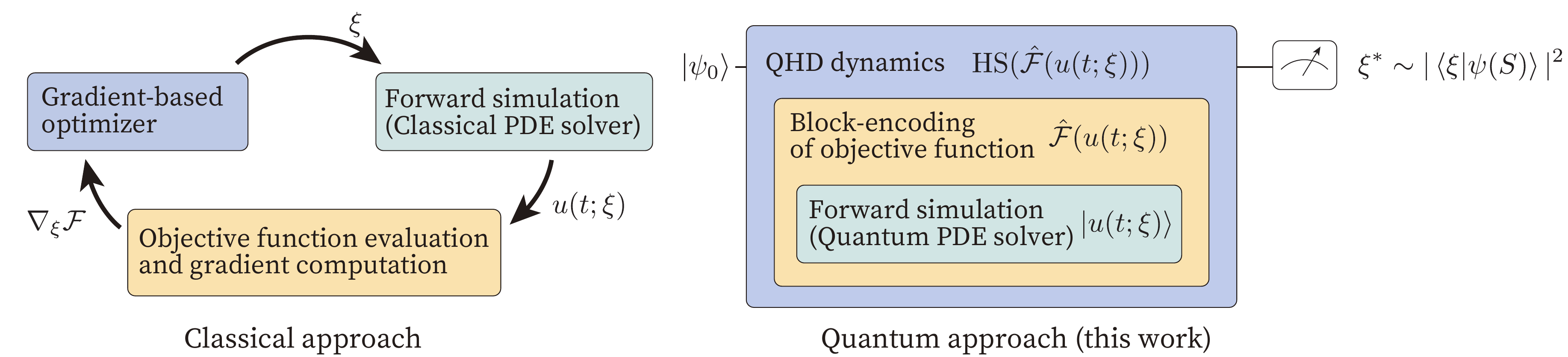}
    \caption{Conceptual diagram of the proposed method (right), compared with the conventional classical approach (left).
    $\text{HS}(\hat{\mathcal{F}}(u(t;\xi))) := \mathcal{T}e^{-\int_0^S (e^{-\nu s'} \nabla_\xi^2/2 + e^{\nu s'} \hat{\mathcal{F}}) \mathrm{d} s'} \ket{\psi_0}$ is Hamiltonian simulation algorithm of the time-dependent Hamiltonian with the objective function as a potential term, where $\nu$ is a constant parameter.} \label{fig:concept}
\end{figure*}

The contribution of this paper is to develop a fully-quantum algorithm for solving PDE-constrained optimization problems, by combining the quantum PDE solver (LCHS, producing the solution $\ket{u(t;\xi)}$) and quantum optimizer (QHD, minimizing the quantized objective function $\hat{\mathcal F}(u(t;\xi))$ with respect to $\xi$). 
A first technical step is to formulate the parameterized PDE family as an extended forward-simulation problem coherent in the design variable, and to compile the corresponding design-dependent forward oracle. Building on this, we explicitly construct the oracle of the objective function $\hat{\mathcal F}(u(t;\xi))$ required by QHD in the form of a block-encoding, in which the PDE solution $\ket{u(t;\xi)}$ is used coherently without intermediate measurement.
Figure~\ref{fig:concept} illustrates the conceptual diagram of the proposed method.
Importantly, the explicit block-encoding of the objective function coherently bridges
a quantum PDE solver and a quantum optimizer, thereby avoiding any classical readout of the output of the quantum PDE solver $\ket{u(t;\xi)}$; the output of our algorithm is a bit string that describes the optimized solution.
Hence, we can construct an end-to-end pipeline for PDE-constrained optimization, avoiding the exponential overhead of classical-quantum hybrid approaches as discussed in Ref.~\cite{catli2026efficient}.

We list our contributions as follows.

\begin{itemize}
    \item \emph{Problem formulation and framework.}
    We formulate PDE-constrained optimization as a target problem for fully coherent quantum algorithms.
    PDE-constrained optimization is a central computational task in scientific computing, inverse problems, parameter calibration, and design optimization, but its quantum-algorithmic treatment has remained largely unexplored.
    We show that this problem can be cast as an extended forward simulation problem, in which the design variables are coherently embedded into the forward PDE solver.
    This leads to a design-dependent forward simulation oracle that serves as the starting point for a fully quantum treatment of PDE-constrained optimization.

    \item \emph{Technical construction of the coherent interface between a quantum PDE solver and a quantum optimizer.}
    We provide explicit constructions of block-encodings for objective functions arising in PDE-constrained optimization.
    These constructions form the crucial interface between a quantum PDE solver and a quantum optimizer by constructing the objective function oracle from the output of the design-dependent quantum PDE solver.
    Since the forward solver naturally produces a subnormalized quantum state whose amplitude of the success flag depends on the forward solution, passing this subnormalization directly to the optimizer would lead to a potentially severe normalization penalty in the subsequent coherent dynamics.

    To mitigate this bottleneck, we further introduce a normalization refinement based on uniform singular value amplification~\cite[Theorem~30]{gilyen2019quantum}.
    This procedure amplifies the relevant singular value range of the extended forward simulation block-encoding uniformly over the design register, without tomography, postselection, or intermediate measurements.
    As a result, the coherent interface from the parameterized PDE solver to the objective oracle becomes a loss-mitigated interface with substantially improved block-encoding normalization.
    This refinement is essential for turning a formally coherent connection between the PDE solver and the optimizer into a quantitatively useful one.

    \item \emph{End-to-end complexity analysis.}
    Because our construction is explicit, we can track the full query and gate complexity of the resulting quantum algorithm, including the dependence on the number of design variables, the discretization size of the PDE, the simulation time, the target precision, and the optimizer-side parameters.
    This detailed scaling analysis identifies the parameter regimes in which the coherent quantum approach can potentially provide an advantage.

    In particular, under the assumption that the optimization landscape is strongly convex, we obtain an end-to-end quantum complexity bound and compare it with conventional classical approaches.
    The better complexity bound of our framework is inherited primarily from the quantum forward simulation component: our method yields polynomial improvements in the discretized system size and an exponential improvement in the spatial dimension dependence associated with high-dimensional PDE discretizations.
\end{itemize}

We also demonstrate how the framework applies to representative PDE-constrained optimization tasks.
Specifically, we consider parameter calibration for the Black--Scholes equation and material-parameter design for the wave equation.
These examples illustrate how the abstract construction can be instantiated in concrete PDE-constrained optimization settings and provide a pathway toward exploiting quantum speedups in practical applications.

We remark that there have been some trials to form a constrained optimization problem from two quantum algorithms (the one for forward simulating the constraint and another for optimization). 
Refs \cite{sato2023quantum, kim2025variational} formulated a PDE-constrained optimization, by dealing with both forward simulation and optimization in the framework of VQAs. 
Quantum simulation-based optimization (QuSO) algorithms~\cite{stein2023exponential, holscher2026quantum} incorporated QSVT-based forward simulation into QAOA-based optimization, where the problem Hamiltonian is constructed using forward solution, and the time evolution by the problem Hamiltonian is implemented using QSVT. 
As the nature of the variational method, all these methods are classical-quantum hybrid algorithms that need multiple measurements to update the optimizer and may suffer from the possible hard-to-train issue, which does not occur in our fully-coherent solver.

The remainder of this paper is organized as follows.
Section~\ref{sec:preliminary} provides some preliminaries where we formulate our problem of interest and provide primitives that compose our algorithm.
Section~\ref{sec:algo} describes the proposed algorithm and its complexity.
In Section~\ref{sec:application}, we provide two applications to demonstrate how our method can be applied to representative PDE-constrained optimization tasks. 
Finally, we conclude this study in Section~\ref{sec:conlusion}.

\section{Preliminaries} \label{sec:preliminary}
First, we provide some preliminaries on the PDE-constrained optimization, followed by the brief introduction of primitives we use: linear combination of Hamiltonian simulation~\cite{an2023linear, an2023quantum} and quantum Hamiltonian descent~\cite{leng2023quantum, leng2025quantum}.
Throughout this paper, we use the notation $\ket{\cdot}$ to describe a unit vector, which can be a quantum state embedded on a quantum computer.
For example, we describe $\ket{u(t; \xi)} := u(t; \xi) / \| u(t; \xi) \| $.

\subsection{PDE-constrained optimization}

In this study, we tackle optimization problems:
\begin{equation}
\begin{aligned}
    &\min_{\xi \in [0, 1]^M} && \mathcal{F}(u (t; \xi)) \\
    &\text{ subject to: } && \difft{u(t; \xi)} = -A(\xi) u(t; \xi), ~ u(0; \xi) = u_0,
\label{eq:ODECO}
\end{aligned}
\end{equation}
where we assume that the admissible space of design variables is normalized as $\Xi_\mathrm{ad} = [0, 1]^M$, without loss of generality.
The constraint is the governing equation of the system of interest, which is an ODE derived from the discretization of a PDE with $u(t) \in \mathbb{C}^N$ a discretized state variables, $u_0 \in \mathbb{C}^N$ an initial state, and $A(\xi) \in \mathbb{C}^{N \times N}$ a design-dependent coefficient matrix whose real part $\Re (A(\xi))$ is assumed to be positive semidefinite.
Since the ODE is a linear system, we assume that $\| u_0 \| = 1$ without loss of generality.
The number of degrees of freedom $N$ of the ODE is determined by the spatial discretization of PDEs.
For solving a PDE on a $d$-dimensional domain with the characteristic length $\ell$, we discretize the domain onto grids with grid spacing $h_x$, which makes the number of degrees of freedom $N$ of the ODE scale as $N \in \Theta ((\ell / h_x)^d)$.
In this study, we do not specify the spatial discretization scheme and use the same notation $u$ for the discretized state variables as that for the original state field $u(x,t)$ for simplicity, unless we need to distinguish them explicitly.

Here, we assume the coefficient matrix can be decomposed as follows:
\begin{align}
    A(\xi) := A_0 + \sum_{\mu=1}^M A_\mu \xi_\mu, \label{eq:A_xi}
\end{align}
where $A_\mu \in \mathbb{C}^{N \times N}$ for $\mu = 1, \dots, M$ with $M$ the number of design variables.
To ensure that $\Re (A(\xi))$ is positive semidefinite, we assume that $\Re (A_\mu)$ is positive semidefinite for all $\mu$.

We consider two types of objective functions.
We first introduce their physical (unscaled) forms. The first is a quadratic objective:
\begin{align}
    \mathcal{F}_\mathrm{quad}^\mathrm{base}(u (\xi) ) &:= u^\dagger(T; \xi) P u(T; \xi),
\end{align}
and the second is a discrepancy objective from a reference state $u_\mathrm{ref}(x)$:
\begin{align}
    \mathcal{F}_\mathrm{err}^\mathrm{base}(u (\xi)) &:= \left( u(T; \xi) - u_\mathrm{ref} \right)^\dagger P \left( u(T; \xi) - u_\mathrm{ref} \right),
\end{align}
where $T$ is a target time, $u_\mathrm{ref} \in \mathbb{C}^{N}$ is a discretized reference state, and $P \in \mathbb{R}^{N \times N}$ is a projector.
Since the convergence guarantee of QHD discussed later is naturally
formulated in terms of an additive error in the objective value, i.e., $| \mathcal{F}(u(\xi^\ast)) - \min_{\xi } \mathcal{F}(u(\xi)) | \leq \epsilon $, the scale of the objective function determines the meaning of the accuracy parameter $\epsilon$.
Indeed, if the physical objective has a characteristic scale $S_F \ll 1$,
then a fixed additive tolerance in the unscaled objective may correspond
to a large error relative to the physically relevant variation of the
objective.
To make the additive-error guarantee comparable across problems, and to
interpret $\epsilon$ as an accuracy relative to a known objective scale,
we introduce a normalization factor $S_F$ and optimize the normalized objective
\begin{align}
    \mathcal{F}(u(\xi))
    := \frac{1}{S_F}\mathcal{F}^{\mathrm{base}}(u(\xi)).
\end{align}
An $\epsilon$-accurate solution for the normalized objective then
corresponds to an $S_F\epsilon$-accurate solution for the original
physical objective. 
Since $S_F$ is independent of the design variable,
this rescaling does not change the optimum.
We take $S_F=\gamma_u^2$, where $\gamma_u$ is a design-independent upper bound
satisfying
\begin{align}
    \gamma_u \geq \max_{\xi }
    \|u(T;\xi)\|.
\end{align}
We then define
\begin{align}
\label{Eq:cost1}
    \mathcal{F}_\mathrm{quad}(u(\xi))
    &:= \frac{1}{\gamma_u^2}
    \mathcal{F}_\mathrm{quad}^\mathrm{base}(u(\xi)).
\end{align}
Because $P$ is a projector, this choice ensures
$\mathcal{F}_\mathrm{quad}(u(\xi)) \leq 1$ for all admissible $\xi$.
This normalization is particularly useful for dissipative systems, where
$\|u(T;\xi)\|$ may be small, and it also yields a tighter
subnormalization factor for the corresponding block-encoding in
Sec.~\ref{sec:block-encoding}.
For the discrepancy objective, we also define
\begin{align}
\label{Eq:cost2}
    \mathcal{F}_\mathrm{err}(u(\xi))
    &:=
    \frac{1}{\gamma_u^2}
    \mathcal{F}_\mathrm{err}^\mathrm{base}(u(\xi)).
\end{align}
Expressing $P = \sum_{j \in \mathcal{J}} \ket{j} \! \bra{j}$ with an index set $\mathcal{J} \subseteq [N]$, we can also express these objective functions as
\begin{align}
    \mathcal{F}_\mathrm{quad} &= \frac{1}{\gamma_u^2} \sum_{j \in \mathcal{J}} \| u_j (T; \xi) \|^2, \\
    \mathcal{F}_\mathrm{err} &= \frac{1}{\gamma_u^2} \sum_{j \in \mathcal{J}} \| u_j (T; \xi) - u_{\mathrm{ref}, j} \|^2,
\end{align}
where the subscript $j$ represents the $j$-th component of the vector.

\subsection{Quantum algorithm for ordinary differential equations}
Here, we briefly introduce LCHS for simulating time-independent ODEs in Eq.~\eqref{eq:ODECO} on a quantum computer.
It should be noted that the formulation of LCHS is capable of handling time-dependent ODEs even with an inhomogeneous part.
Based on the LCHS~\cite{an2023linear, an2023quantum}, an analytic form of the solution for the ODE $du/dt=-Au$ with $u(0)=\ket{u(0)}$ is given as follows:
\begin{align}
    u(t) &= \int_\mathbb{R} \frac{f(k)}{(1 - ik)} e^{- i \left( H + k L \right) t } u(0) ~ \mathrm{d}k, \label{eq:lchs}
\end{align}
where $L :=(A + A^\dagger)/2$ and $H := -i(A - A^\dagger)/2$ are Hermitian operators with the requirement of $L$ being positive semidefinite, and $f(k)$ is a kernel function satisfying certain conditions.
By approximating the integration by numerical integration with discretized $k$, we can use LCU to prepare the state of Eq.~\eqref{eq:lchs} on a quantum computer as 
\begin{equation}
\label{eq:linear solution}
     \frac{1}{\alpha_\mathrm{LCHS}} \exp \left( - A t \right) \ket{u(0)} \! \ket{0}^{\otimes a_\mathrm{LCHS}} + \ket{\perp}, 
\end{equation}
where the first term includes the solution vector $u(t) = \| u(t) \| \ket{u(t)} = \exp \left( - A t \right) \ket{u(0)}$. 
Also, $\ket{\perp}$ represents a vector orthogonal to the first term, $a_\mathrm{LCHS}$ is the number of ancillary qubits, and $\alpha_\mathrm{LCHS}$ is the $\ell_1$ norm of the vector whose entries are kernel values $f(k) / (1 - ik)$ on the discretized $k$ space.
This value is fixed once the kernel, quadrature rule, and discretization range are determined, and is independent from the quantum state. 
That is, given a LCHS error $\varepsilon$, the LCHS can be viewed as $( \alpha_\mathrm{LCHS}, a_\mathrm{LCHS}, \varepsilon)$-block-encoding of $\exp(-A t)$:
\begin{align} \label{eq:LCHS}
    U_\mathrm{LCHS}(A) := \begin{pmatrix}
        \frac{\exp (- A t)}{ \alpha_\mathrm{LCHS} } & \ast \\
        \ast & \ast
    \end{pmatrix},
\end{align}
which we will use a subroutine for the PDE-constrained optimization.

When using the kernel function of the form $f(k) = 1 / (2 \pi e^{-2^\beta} e^{(1+ik)^\beta})$ for $\beta \in (0, 1)$, the LCHS requires $\tilde{\mathcal{O}}((\| u(0) \| / \| u(T) \|) \alpha_A T \log (1/\varepsilon)^{1/\beta})$ queries to $A$ (formally, queries to $H + kL$) with $\alpha_A \geq \| A \|$ to achieve error at most $\varepsilon$, which is nearly optimal scaling~\cite{an2023quantum}.
Recently, this scaling has been further improved to the optimal scaling $\tilde{\mathcal{O}}((\| u(0) \| / \| u(T) \|) \alpha_A T \log (1/\varepsilon))$~\cite{low2025optimal}.
On the other hand, the simplest and original implementation of the LCHS is based on Trotterization with the kernel $f(k) = 1 / (\pi(1+ik))$, which requires $\tilde{\mathcal{O}}((\| u(0) \| / \| u(T) \|) (\alpha_A T)^{1 + 1/p} / \varepsilon^{1 + 1/p})$ queries to Hamiltonian simulation oracles $e^{-iH\tau}$ and $e^{-iL\tau}$ where $p$ is the order of the product formula and $\tau$ is a time increment~\cite{an2023linear}.
For both cases, the factor $\| u(0) \| / \| u(T) \|$ comes from amplitude amplification.
In the previous paper~\cite{an2023linear}, the original implementation also requires additional overhead $\| u(0) \| / \| u(T) \|)^{1 + 1/p}$ to bound the relative Trotter error. 
In our study, we can omit these factors because we evaluate the objective function using an unnormalized state.

Note that the solution of the ODE is prepared as the amplitude of a quantum state. 
Hence, we can only extract partial information of the solution by setting some meaningful observables, which is non-trivial for practical use case.

\subsection{Quantum Hamiltonian descent}

Quantum Hamiltonian descent (QHD) is a quantum analogue of the continuous-time limit of gradient descent algorithm for continuous optimization~\cite{leng2023quantum, leng2025quantum}. 
QHD encodes an objective function $\mathcal{F}(\xi)$ as the potential in the form of time-dependent (infinite-dimensional) quantum Hamiltonian $\mathcal{H}(t)$ as 
\begin{align}
    \mathcal{H}(s) := \frac{1}{\lambda(s)} \sum_{\mu = 1}^{M} \frac{1}{2} \ddiff{}{\xi_\mu} + \lambda(s) \mathcal{F}(\xi), \label{eq:QHDHam}
\end{align}
where $s$ is time and $\lambda(s) > 0$ is a monotonically increasing function. 
Notably, the quantum state $\psi(s, \xi)$ evolved from an initial state $\psi(0, \xi)$ by the Schr\"{o}dinger equation 
\begin{align}
    \difft{\psi(s, \xi)} = -i \mathcal{H}(s) \psi(s, \xi), \label{eq:schrodinger_eq}
\end{align} 
concentrates near low-energy regions.
That is, after a sufficiently long time $s$, sampling $\xi^\ast$ from the probability distribution $| \psi(s, \xi) |^2$ yields an approximate solution to the minimizer of $\mathcal{F}(\xi)$ with high probability.
When choosing $\lambda (s) = e^{\nu s}$ with a parameter $\nu > 0$, this dynamics governed by the Schr\"{o}dinger equation corresponds to the frictional dynamics with friction coefficient of $\nu$; that is, the quantum counterpart of the classical dynamics described by Newton's equation of motion $\ddot{\xi} + \nu \dot{\xi} + \nabla \mathcal{F} = 0$.

QHD is designed to be implementable on both gate-based quantum computers and analog hardware~\cite{leng2023quantum}.
Under the phase oracle access to the objective function, the previous work~\cite{catli2026efficient} showed that the quantum algorithm with $\lambda (s) = e^{\nu s}$ exhibits exponentially better scaling with the condition number of the Hessian than a vanilla gradient descent for strongly convex optimization problems.
The previous work~\cite{leng2025quantum} also showed the exponentially fast convergence of QHD with $\lambda (s) = e^{\nu s}$ for a strongly convex function and proposed other choices of $\lambda(s)$: $\lambda (s) = s^3$ for a convex function, and $\lambda (s) = s^{1/3}$ for a non-convex function.
In the work~\cite{leng2025quantum}, the authors also provided numerical results exhibiting the better convergence of discrete-time QHD by the product formula.

An important caveat is that, to implement the QHD on a quantum computer, the oracle access to the objective function is required, and such oracle construction is in general non-trivial. 
This paper proposes the explicit construction of the block-encoding of the objective function in the subsequent section.

\section{Algorithm} \label{sec:algo}

In this study, we propose a full quantum algorithm for solving PDE-constrained optimization problems by using QHD as an optimizer.
Here, we briefly explain our proposed method, followed by the explicit block-encoding, which is an essential part of our approach and complexity analysis.

\subsection{Overview} \label{sec:overview}
In QHD, we need to solve the Schr\"{o}dinger equation for the time-dependent Hamiltonian $\mathcal{H}(s)$ in Eq.~\eqref{eq:QHDHam} in which the objective function of the PDE-constrained optimization is encoded into the potential term $\mathcal{F}(\xi)$.
First, we discretize the Schr\"{o}dinger equation so that it can be dealt with on qubits.
Specifically, we discretize each design variable $\xi_\mu~(\mu=1,\ldots,M)$ into $m$ bits $\xi_\mu^{(m)}, \dots, \xi_\mu^{(1)}$, according to
\begin{align}
    \xi_\mu = h_\xi \sum_{b=1}^{m} \xi_\mu^{(b)} 2^{b-1},
\end{align}
where $h_\xi$ is the interval of discretized design variables, i.e., $h_\xi = 1/(2^m - 1)$.
Hereinafter, we use the notation 
\begin{equation}
    \ket{\xi} = \ket{\xi_M, \dots, \xi_1} 
      = \ket{\xi_M^{(m)}, \dots, \xi_M^{(1)}, \dots, \xi_1^{(m)}, \dots, \xi_1^{(1)}}
\end{equation}
to represent the design variables $\xi$ encoded in the computational basis of a $Mm$-qubit system.
The quantum state $\psi(s, \xi)$ is represented in terms of the ket vector on the computational basis as $\ket{\psi(s)} = \sum_\xi \psi(s, \xi) \ket{\xi}$.
By this discretization, the second-order derivative $\partial^2 / \partial \xi_\mu^2$ in the Hamiltonian is replaced by the finite difference operator $\Lambda \in \mathbb{R}^{2^m \times 2^m}$ defined as 
\begin{align}
    \Lambda := \frac{1}{h_\xi^2} \begin{pmatrix}
        -2 & 1 &  &  &  \\
        1 & -2 & 1 &  &  \\
         & \ddots & \ddots & \ddots \\
          & & 1 & -2 & 1 \\
         & & & 1 & -2 
    \end{pmatrix}.
    \label{eq:finitedifferenceop}
\end{align}
Thus, the discrete Laplace operator for the Laplacian $\sum_{\mu = 1}^{M} \partial^2 / \partial \xi_\mu^2$  is replaced with
\begin{align}
    \Lambda^{(M)} := \sum_{\mu=1}^M I^{\otimes (M - \mu)} \otimes  \Lambda \otimes I^{\otimes (\mu-1)},
\end{align}
where $I$ is an identity matrix of size $2 \times 2$.
The time-dependent Hamiltonian after discretization is then obtained, as follows:
\begin{align}
    \mathcal{H}(s) = \frac{1}{2 \lambda(s)} \Lambda^{(M)} + \lambda(s) \hat{\mathcal{F}}, 
\label{eq:QHDHam_discrete}
\end{align}
where $\hat{\mathcal{F}} := \sum_{\xi} \mathcal{F}(u(\xi)) \ket{\xi} \! \bra{\xi}$.
In the present study, we choose $\lambda(s) = e^{\nu s}$.

The proposed method for PDE-constrained optimization consists of the following steps.
\begin{enumerate}
    \item Set the friction parameter $\nu$ and the simulation time $S$.
    
    \item Prepare an initial state $\ket{\psi(0)} = U_{\psi_0} \ket{0}^{\otimes Mm}$ representing the initial probability distribution of design variables $\xi$ by a state preparation oracle $U_{\psi_0}$. This study uses the uniform superposition as the initial state, i.e., $\ket{\psi(0)} = 2^{-Mm/2}\sum_\xi \ket{\xi}$, which can be prepared by the Hadamard gate.
    
    \item Encode the PDE solution $\ket{u(t)}$ coherently into the objective function of the PDE-constrained optimization, $\mathcal{F}(\xi)$; then, 
    perform the time-dependent Hamiltonian simulation of $\mathcal{H}(s)$ in Eq.~\eqref{eq:QHDHam_discrete} from time $0$ to $S$. 
    
    \item Obtain a solution $\xi^\ast \sim  | \braket{\xi}{\psi(S)} |^2$ by measurement on the computational basis.
\end{enumerate}
The key component of the QHD is the step 3, i.e. the implementation of the time evolution of the time-dependent Hamiltonian $\mathcal{H}(s)$, which contains the PDE solution $\ket{u(t)}$. 
The step 3 has three layers: first, we construct a design-coherent forward simulation solver for the parameterized PDE family; second, we convert its output into an oracle of the objective function; third, we use that oracle within the optimizer dynamics.

To implement the time evolution $U_\mathcal{H}(s, 0) := \mathcal{T} e^{-i\int_0^s \mathcal{H}(s') \mathrm{d} s'}$ where $\mathcal{T}$ is a time-ordering operator, we employ two approaches: 
\begin{itemize}
    \item the interaction picture simulation using a truncated Dyson series;
    \item a simple product formula.
\end{itemize}
The former is effective for accurately simulating the time-dependent Hamiltonian, which ensures convergence to the optimal solution for strongly convex-optimization problems, while the latter provides a simple implementation despite an unfavorable error. 
In either case, we divide the time interval into short ones $0 := s_0 < s_1 < \cdots < s_{N_\mathrm{step} - 1} < s_{N_\mathrm{step}} =: S$ with uniform interval $\Delta s$, i.e., $s_j = j\Delta s$.

In the interaction picture simulation, we transform the time-dependent Hamiltonian $\mathcal{H}(s)$ into the interaction picture one: 
\begin{align}
    \mathcal{H}_\mathrm{I}(s) &:= 
    e^{i \int_0^s  e^{\nu s'} \hat{\mathcal{F}} \mathrm{d} s'} \frac{e^{-\nu s}}{2} \Lambda^{(M)} e^{-i \int_0^s e^{\nu s'} \hat{\mathcal{F}} \mathrm{d} s' } \nonumber \\
    &= e^{i \frac{1}{\nu} (e^{\nu s} - 1) \hat{\mathcal{F}}} \frac{e^{-\nu s}}{2} \Lambda^{(M)} e^{-i \frac{1}{\nu} (e^{\nu s} - 1) \hat{\mathcal{F}}}, \label{eq:QHD_ham_I}
\end{align}
where the quantum state is evolved as
\begin{align}
    \ket{\psi(s)} = \mathcal{T} e^{-i \int_0^s e^{\nu s'} \hat{\mathcal{F}} \mathrm{d} s'} ~ \mathcal{T} e^{-i\int_0^s \mathcal{H}_I(s') \mathrm{d} s'} \ket{\psi(0)}.
\end{align} 
Then we approximate the time evolution $\mathcal{T} e^{-i\int_0^s \mathcal{H}_I(s') \mathrm{d} s'}$ by the multiple short-time evolution from time $s_j$ to $s_{j+1}$, denoted by $U_\mathrm{TDS}(s_{j+1}, s_j) :=\mathcal{T} e^{-i \int_{s_j}^{s_{j+1}} \mathcal{H}_\mathrm{I}(s') \mathrm{d} s'} $, by the truncated Dyson series~\cite{low2018hamiltonian}.
The total approximated time evolution $U_\mathcal{H}(S, 0)$ is given as
\begin{align}
    &U_\mathcal{H}(S, 0) \nonumber \\
    &\approx e^{-i \frac{1}{\nu} (e^{\nu S}-1)\hat{\mathcal{F}}} U_\mathrm{TDS}(s_{N_\mathrm{step}}, s_{N_\mathrm{step} - 1}) \cdots U_\mathrm{TDS}(s_1, s_0).
\end{align}

In the simple product formula, we approximate the short-time evolution of the Hamiltonian $\mathcal{H}(s)$~\cite{leng2025quantum} as
\begin{align}
    U_\mathcal{H}(s_{j+1}, s_j)
    &\approx e^{- i \frac{1}{2} e^{-\nu s_j} \Lambda^{(M)} \Delta s} e^{ - i e^{\nu s_j} \hat{\mathcal{F}} \Delta s}  \nonumber \\
    &=: U_\mathrm{PF}(s_{j+1}, s_j).
\end{align}
Then, we approximate the time evolution $U_\mathcal{H}(S, 0)$, as
\begin{align}
    U_\mathcal{H}(S, 0) \approx U_\mathrm{PF}(s_{N_\mathrm{step}}, s_{N_\mathrm{step} - 1}) \cdots U_\mathrm{PF}(s_1, s_0).
\end{align}

In either case, we need to implement the time evolution of the objective function $e^{-i \hat{\mathcal{F}} s}$ for an arbitrary time increment $s$. 
In the interaction picture simulation, we need to implement the Hamiltonian by some means; here we employ the block-encoding of $\mathcal{H}_I(s)$, as an access model. 
For the product formula, we need the time evolution of the discrete Laplace operator $\Lambda^{(M)}$, which is given in Appendix~\ref{sec:BE_PDE}.

Our complexity analysis focuses on the interaction picture simulation under the strong-convexity assumption, even though objective functions of PDE-constrained optimization are often not strongly convex.
This is because this assumption allows a rigorous end-to-end convergence for the QHD dynamics. 
This assumption should not be interpreted as making the underlying PDE-constrained optimization problem easy since the dominant computational burden remains the repeated solution of large forward and adjoint PDE problems.
When strong convexity is absent, such convergence guarantees can disappear, and the practical value of executing a highly accurate time-dependent Hamiltonian simulation may diminish. 
This motivates a more heuristic stance: rather than expending resources to suppress simulation error, we are interested in whether a coarse but simple time evolution is sufficient for the optimizer to provide useful optimized solutions. 
This is the reason why we consider the simple product formula as a minimal implementation, which is numerically examined later.
In short, we use the truncated Dyson series to derive explicit complexity bounds for the strongly convex case, and we use a product formula as a simple baseline whose practical value we confirm by experiment.

\subsection{Design-dependent forward simulation and block-encoding of the objective function} 
\label{sec:block-encoding}

As mentioned in the previous subsection, our primal problem is how to implement the time evolution of the objective function $e^{-i \hat{\mathcal{F}} s}$, where again $\hat{\mathcal{F}} = \sum_{\xi} \mathcal{F}(u(\xi))\ketbra{\xi}$. 
Here, we construct a block-encoding of the objective function $\mathcal{F}(u(\xi))$ for implementing its time evolution.
Our goal is to implement a unitary $U_\mathrm{obj}$ which satisfies
\begin{align} \label{eq:BE_obj}
    &(I^{\otimes Mm} \otimes \bra{0}^{\otimes a_\mathrm{obj}}) U_\mathrm{obj} (I^{\otimes Mm} \otimes \ket{0}^{\otimes a_\mathrm{obj}}) \nonumber \\
    &= \frac{1}{\alpha_\mathrm{obj}}
    \left(\hat{\mathcal{F}} - \delta_\mathrm{obj} I^{\otimes Mm} \right),
\end{align}
where $a_\mathrm{obj}$ is the number of ancillary qubits for the block-encoding, $\alpha_\mathrm{obj}$ is a normalization constant so that the objective function can be encoded into a unitary, and $\delta_\mathrm{obj}$ is an offset.

As a first step, we lift the parameterized PDE family to a single extended forward simulation problem on the joint design-state Hilbert space. This design-coherent formulation is the starting point that makes the PDE solve coherently in the design variable and therefore usable for the subsequent oracle construction.
Specifically, we aim at implementing a design-dependent forward simulation oracle $U_\mathrm{for}$ that satisfies
\begin{align} \label{eq:U_for}
    & U_\mathrm{for} \ket{\xi} \! \ket{0}^{\otimes n} \! \ket{0}^{\otimes a_\mathrm{for}} \nonumber \\
    &= \frac{\| u (t; \xi) \|}{\alpha_\mathrm{for}} \ket{\xi} \! \ket{u (t; \xi)} \! \ket{0}^{\otimes a_\mathrm{for}} + \sqrt{1 - \frac{\| u (t; \xi) \|^2 }{\alpha_\mathrm{for}^2}} \ket{\xi} \! \ket{\chi},
\end{align}
as a direct extension of Eq.~\eqref{eq:linear solution} so that the space of design parameters is included in the dynamics.
Here, $\ket{u (t; \xi)} := (1 / \| u (t; \xi) \|) \sum_{j=0}^{2^n-1} u_j (t; \xi) \ket{j}$ is the normalized solution to the linear equation \eqref{eq:ODECO}, 
represented in terms of the computational basis $\{\ket{j}\}$, $\ket{\chi}$ is a state orthogonal to the state $\ket{u (t; \xi)} \! \ket{0}^{\otimes a_\mathrm{for}}$, $a_\mathrm{for}$ is the number of ancillary qubits, $n := \log_2 N$, and $\alpha_\mathrm{for}$ is a constant for normalization. 
We confirm $\alpha_\mathrm{for} = \alpha_\mathrm{LCHS}$ and 
$a_\mathrm{for} = a_\mathrm{LCHS} + \lceil \log_2(Mm + 1) \rceil$ as follows.
We consider the following ODE for an extended state variable $\tilde{u}(t) \in \mathbb{C}^{2^{(Mm+n)}}$:
\begin{align}
    \difft{\tilde{u} (t)} = - \tilde{A}(\xi) \tilde{u} (t),
\end{align}
where $\tilde{A}(\xi) := \sum_\xi \left( \ket{\xi} \! \bra{\xi} \otimes A(\xi) \right)$. This dynamics evolves an initial state $\ket{\tilde{u}(0)} := \ket{\psi} \! \ket{u_0}$ with arbitrary quantum state $\ket{\psi} = \sum_\xi \psi_\xi \ket{\xi} \in \mathbb{C}^{2^{Mm}}$.
Since the analytic form of the solution for this ODE is given as
\begin{align}
    &\tilde{u} (t) \nonumber \\
    &= e^{- \tilde{A}(\xi) t} \tilde{u} (0) \nonumber \\
    &= \prod_{\xi} \left( \ket{\xi} \! \bra{\xi} \otimes e^{-A(\xi) t} + \left( I^{\otimes Mm} - \ket{\xi} \! \bra{\xi} \right) \otimes I^{\otimes n} \right) \tilde{u}(0) \nonumber \\
    &= \sum_\xi \psi_\xi \ket{\xi} \otimes \| u (t; \xi) \| \ket{u(t; \xi)},
\end{align}
the forward simulation oracle can be implemented using the LCHS oracle $U_\mathrm{LCHS}(\tilde{A})$ with a state preparation oracle $U_{u_0}$ acting as $\ket{u_0} = U_{u_0} \ket{0}^{\otimes n}$; that is, $U_\mathrm{for} = U_\mathrm{LCHS} (\tilde{A}) (I^{\otimes Mm} \otimes U_{u_0} \otimes I^{\otimes a_\mathrm{for}})$.
To implement the LCHS oracle $U_\mathrm{LCHS} (\tilde{A})$ for the design-dependent matrix $\tilde{A}$, we use the following relation:
\begin{align}
    \tilde{A} &= \sum_\xi \left( \ket{\xi} \! \bra{\xi} \otimes A(\xi) \right) \nonumber \\
    &= I^{\otimes Mm} \otimes A_0 \nonumber \\
    &\quad + \sum_{\xi} \left( \ket{\xi} \! \bra{\xi} \otimes \left( \sum_{\mu=1}^M A_\mu\sum_{b=1}^{m} h_\xi 2^{b-1} \xi_\mu^{(b)} \right) \right) \nonumber \\
    &= I^{\otimes Mm} \otimes A_0 + \sum_{\mu=1}^M \sum_{b=1}^{m} h_\xi 2^{b-1}  \left( \ket{1} \! \bra{1}_{\mu, b} \otimes A_\mu \right), \label{eq:A_decomp}
\end{align}
where $\ket{1} \! \bra{1}_{\mu, b}$ acts on the $b$-th qubit for the $\mu$-th design variable.
Thus, a block-encoding of $\tilde{A}$ can be obtained via linear combination of block-encoded matrices of ${A_\mu}$ \cite[Lemma~52]{gilyen2019quantum} for $Mm + 1$ terms in the above equation.
To implement the LCU for $Mm + 1$ terms, we require $\lceil \log_2(Mm + 1) \rceil$ ancilla qubits.
Therefore, the total number of ancillary qubits for the extended forward simulation is given by the sum of those required for LCHS and those required to construct $\tilde{A}$, resulting in $a_\mathrm{for} = a_\mathrm{LCHS} + \lceil \log_2(Mm + 1) \rceil$.
Note that the ancilla qubits required for block-encoding each $A_\mu$, which depend on the specific PDE under consideration, are not included here.
Based on Eqs.~\eqref{eq:LCHS} and \eqref{eq:U_for}, we also obtain the relation of $\alpha_\mathrm{for} = \alpha_\mathrm{LCHS}$.

\begin{figure*}[t]
    \centering
    \begin{quantikz}[transparent]
        & \qw \qwbundle{Mm} & \qw & \gate[wires=3]{U_\mathrm{LCHS}(\tilde{A})} & \qw & \gate[wires=3]{U_\mathrm{LCHS}(\tilde{A})^\dagger} & \qw & \qw \\
        \lstick[1]{$\ket{0}^{\otimes n}$} & \qw \qwbundle{n} & \gate[wires=1]{U_{u_0}} & \qw & \gate[wires=2]{R_P} & \qw & \gate[wires=1]{U_{u_0}^\dagger} & \rstick[1]{$\bra{0}^{\otimes n}$} \\
        \lstick[1]{$\ket{0}^{\otimes a_\mathrm{for}}$} & \qw \qwbundle{a_\mathrm{for}} & \qw & \qw & \qw & \qw & \qw & \rstick[1]{$\bra{0}^{\otimes a_\mathrm{for}}$}
    \end{quantikz}
    \caption{Quantum circuit for block-encoding of $\mathcal{F}_\mathrm{quad}$.} \label{fig:qc_quad}
\end{figure*}
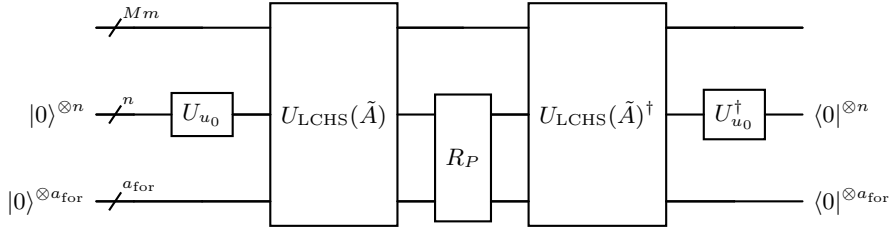

\begin{figure*}[t]
    \centering
    \begin{quantikz}[transparent, column sep = 0.6em]
       & \qw \qwbundle{Mm} & \qw & \qw & \gate[wires=3]{U_\mathrm{LCHS}(\tilde{A})} & \qw & \qw & \qw & \qw & \qw & \gate[wires=3]{U_\mathrm{LCHS}(\tilde{A})^\dagger} & \qw & \qw & \qw & \qw \\
      \lstick[1]{$\ket{0}^{\otimes n}$} & \qw \qwbundle{n} & \gate[wires=1]{U_\mathrm{ref}} & \gate[wires=1]{U_{u_0}} & \qw & \qw & \qw & \gate[wires=3]{R'_P} & \qw & \qw & \qw & \gate[wires=1]{U_{u_0}^\dagger} & \gate[wires=1]{U_\mathrm{ref}^\dagger} & \qw & \rstick[1]{$\bra{0}^{\otimes n}$}\\
      \lstick[1]{$\ket{0}^{\otimes a_\mathrm{for}}$} & \qw \qwbundle{a_\mathrm{for}} & \qw & \qw & \qw & \qw & \qw & \qw & \qw & \qw & \qw & \qw & \qw & \qw & \rstick[1]{$\bra{0}^{\otimes a_\mathrm{for}}$}\\
      \lstick[1]{$\ket{0}$} & \gate{U_\mathrm{PREP}} & \ctrl{-2} & \octrl{-2} & \octrl{-1} & \gate{Z} & \gate{U_\mathrm{PREP}^\dagger} & \qw & \gate{U_\mathrm{PREP}} & \gate{Z} & \octrl{-1} & \octrl{-2} & \ctrl{-2} & \gate{U_\mathrm{PREP}^\dagger} & \rstick[1]{$\bra{0}$}
    \end{quantikz}
    \caption{Quantum circuit for block-encoding of $\mathcal{F}_\mathrm{err}$.} \label{fig:qc_err}
\end{figure*}
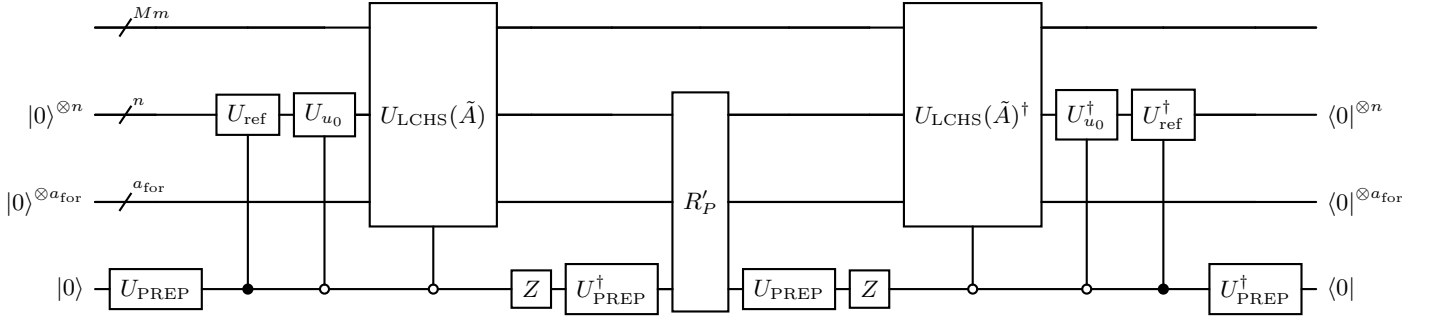

We are now ready to construct block-encodings of the objective functions given by Eqs. \eqref{Eq:cost1} and \eqref{Eq:cost2}. 
Recall that both these quantities are defined in terms of the projector $P = \sum_{j \in \mathcal{J}} \ket{j} \! \bra{j}$. 
Then, by applying the reflection $R_P := 2 P \otimes \ket{0} \! \bra{0}^{\otimes a_\mathrm{for}} - I$, where $\ket{0} \! \bra{0}^{\otimes a_\mathrm{for}}$ is for the success flag of LCHS, we have 
\begin{align}
    &(I^{\otimes Mm} \otimes R_P) U_\mathrm{for} \ket{\xi} \! \ket{0}^{\otimes n} \! \ket{0}^{\otimes a_\mathrm{for}} \nonumber \\
    &= \frac{\| u (t; \xi) \|}{\alpha_\mathrm{for}} \ket{\xi} (2P - I^{\otimes n}) \ket{u (t; \xi)} \! \ket{0}^{\otimes a_\mathrm{for}} \nonumber \\
    & \quad - \sqrt{1 - \frac{\| u (t; \xi) \|^2 }{\alpha_\mathrm{for}^2}} \ket{\xi} \! \ket{\chi}.
\end{align}
Using this equation, we can construct a unitary $U_\mathrm{obj, quad} := U_\mathrm{for}^\dagger (I^{\otimes Mm} \otimes R_P) U_\mathrm{for}$ which is the $(\alpha_\mathrm{for}^2 \gamma_u^{-2} / 2, n+a_\mathrm{for}, 2 \alpha_\mathrm{for} \gamma_u^{-2} \varepsilon)$-block-encoding of the objective function $\mathcal{F}_\mathrm{quad}$ as follows:
\begin{widetext}
\begin{align}
    \label{eq:BE_quad}
    &\left( I^{\otimes Mm} \otimes \bra{0}^{\otimes (n+a_\mathrm{for})} \right) U_\mathrm{for}^\dagger R_P U_\mathrm{for} \left( I^{\otimes Mm} \otimes \ket{0}^{\otimes (n + a_\mathrm{for})} \right) \nonumber \\
    &=\sum_{\xi, \xi'} \left( \ket{\xi'} \! \bra{\xi'} \otimes \bra{0}^{\otimes (n+a_\mathrm{for})} \right) U_\mathrm{for}^\dagger R_P U_\mathrm{for} \left( \ket{\xi} \! \bra{\xi} \otimes \ket{0}^{\otimes (n + a_\mathrm{for})} \right) \nonumber \\
    &=\sum_{\xi} \left( \frac{\| u (t; \xi) \|^2}{\alpha_\mathrm{for}^2} \bra{u (t; \xi)} (2P - I^{\otimes n}) \ket{u (t; \xi)} - 1 + \frac{\| u (t; \xi) \|^2 }{\alpha_\mathrm{for}^2} \right) \ket{\xi} \! \bra{\xi}  \nonumber \\
    &=\sum_{\xi} \left( \frac{2 \| u (t; \xi) \|^2}{\alpha_\mathrm{for}^2} \bra{u (t; \xi)} P \ket{u (t; \xi)} - 1 \right) \ket{\xi} \! \bra{\xi}  \nonumber \\
    &= \frac{2 \gamma_u^2 \hat{\mathcal{F}}_\mathrm{quad}(u(\xi))}{ \alpha_\mathrm{for}^2} - 1,
\end{align}
\end{widetext}
where we regard $\ket{0}^{\otimes n}$ as ancilla qubits and the error $\varepsilon$ comes from the error of LCHS.
We use the formula~\cite[Lemma~53]{gilyen2019quantum} to derive the error of the block-encoding $2 \alpha_\mathrm{for} \gamma_u^{-2} \varepsilon$.
Figure~\ref{fig:qc_quad} illustrates the corresponding quantum circuit.
This block-encoding corresponds to our goal in Eq.~\eqref{eq:BE_obj} with $\alpha_\mathrm{obj} = \delta_\mathrm{obj} = \alpha_\mathrm{for}^2 \gamma_u^{-2} / 2$.

Next, for encoding $\mathcal{F}_\mathrm{err}$, we assume a reference state preparation oracle $U_\mathrm{ref}$ that prepares 
\begin{align}
    U_\mathrm{ref} \ket{0}^{\otimes n} = \ket{u_\mathrm{ref}},
\end{align}
where $\ket{u_\mathrm{ref}} = u_\mathrm{ref} / \| u_\mathrm{ref} \|$ is a normalized reference state that can be encoded into a quantum state.
Then, we consider LCU of $\alpha_\mathrm{for} U_\mathrm{for} - \| u_\mathrm{ref} \| U_\mathrm{ref}$ by a preparation oracle $U_\mathrm{PREP}$ acting as
\begin{align}
    U_\mathrm{PREP} \ket{0} &= \frac{1}{\sqrt{\alpha_\mathrm{for} + \| u_\mathrm{ref} \|}} \left( \sqrt{\alpha_\mathrm{for}} \ket{0} + \sqrt{\| u_\mathrm{ref} \|} \ket{1} \right), \label{eq:LCU_PREP}
\end{align}
and a select oracle $U_\mathrm{SEL}$ defined as
\begin{align}
    U_\mathrm{SEL} &:=  U_\mathrm{for} \otimes \ket{0} \! \bra{0} + I^{\otimes Mm} \otimes U_\mathrm{ref} \otimes I^{\otimes a_\mathrm{for}} \otimes \ket{1} \! \bra{1},
\end{align}
which prepares the quantity $u(t; \xi) - u_\mathrm{ref}$ as follows.
\begin{widetext}
\begin{align}
    &(I^{\otimes (Mm + n + a_\mathrm{for})} \otimes U_\mathrm{PREP}^\dagger )(I^{\otimes (Mm + n + a_\mathrm{for})} \otimes Z) U_\mathrm{SEL} (I^{\otimes (Mm + n + a_\mathrm{for})} \otimes U_\mathrm{PREP}) \ket{\xi} \! \ket{0}^{\otimes (n + a_\mathrm{for} + 1)} \nonumber \\
    &= \left( \frac{1}{\alpha_\mathrm{LCU}} \left( \alpha_\mathrm{for} U_\mathrm{for} - \| u_\mathrm{ref} \| I^{\otimes Mm} \otimes U_\mathrm{ref} \otimes I^{\otimes a_\mathrm{for}} \right) \ket{\xi} \! \ket{0}^{\otimes (n + a_\mathrm{for})} \right) \ket{0} + \ket{\perp} \nonumber \\
    &= \frac{1}{\alpha_\mathrm{LCU}} \ket{\xi} \! \left( \| u (t; \xi) \| \ket{u (t; \xi)} - \| u_\mathrm{ref} \| \ket{u_\mathrm{ref}} \right) \! \ket{0}^{\otimes a_\mathrm{for} + 1} \nonumber \\
    & \quad + \ket{\perp}, \label{eq:LCU}
\end{align}
\end{widetext}
where $\alpha_\mathrm{LCU} = \alpha_\mathrm{for} + \| u_\mathrm{ref} \|$.
Note that the preparation oracle in Eq.~\eqref{eq:LCU_PREP} requires the norm of the reference state $\| u_\mathrm{ref} \|$, which we assume is computed classically, since $u_\mathrm{ref}$ is given as classical data.
With this unitary denoted by $U_\mathrm{LCU} := (I \otimes U_\mathrm{PREP}^\dagger) (I \otimes Z) U_\mathrm{SEL} (I \otimes U_\mathrm{PREP})$, and a reflection operator $R'_P := 2 P \otimes \ket{0} \! \bra{0}^{\otimes (a_\mathrm{for} + 1)} - I^{\otimes (n + a_\mathrm{for} + 1)}$, we construct a unitary $U_\mathrm{obj, err} := U_\mathrm{LCU}^\dagger (I^{\otimes Mm} \otimes R'_P) U_\mathrm{LCU}$ which is the $(\alpha_\mathrm{LCU}^2 \gamma_u^{-2} / 2, n+a_\mathrm{for}+1, 2 \alpha_\mathrm{LCU} \gamma_u^{-2} \varepsilon)$-block-encoding of the objective function $\mathcal{F}_\mathrm{err}$, as follows:
\begin{widetext}
\begin{align}
    &\left( I^{\otimes Mm} \otimes \bra{0}^{\otimes (n + a_\mathrm{for} + 1)} \right) U_\mathrm{LCU}^\dagger (I^{\otimes Mm} \otimes R'_P) U_\mathrm{LCU} \left( I^{\otimes Mm} \otimes \ket{0}^{\otimes (n + a_\mathrm{for} + 1)} \right) \nonumber \\
    &=\sum_{\xi, \xi'} \left( \ket{\xi'} \! \bra{\xi'} \otimes \bra{0}^{\otimes (n + a_\mathrm{for} + 1)} \right)  U_\mathrm{LCU}^\dagger (I^{\otimes Mm} \otimes R'_P) U_\mathrm{LCU} \left( \ket{\xi} \! \bra{\xi} \otimes \ket{0}^{\otimes (n + a_\mathrm{for} + 1)} \right) \nonumber \\
    &= \sum_{\xi, \xi'} \left( \frac{2 \left( u (t; \xi) - u_\mathrm{ref} \right)^\dagger P \left( u (t; \xi) - u_\mathrm{ref} \right)}{\alpha_\mathrm{LCU}^2 } - 1 \right) \ket{\xi} \! \bra{\xi} \nonumber \\
    &= \frac{2 \gamma_u^2 \hat{\mathcal{F}}_\mathrm{err}(\xi)}{\alpha_\mathrm{LCU}^2 } - 1,
\end{align}
\end{widetext}
where we use the formula~\cite[Lemma~53]{gilyen2019quantum} to derive the error of the block-encoding $2 \alpha_\mathrm{LCU} \gamma_u^{-2} \varepsilon$.
Figure~\ref{fig:qc_err} illustrates the corresponding quantum circuit.
This block-encoding corresponds to that in Eq.~\eqref{eq:BE_obj} with $\alpha_\mathrm{obj} = \delta_\mathrm{obj} = \alpha_\mathrm{LCU}^2 \gamma_u^{-2} / 2$.
Now, by QSVT combined with oblivious amplitude amplification (OAA)~\cite{martyn2021grand}, we obtain the time evolution $e^{-i \hat{\mathcal{F}} t}$ for both cases $\hat{\mathcal{F}}=\hat{\mathcal{F}}_{\rm quad}$ and $\hat{\mathcal{F}}= \hat{\mathcal{F}}_{\rm err}$.

\subsection{Improving subnormalization of the block-encodings}
\label{sec:usva}
Recall that we define the normalized quadratic objective function as
\begin{align}
    \mathcal{F}_{\mathrm{quad}}(u(\xi)) &:= \frac{1}{\gamma_u^2} u^\dagger(t;\xi) P u(t;\xi),
\end{align}
where $\gamma_u$ is a design-independent upper bound satisfying
\begin{align}
    \gamma_u \ge \max_{\xi} \|u(t;\xi)\|.
\end{align}
Then, the baseline construction in Eq.~\eqref{eq:BE_quad} shows that the block-encoding normalization of $\mathcal{F}_{\mathrm{quad}}$ includes the inverse of the upper bound of the success probability of the forward simulation.
This can be unfavorable when the solution norm becomes small, for example, in highly dissipative systems.

A natural idea is to apply amplitude amplification to the LCHS-based
forward simulation. 
However, ordinary amplitude amplification is not directly suitable in our setting.
Indeed, for an input design state $\ket{\psi} = \sum_{\xi} \psi_{\xi} \ket{\xi}$, the success probability of the extended forward simulation depends on
\begin{align}
    \|\tilde{u}(t)\|^2 = \sum_{\xi} |\psi_{\xi}|^2 \|u(t;\xi)\|^2,
\end{align}
which is not assumed to be known a priori during the coherent QHD dynamics.
Therefore, instead of ordinary amplitude amplification, we use the uniform singular value amplification (USVA)~\cite[Theorem~30]{gilyen2019quantum}, applied directly to the extended forward simulation oracle.

To describe this, define the input and output projectors
\begin{align}
    \Pi_{\mathrm{in}}
    &:=
    I^{\otimes Mm}\otimes
    |0\rangle\langle 0|^{\otimes (n+a_{\mathrm{for}})},
    \\
    \Pi_{\mathrm{out}}
    &:=
    I^{\otimes (Mm+n)}\otimes
    |0\rangle\langle 0|^{\otimes a_{\mathrm{for}}}.
\end{align}
From Eq.~\eqref{eq:U_for}, the nonzero singular values of the projected operator
$\Pi_{\mathrm{out}} U_{\mathrm{for}} \Pi_{\mathrm{in}}$
are given by
\begin{align}
    \sigma_{\xi}
    =
    \frac{\|u(t;\xi)\|}{\alpha_{\mathrm{for}}}
    =\frac{\|u(t;\xi)\|}{\alpha_{\mathrm{LCHS}}}.
\end{align}
Let $\delta \in (0,1)$ and define
\begin{align}
    \beta_u := \frac{\gamma_u}{1-\delta}.
\end{align}
Then, since $\|u(t;\xi)\|\le \gamma_u$, we have
\begin{align}
    \sigma_{\xi}
    \le
    \frac{\gamma_u}{\alpha_{\mathrm{for}}}
    =
    \frac{(1-\delta)\beta_u}{\alpha_{\mathrm{for}}}.
\end{align}
Hence, by applying USVA~\cite[Theorem~30]{gilyen2019quantum} to the linear transformation
\begin{align}
    x \mapsto \frac{\alpha_{\mathrm{for}}}{\beta_u} x
\end{align}
on the interval $|x|\le \gamma_u/\alpha_{\mathrm{for}}$,
we obtain a unitary $U_{\mathrm{for}}^{(\mathrm{USVA})}$ using
\begin{align}
    \mathcal{O}\!\left(
        \frac{\alpha_{\mathrm{for}}}{\beta_u \delta}
        \log\!\left(
            \frac{\alpha_{\mathrm{for}}}{\beta_u \varepsilon_{\mathrm{USVA}}}
        \right)
    \right)
\end{align}
queries to $U_{\mathrm{for}}$ and $U_{\mathrm{for}}^\dagger$, and a single ancilla qubit, such that
\begin{align}
    & U_{\mathrm{for}}^{(\mathrm{USVA})}
    \ket{\xi} \! \ket{0}^{\otimes n} \! \ket{0}^{\otimes a_{\mathrm{for}}} \! \ket{0} \nonumber \\
    &= \frac{\|u(t;\xi)\|}{\beta_u}
    \ket{\xi} \! \ket{u(t;\xi)} \! \ket{0}^{\otimes a_{\mathrm{for}}}  \! \ket{0}\nonumber \\
    & + \sqrt{1 - \frac{\|u(t;\xi)\|^2}{\beta_u^2}} \ket{\xi} \! \ket{\chi_\mathrm{USVA}},
\end{align}
in the sense of $(\varepsilon + \varepsilon_{\mathrm{USVA}})$ approximation where $\ket{\chi_\mathrm{USVA}}$ is orthogonal to $\ket{u(t;\xi)} \! \ket{0}^{\otimes (a_{\mathrm{for}}+1)}$.

Using this unitary $U_\mathrm{for}^{(\mathrm{USVA})}$, we define
\begin{align}
    U_{\mathrm{obj,quad}}^{(\mathrm{USVA})} :=\left(U_{\mathrm{for}}^{(\mathrm{USVA})} \right)^\dagger
    \left(I^{\otimes Mm}\otimes R'_P\right)
    U_{\mathrm{for}}^{(\mathrm{USVA})},
\end{align}
which gives $((1 - \delta)^{-2} / 2, n + a_{\mathrm{for}} + 1, 4 \gamma_u (1 - \delta)^{-1} \varepsilon)$ block-encoding of $\hat{\mathcal{F}}_{\mathrm{quad}}$ with an offset $\delta_\mathrm{obj} = (1 - \delta)^{-2} / 2$ where we set $\varepsilon_\mathrm{USVA} = \varepsilon$.
Thus, the subnormalization constant is changed from $\alpha_\mathrm{for}^{2} \gamma_u^{-2} / 2 $ of the baseline construction to $(1-\delta)^{-2} / 2$ using
$\mathcal{O}\!\left(
        \frac{\alpha_{\mathrm{for}}(1-\delta)}{\gamma_u \delta}
        \log\!\left(
            \frac{\alpha_{\mathrm{for}}(1-\delta)}{\gamma_u \varepsilon_{\mathrm{USVA}}}
        \right)
    \right)$
additional query of $U_{\mathrm{for}}$ and $U_{\mathrm{for}}^\dagger$.
Hence, by choosing $\delta \in \Theta(1)$, more specifically, $\delta = 1/2$, since $\alpha_\mathrm{for} = \alpha_\mathrm{LCHS} \in \mathcal{O}(1)$ holds~\cite{an2023quantum}, this USVA-based approach exhibits a nearly quadratic improvement regarding the number of queries of $U_{\mathrm{for}}$.
We emphasize that USVA does not eliminate the intrinsic dependence
on the scale of the forward solution.
Its role here is to avoid an additional quadratic normalization
penalty when the forward simulation output is converted into the
quadratic objective block-encoding.
Specifically, the relevant norm-dependent overhead is improved from
approximately $\mathcal{O}(\gamma_u^{-2})$ to $\mathcal{O}(\gamma_u^{-1})$ up to
polylogarithmic factors.
Consequently, if the relevant forward solution scale becomes
exponentially small, the resulting end-to-end algorithm can still be inefficient.

Note that the USVA must be performed at the extended forward simulation.
Once we obtain the $(\alpha_\mathrm{for}^2 \gamma_u^{-2} / 2 , n + a_\mathrm{for}, 2 \alpha_\mathrm{for} \varepsilon)$-block-encoding of the objective function, the USVA requires $\mathcal{O}(\gamma_u^{-2})$ queries to the block-encoding, which brings no improvement.

This USVA-based construction also works for encoding $\mathcal{F}_\mathrm{err}$.
By replacing $U_\mathrm{for}$ with $U_\mathrm{for}^{(\mathrm{USVA})}$, the coefficient $\alpha_\mathrm{for}$ in LCU changes to $\gamma_u / (1 - \delta)$.
Consequently, the subnormalization factors changes from $\alpha_{\mathrm{obj,err}}^{(\mathrm{base})} = (\alpha_\mathrm{for} + \| u_\mathrm{ref} \|)^2 \gamma_u^{-2} / 2$ to $\alpha_{\mathrm{obj,err}}^{(\mathrm{USVA})} = (\gamma_u / (1 - \delta) + \| u_\mathrm{ref} \|)^2 \gamma_u^{-2} / 2$.
Given that $\| u_\mathrm{ref} \|$ scales with $\gamma_u$ and $\alpha_\mathrm{for} \in \mathcal{O}(1)$, the subnormalization improves from $\alpha_{\mathrm{obj,err}}^{(\mathrm{base})} \in \mathcal{O}(\gamma_u^{-2})$ to $\alpha_{\mathrm{obj,err}}^{(\mathrm{USVA})} \in \mathcal{O}(1)$.

\subsection{Complexity analysis} \label{sec:complexity}

In this subsection, we derive the query and gate complexities of the block-encoding of the objective function, followed by the total gate complexity of our algorithm. 
Then, we compare it with that of a classical counterpart.
We also discuss the space complexity of our algorithm.

\subsubsection{Gate complexity of block-encoding} \label{sec:complexity_obj}

Here, we derive the gate complexity for the block-encoding of the objective functions $\hat{\mathcal{F}}$.
First, we derive the query complexity of the objective function to the coefficient matrices $A_\mu$.
Since the block-encoding of the objective function requires a single use of $U_\mathrm{for}^{(\mathrm{USVA})}$ and $(U_\mathrm{for}^{(\mathrm{USVA})})^\dagger$ for $\mathcal{F}_\mathrm{quad}$, or their controlled version for $\mathcal{F}_\mathrm{err}$, and the unitary $U_\mathrm{for}^{(\mathrm{USVA})}$ requires $\mathcal{O}(\gamma_u^{-1} \log\!\left( \gamma_u^{-1} \varepsilon^{-1} \right))$ queries to $U_\mathrm{LCHS}(\tilde{A})$ with $\delta = 1/2$ and $\alpha_\mathrm{for} \in \mathcal{O}(1)$, it suffices to evaluate the query complexity to $A_\mu$ for implementing $U_\mathrm{LCHS}(\tilde{A})$.

Based on the previous work~\cite{an2023quantum}, $U_\mathrm{LCHS}(\tilde{A})$ without amplitude amplification requires $\tilde{\mathcal{O}}(\tilde{\alpha}_A T \log (1/\varepsilon)^{1/\beta})$ queries to $\tilde{A}$ where $\tilde{\alpha}_A \geq \| \tilde{A} \|$.
Note that our method does not include amplitude amplification because the objective function oracle is used coherently inside the QHD dynamics, and amplitude amplification would require prior knowledge of the success probability of the extended forward simulation, or require amplitude estimation with intermediate measurements.
Instead, as discussed in Sec.~\ref{sec:usva}, we introduce a uniform singular value amplification (USVA), applied directly to the extended forward simulation oracle.
Since Eq.~\eqref{eq:A_decomp} means that $\| \tilde{A} \| \leq (M+1) \max(\| A_0 \|, \max_\mu \| A_\mu \|)$, the query complexity of $U_\mathrm{LCHS}(\tilde{A})$ results in $\tilde{\mathcal{O}}( \alpha_A M T \log (1/\varepsilon)^{1/\beta})$ with $\alpha_A \geq \max (\| A_0 \|, \max_\mu \| A_\mu \|)$, which is the query complexity for the block-encoding of the objective functions.

When we assume that PDEs are defined on the $d$-dimensional space and use $n_x := n / d$ qubits for each dimension, the block-encoding of a finite difference operator defined in Appendix~\ref{sec:BE_PDE} requires $\mathcal{O}(d n_x^2)$ single and two-qubit gates, and its norm is $\mathcal{O}(d/h_x^\eta)$ with $h_x$ the interval of spatial discretization, and $\eta$ the order of spatial derivative in the coefficient matrices, as discussed in Appendix~\ref{sec:BE_PDE}.
Note that $\eta$ can differ from the order of spatial derivative in the original PDE depending on how PDE is transformed into a first-order ODE in time.
Assuming that coefficients of PDEs can be block-encoded by $\mathcal{O}(n)=\mathcal{O}(d n_x)$ gates and the dominant coefficient of the PDE is $\alpha_\mathrm{PDE}$ (see Appendix~\ref{sec:BE_C}), we can obtain the block-encoding of coefficient matrices with $\alpha_A \in \mathcal{O}(\alpha_\mathrm{PDE} d/h_x^\eta)$ using $\mathcal{O}(d n_x^2 + d n_x)$ gates.
Furthermore, the block-encoding of $\tilde{A}$ in Eq.~\eqref{eq:A_decomp} requires $Mm$ times applications of block-encodings of $A_\mu$.

Other than queries to the coefficient matrices, the block-encoding of the objective function includes 
\begin{enumerate}
    \item $\mathcal{O}(\gamma_u^{-1} \log\!\left( \gamma_u^{-1} \varepsilon^{-1} \right))$ queries to the state preparation oracle of initial forward solution by unitary $U_{u_0}$ for $\ket{u_0} = U_{u_0} \ket{0}$ for the construction of $U_\mathrm{for}^{(\mathrm{USVA})}$. We assume that the initial state of forward simulation $u_0$ can be efficiently prepared so that $U_{u_0}$ can be implemented using $\mathcal{O}(\mathrm{polylog}(N))$ single and two-qubit gates.
    \item A single use of the reflection operator $R_P$ or $R_P'$. Here, we assume that the projector $P$ is represented as $P = P_n \otimes \cdots \otimes P_1$ for $P_{\cdot} \in \{ \ket{0} \! \bra{0}, \ket{1} \! \bra{1}, I \}$. Then we can implement $R_P$ using at most $(n + a_\mathrm{for})$-controlled $Z$ gate, which is decomposed into $\mathcal{O}(n + a_\mathrm{for})$ single and two-qubit gates. 
    \item  Two uses of reference state preparation by unitary $U_\mathrm{ref}$ for $\ket{u_\mathrm{ref}} = U_\mathrm{ref} \ket{0}$ when the objective function has the form of $\mathcal{F}_\mathrm{err}$. We assume that the reference state can be efficiently prepared so that $U_\mathrm{ref}$ can be implemented using $\mathcal{O}(\mathrm{polylog}(N))$ single and two-qubit gates.
\end{enumerate}
Thus, the gate complexity of the block-encoding of the objective function results in
\begin{align}
    \tilde{\mathcal{O}} \left( \left( \frac{\alpha_\mathrm{PDE} d^2(n_x^2 + n_x) M^2m T}{h_x^\eta} \log (1/\varepsilon)^{1/\beta} + \mathrm{polylog}(N) \right) \right. \nonumber \\
    \times \left.
        \frac{1}{\gamma_u}
        \log\!\left(
            \frac{ 1 }{\gamma_u \varepsilon}
        \right) \right).
\end{align}
When PDEs are nondimensionalized, which means that $h_x \sim \mathcal{O}(1 / N_x)$, we obtain the gate complexity of 
\begin{align}\label{eq:alpha-obj}
    \tilde{\mathcal{O}}\left (   
        \frac{ \alpha_\mathrm{PDE} d^2 N_x^{\eta} M^2m T }{\gamma_u} \log (1/\varepsilon)^{1+1/\beta} \right),
\end{align}
where $\tilde{\mathcal{O}}$ hinders poly-logarithmic factors in all parameters other than $\varepsilon$.

In the case of the simple implementation by Trotterization, a similar calculation derives the query complexity for the block-encoding of the objective function as $\tilde{\mathcal{O}}(\alpha_A M T^{1+1/p} / (\gamma_u \varepsilon^{1+1/p}))$ where $p$ is the order of the product formula.
This means that the gate complexity is $\tilde{\mathcal{O}}(\alpha_\mathrm{PDE} d^2 M^2 m N_x^{\eta} T^{1+1/p} /(\gamma_u \varepsilon^{1+1/p}))$.

\subsubsection{Gate complexities of our algorithm} \label{sec:gate_complexity}

In addition to the query to any of the coefficient matrices, our algorithm includes several other components.
Here, we describe these components and derive the gate complexity for our algorithm to simulate the time evolution of $\mathcal{H}(s)$ within an error $\epsilon$.
First, we define the key component $\texttt{HAM-T}_j$, which is a block-encoding of
\begin{align}
    \sum_{q=0}^{Q-1} \ket{q} \!\bra{q} \otimes \frac{1}{\alpha_\Lambda} \mathcal{H}_\mathrm{I}\left(s_j + \frac{q\Delta s}{Q} \right),
\end{align}
where $Q \in \mathcal{O}(\frac{S}{\epsilon} (\alpha_\Lambda + \alpha_\mathrm{obj} e^{\nu S} + \nu))$, and $\alpha_\Lambda = 5M / 2 h_\xi^2$, requiring $a_\mathrm{HT}= 4 + 2a_\mathrm{obj} + 2\lceil \log ((1 / \nu) (16 \alpha_\Lambda N_\mathrm{step} e^{\nu S}) / \epsilon ) \rceil + \lceil \log(5M) \rceil + \lceil \log(Q) \rceil$ ancillary qubits, as explained in Ref.~\cite{low2018hamiltonian, catli2026efficient} and Appendix~\ref{sec:BE_HAM-T}.
The components we use for our algorithm are as follows.
\begin{enumerate}
    \item A single use of state preparation of design variables by unitary $U_{\psi_0}$ for $\ket{\psi(0)} = U_{\psi_0} \ket{0}^{\otimes Mm}$. We use $Mm$ Hadamard gates to prepare the state $\ket{\psi(0)} = 2^{-Mm/2}\sum_\xi \ket{\xi}$. 
    \item $N_\mathrm{step} \in \mathcal{O}(\alpha_\Lambda S)$ segments of interaction picture simulation by truncated Dyson series. 
    As described in Ref.~\cite{catli2026efficient} , this part requires
    \begin{enumerate}
        \item A single execution of $e^{-i \frac{1}{\nu} (e^{\nu S} - 1) \hat{\mathcal{F}}}$,
        \item $\mathcal{O}(\alpha_\Lambda S \frac{\log (\alpha_\Lambda S / \epsilon)}{\log \log (\alpha_\Lambda S / \epsilon)})$ queries to $\texttt{HAM-T}_j$ oracle,
        \item $\mathcal{O} \left( \alpha_\Lambda S a_\mathrm{HT} \frac{ \log( \alpha_\Lambda S / \epsilon)}{\log \log (\alpha_\Lambda S / \epsilon)} \right)$ primitive gates.
    \end{enumerate}
\end{enumerate}
Since the number of quantum gates for the first component is negligible compared with that for the second one, we focus on the gate complexity of the second component.
The number of single and two-qubit gates required for each component (a), (b), and (c) is, respectively, 
\begin{widetext}
\begin{align}
    & \text{(a):} && \tilde{\mathcal{O}}\left ( \frac{1}{\nu \gamma_u} e^{\nu S} \alpha_\mathrm{PDE} d^2 N_x^{\eta} M^2 m T \log \left( \frac{e^{\nu S} / \nu}{\epsilon} \right)^{2 + \frac{1}{\beta}} \right), \\
    & \text{(b):} &&\tilde{\mathcal{O}} \Bigg( \Bigg( \frac{1}{\nu \gamma_u} e^{\nu S} \alpha_\mathrm{PDE} d^2 N_x^{\eta} M^2 m T \log \left( \frac{ e^{\nu S} / \nu} {h_\xi^4 \epsilon} \right)^{3+\frac{1}{\beta}} + \log \left( \frac{M/h_\xi^2 + e^{\nu S} + \nu}{\epsilon} \right) \Bigg) \frac{M S}{h_\xi^2} \log \left( \frac{1}{ \epsilon} \right) \Bigg), \\
    & \text{(c):} &&\tilde{\mathcal{O}} \left( \frac{M S}{h_\xi^2} \left(\log ( \frac{ e^{\nu S} + \nu}{\epsilon} )^2 +  \log ( \frac{ e^{\nu S} / \nu }{ \epsilon} )^2  \right) \right).
\end{align}
\end{widetext}
where we regard $\alpha_\mathrm{obj} \in \mathcal{O}(1)$, which is problem-independent, and $\tilde{\mathcal{O}}$ hinders poly-logarithmic factors in all parameters.
Obviously, the component (b) is dominant and is the total gate complexity of our algorithm.
Although it is difficult to predetermine an appropriate optimization time $S$ in general for PDE-constrained optimization, we restrict the time to $S = \log(a / \epsilon)/\nu$ with $a$ a constant determined by the expected value and variance of the initial parameter position, as well as the maximum eigenvalue of the Hessian of the objective function, which ensures that the optimized solution $\xi^\ast$ sampled from the final state satisfies $| \mathcal{F}(u(\xi^\ast)) - \min_{\xi} \mathcal{F}(u(\xi))| \leq \epsilon$ with at least constant probability when the objective function is strongly convex and smooth~\cite{catli2026efficient}.
Then, we obtain the total gate complexity of our algorithm as
\begin{align}
    \tilde{\mathcal{O}} \Bigg( \frac{\alpha_\mathrm{PDE} d^2 N_x^{\eta} M^3 T}{ \gamma_u h_\xi^2 \epsilon} \Bigg),
    \label{eq:quantumgates}
\end{align}
where we regard $\nu$ as a constant parameter.

One limitation of this analysis is that we can ensure the convergence when the objective function is strongly convex.
In general cases, the objective function can often be non-convex, where we may not need to focus on the precise simulation of the time evolution by the time-dependent Hamiltonian $\mathcal{H}(s)$.

\begin{table*}[t]
    \centering
    \begin{tabular}{c|c|c}
                        & Classical & Quantum \\
        \hline
        Gate complexity & $\tilde{\mathcal{O}} \left(
        \frac{\alpha_\mathrm{PDE}^{1+\frac{2}{p}} N_x^{ \left( \eta + \frac{2\eta }{p} + d \right)} M^{2+\frac{5}{2p}} T^{1+\frac{2}{p}} }{ \gamma_u^{\frac{1}{p}} \epsilon^{\frac{1}{2p}} }\right)$ & $\tilde{\mathcal{O}} \Bigg( \frac{\alpha_\mathrm{PDE} d^2 N_x^\eta M^3 T}{\gamma_u h_\xi^2 \epsilon} \Bigg)$ \\
        Space complexity & $\mathcal{O} \left( \frac{\alpha_\mathrm{PDE}^{1 + \frac{2}{p}} d^{1 + \frac{2}{p}} N_x^{\eta (1 + \frac{2\eta}{p}) + d} M^{1 + \frac{5}{2p}} T^{1 + \frac{2}{p}}}{\gamma_u^\frac{1}{p} \epsilon^\frac{1}{2p}} \right)$ & $M \lceil \log_2 ( 1/ h_\xi + 1) \rceil + \tilde{\mathcal{O}} \left( \eta d \log N_x +  \log \left( \frac{\alpha_\mathrm{PDE} d M T}{h_\xi \epsilon} \right) \right)$
    \end{tabular}
    \caption{Comparison of gate and space complexities between the proposed method and the conventional classical algorithm for strongly convex objective functions to achieve tolerance $\epsilon$. For parameters regarding the underlying PDE, $N_x := N^{1/d}$ is the number of degrees of freedom per spatial dimension with $N$ the total number of degrees of freedom, $d$ is the spatial dimension, $T$ is the final time, $\alpha_\mathrm{PDE}$ is the dominant coefficient, $\eta$ is the order of spatial derivative, and $\gamma_u \ge \| u(T; \xi) \|$ is an upper bound of the norm of PDE solution. For parameters regarding optimization, $M$ is the number of design variables, and $h_\xi$ is the interval of discretized design variables. $p$ denotes the order of the time integration scheme used for both the classical PDE solver and gradient evaluation. The gate complexity for the classical algorithm corresponds to the number of arithmetic operations.} \label{tbl:complexity}
\end{table*}

\subsubsection{Classical counterpart} \label{sec:classical_counterpart}
Here, we briefly discuss the computational complexity of a classical method for PDE-constrained optimization.
A detailed derivation is given in Appendix~\ref{sec:classical_adjoint}.
We consider a classical gradient-based optimizer combined with the adjoint variable method.
For a fixed design parameter $\xi$, the method first solves the forward problem
$\mathrm{d}u(t;\xi) / \mathrm{d}t =-A(\xi)u(t;\xi)$ and then solves the adjoint problem
\begin{align}
    -\difft{v(t)}+A(\xi)^\dagger v(t)=0,
\end{align}
backward in time with the terminal condition
$v(T)=-\partial \mathcal{F}/\partial u(T)$.
The gradient is then evaluated as
\begin{align}
    \frac{\partial \mathcal{F}}{\partial \xi_\mu}
    =
    2\mathrm{Re}
    \left[
        \int_0^T v^\dagger(t) A_\mu u(t) \mathrm{d}t
    \right].
\end{align}

We use the same accuracy convention as in the quantum algorithm: an $\epsilon$-accurate solution for the normalized objective corresponds to a $\gamma_u^2\epsilon$-accurate solution for the physical objective.
Under the assumption that the objective function is smooth and strongly convex, it suffices to evaluate the gradient within $\mathcal{O}(\gamma_u\sqrt{\epsilon})$ error in the Euclidean norm, and the number of optimization iterations is $\mathcal{O}(\log(1/\epsilon))$~\cite{friedlander2012hybrid,devolder2013first}.
Thus, each gradient component needs to be evaluated within additive error
\begin{align}
    \epsilon_g
    \in
    \mathcal{O}
    \left(
        \gamma_u \sqrt{\frac{\epsilon}{M}}
    \right).
\end{align}

Suppose that a $p$-th-order time integrator is used for the forward and
adjoint simulations, and that a quadrature rule of the same order is used for the time integral in the gradient formula.
The error in each gradient component, including both the quadrature error and the errors from inexact
forward and adjoint states, is bounded by
\begin{align}
    \mathcal{O}
    \left(
        \frac{\|A\|^{p+2}T^{p+2}}{N_t^p}
    \right),
\end{align}
where $N_t$ is the number of time steps.
Matching this error with $\epsilon_g$ gives
\begin{align}
    N_t
    \in
    \Theta
    \left(
        \frac{
            \|A\|^{1+\frac{2}{p}}
            M^{\frac{1}{2p}}
            T^{1+\frac{2}{p}}
        }{
            \gamma_u^{\frac{1}{p}}
            \epsilon^{\frac{1}{2p}}
        }
    \right).
\end{align}

Assuming that each $A_\mu$ is $s$-sparse with $s\in\mathcal{O}(1)$, the matrix
$A(\xi)$ is $\mathcal{O}(M)$-sparse.
Therefore, the forward and adjoint simulations, together with the evaluation
of all $M$ gradient components, require
\begin{align}
    \mathcal{O}
    \left(
        \frac{
            \|A\|^{1+\frac{2}{p}}
            N_x^d
            M^{1+\frac{1}{2p}}
            T^{1+\frac{2}{p}}
        }{
            \gamma_u^{\frac{1}{p}}
            \epsilon^{\frac{1}{2p}}
        }
    \right)
\end{align}
arithmetic operations per optimization iteration.
Using
$\|A\|\leq (M+1)\alpha_A$,
$\alpha_A=\mathcal{O}(\alpha_{\mathrm{PDE}}dN_x^{\eta})$,
and multiplying by the logarithmic number of optimization iterations, we
obtain the total classical arithmetic cost
\begin{align}
    \tilde{\mathcal{O}}
    \left(
        \frac{
            \alpha_\mathrm{PDE}^{1+\frac{2}{p}}
            d^{1+\frac{2}{p}}
            N_x^{\eta
            \left(
                1+\frac{2}{p}
            \right) + d}
            M^{2+\frac{5}{2p}}
            T^{1+\frac{2}{p}}
        }{
            \gamma_u^{\frac{1}{p}}
            \epsilon^{\frac{1}{2p}}
        }
    \right).
    \label{eq:classicalgates}
\end{align}

Therefore, by comparing Eq.~\eqref{eq:quantumgates} and
Eq.~\eqref{eq:classicalgates}, our algorithm can
potentially provide a polynomial speedup in the PDE system size $N_x$, whose classical scaling is $N_x^{\eta (1+\frac{2}{p}+\frac{d}{\eta})}$, an exponential speedup in
the spatial dimension $d$, and a polynomial speedup of degree
$1+2/p$ in the evolution time $T$.
These speedups are mainly inherited from those in the forward PDE simulation.
Also, the complexity bound of our algorithm exhibits a better dependence on the number of design variables $M$, when $p < 5/2$, so the advantage with respect to $M$ depends on the order $p$ of the time integrator of forward and adjoint simulations, as well as the order of the time integration for evaluating gradients.
Consequently, our algorithm is particularly suitable for large-scale and high-dimensional PDE-constrained optimization problems with a moderate number of design parameters.

\subsubsection{Space complexity} \label{sec:space_complexity}
The number of qubits required for our algorithm is $Mm + a_\mathrm{HT}$ where $Mm$ and $a_\mathrm{HT}$ are the number of system and ancillary qubits for $\texttt{HAM-T}_j$, respectively.
Given that the $a_\mathrm{LCHS}$ ancillary qubits for LCHS are $a_\mathrm{LCHS} = \mathcal{O}( \mathrm{log} ( \tilde{\alpha}_A T \mathrm{log} (1 / \varepsilon)^{1 + 1/ \beta} )) = \mathcal{O}( \mathrm{log} ( \alpha_\mathrm{PDE} d N_x^{\eta} M T \mathrm{log} (1 / \varepsilon)^{1 + 1/ \beta} ))$~\cite{an2023quantum}, the total number of qubits scales as
\begin{align}
    &M \lceil \log_2 ( 1/ h_\xi + 1) \rceil + \tilde{\mathcal{O}} \left( \eta d \log N_x +  \log \left( \frac{\alpha_\mathrm{PDE} d M T}{h_\xi \epsilon} \right) \right),
\end{align}
where we explicitly describe $Mm = M \lceil \log_2 ( 1/ h_\xi + 1) \rceil$ qubits for design variables.

On the other hand, in the classical adjoint-variable method, the forward
trajectory must be available when evaluating the adjoint gradient.
Thus, a straightforward implementation requires
$\mathcal{O}(d N_x N_t)$ space for storing the forward and adjoint state vectors,
$\mathcal{O}(M d N_x)$ space for the coefficient matrices $A_\mu$, and
$\mathcal{O}(M)$ space for the design variables, which results in the total space complexity as
\begin{equation}
    \mathcal{O} \left( \frac{\alpha_\mathrm{PDE}^{1 + \frac{2}{p}} d^{1 + \frac{2}{p}} N_x^{\eta (1 + \frac{2}{p}) + d} M^{1 + \frac{5}{2p}} T^{1 + \frac{2}{p}}}{\gamma_u^\frac{1}{p} \epsilon^\frac{1}{2p}} \right).
\end{equation}
Therefore, our algorithm has an exponential space advantage over conventional classical approaches with respect to $N_x$ and $T$.

\begin{table*}[t]
\centering
\caption{
Spatial discretization dependence of representative quantum
algorithms for differential equations relevant to the present work.
Here, $u_r := \| u_0 \| / \| u(T) \| \geq \gamma_u^{-1}$ and $\alpha_A \in \mathcal{O}(\alpha_\mathrm{PDE} d/h_x^\eta) = \mathcal{O}(\alpha_\mathrm{PDE} d N_x^\eta)$, where $N_x$ denotes the number of spatial degrees
of freedom per dimension. $\eta$ is the order of spatial derivative in Hamiltonian, which is unity for the wave equation. For the present work, only the dependence on PDE-side parameters is displayed.
The full end-to-end complexity including the optimizer-side parameters is given in Eq.~\eqref{eq:quantumgates} and Table~\ref{tbl:complexity}.
}
\label{tab:pde_complexity_comparison}
\begin{tabular}{llll}
\hline\hline
Problem class & Method & Query complexity & Gate complexity \\
\hline
Linear ODE solution preparation
& \begin{tabular}{l}
    LCHS~\cite{an2023linear, an2023quantum, low2025optimal} \\ 
    Schr\"{o}dingerization~\cite{jin2024quantum, jin2023aquantum, jin2026schrodingerization}
    \end{tabular}
& $\widetilde{\mathcal{O}}(\alpha_A u_r T)$ & \begin{tabular}{l}
    $\widetilde{\mathcal{O}}(\alpha_\mathrm{PDE} d^2 N_x^\eta u_r T)$ \\
    (with this work's block-encoding)
\end{tabular} \\
Wave equation solution preparation
& Costa et al.~\cite{costa2019quantum}
& N/A & $\widetilde{\mathcal{O}}(d^2 N_x T)$ \\
\hline
PDE-constrained optimization
& Present work
& $\widetilde{\mathcal{O}} \left( \frac{\alpha_A T}{\gamma_u} \right)$ & $\widetilde{\mathcal{O}} \left( \frac{\alpha_\mathrm{PDE} d^2 N_x^\eta T}{\gamma_u} \right)$ \\
\hline\hline
\end{tabular}
\end{table*}

\subsubsection{Comparison of complexities}
Here, we summarize the gate and space complexities of the proposed method compared with the conventional classical method, which are listed in Table~\ref{tbl:complexity}.
As stated in Section~\ref{sec:classical_counterpart}, our algorithm has a better bound against the conventional classical algorithm, and has the potential for providing a polynomial speedup of degree $1 + d / \eta + 2/p$ for the system size of PDEs, $N_x$, an exponential speedup for the spatial dimension of PDEs, $d$, and a polynomial speedup of degree $1+2/p$ for the evolution time of PDEs, $T$.
These speedups are attributed to those in the underlying PDE simulation.
Also, our algorithm exhibits a slightly better complexity bound on the number of design variables $M$, when $p < 5/2$.

To clarify the spatial discretization dependence, recall that $N_x:=N^{1/d}$ denote the number of degrees of freedom per
spatial direction.
For the finite difference discretization considered here,
the block-encoding normalization of the PDE operator scales as
$\alpha_A \in \mathcal{O}(\alpha_\mathrm{ PDE} d/h_x^{\eta})
=\mathcal{O}(\alpha_\mathrm{PDE} d N_x^\eta)$.
Hence, the polynomial dependence on $N_x$ in our quantum complexity is inherited from the underlying forward PDE simulation rather than introduced by the coherent interface to the optimizer.
For fixed spatial dimension $d$, such a polynomial improvement with respect to the spatial resolution is consistent with previous complexity analyses of quantum PDE algorithms. 
In particular, Montanaro and Pallister~\cite{montanaro2016quantum} showed that the quantum finite element method provides a polynomial speedup for fixed $d$, with the advantage increasing with the spatial dimension, while Costa et al.~\cite{costa2019quantum} obtained a potential polynomial speedup for the wave equation at fixed $d$ and an exponential speedup with respect to $d$. 
In a related direction, Sakamoto and Fujii~\cite{sakamoto2026quantum} showed that short-time simulation of geometrically local classical linear dynamics, a class motivated by PDE discretizations, can be dequantized and therefore does not generally yield an exponential quantum time advantage in that regime. 
Table~\ref{tab:pde_complexity_comparison} summarizes representative quantum PDE and differential-equation algorithms to clarify that the spatial resolution dependence appearing in the present end-to-end complexity is already inherited from the underlying forward PDE simulation and is consistent with representative quantum PDE algorithms under comparable discretization settings.

In addition, as described in Section~\ref{sec:space_complexity}, our algorithm has an exponential space advantage over conventional classical approaches with respect to $N_x$ and $T$.
These results suggest that the proposed framework is most favorable for problems involving large-scale and high-dimensional PDE simulations with a moderate number of design parameters, provided that the additional requirements associated with state preparation, operator access, and solution normalization do not dominate the PDE-side advantage. 
The main assumptions and parameter regimes underlying the complexity analysis and the potential quantum advantage are summarized in Table~\ref{tab:assumptions}.

Note that the above comparison is with a fully classical approach, in which both the PDE simulation and the optimization are performed classically.
Another relevant baseline for our framework is a hybrid approach that uses a quantum PDE solver for each objective evaluation, but still performs the outer optimization loop classically.
Relative to that baseline, our framework can in principle retain the optimizer-side benefits of QHD, while avoiding repeated measurement and readout of the forward-simulation output.

\begin{table*}[t]
\centering
\caption{
Main assumptions and parameter regimes underlying the complexity analysis and the potential quantum advantage of the proposed framework. The strong convexity assumptions are for ensuring the QHD convergence, and our framework can still work without strong convexity.
}
\label{tab:assumptions}
\begin{tabular}{p{0.27\textwidth}p{0.65\textwidth}}
\hline\hline
Assumption / requirement & Role in the present framework \\
\hline

Efficient access to the PDE operators
&
The matrices $A_\mu$ are assumed to admit efficient block-encodings or Hamiltonian-simulation oracles implemented by $\mathcal{O}(\mathrm{polylog}(N))$ gates.
Otherwise, the cost of constructing the PDE oracle can dominate the overall complexity.
See Appendices~\ref{sec:BE_PDE} and \ref{sec:BE_C} for the specific construction in this work.
\\

Efficient state preparation
&
The initial state $u_0$, and the reference state $u_{\rm ref}$ when required, are assumed to be efficiently preparable using $\mathcal{O}(\mathrm{polylog}(N))$ gates.
\\

Efficient objective projector
&
The projector $P$ defining the objective function is assumed to admit an efficient implementation by $\mathcal{O}(\mathrm{polylog}(N))$ gates as described in Sec.~\ref{sec:complexity_obj}.
\\

Sufficient forward-solution norm
&
The complexity retains an inverse dependence on the characteristic solution scale $\gamma_u$.
Thus, an exponentially small relevant solution norm can remove the potential quantum advantage.
\\

Optimization-side parameter dependency
&
A potential advantage requires that the polynomial overheads in
the number of design variables $M$, the design discretization
$h_\xi^{-1}$, and the target accuracy $\epsilon^{-1}$ do not
dominate the advantage inherited from the forward PDE simulation.
\\

Strong convexity and smoothness
&
These assumptions are required only for the rigorous end-to-end
QHD convergence and complexity bound in Eq.~(57).
The block-encoding construction itself does not require strong
convexity.
\\

\hline\hline
\end{tabular}
\end{table*}

\section{Applications} \label{sec:application}
In this section, we provide two applications, one of which is a parameter calibration problem in the Black-Scholes equation, and the other of which is the material parameter design problem of the wave equation.
In the following subsections, each problem and its motivation are first described, followed by numerical experiments.
For numerical experiments, we used Python and its library Qiskit 1.3~\cite{javadi2024quantum}, the open-source toolkit for quantum computation, to implement quantum circuits.
We used the statevector simulation to validate the block-encoding, while we employed the direct calculation of the matrix-vector multiplication for the time-dependent Hamiltonian simulation.
For simplicity of implementation, USVA was not implemented, and a product formula was employed for the time evolution governed by the time-dependent Hamiltonian.
For the numerical experiments, we use the heuristic scale-normalized friction parameter $\nu=\sqrt{\bar{F}}$, as discussed in Appendix~\ref{sec:friction}.
This is an empirical default used when curvature information of the objective function landscape, i.e., the lowest eigenvalue of the Hessian of the objective, is not available a priori;
when such a parameter is known, a curvature-based choice such as that in Ref.~\cite{leng2025quantum, catli2026efficient} is preferable.

\subsection{Parameter calibration in Black-Scholes equation}
\subsubsection{Motivation and formulation}
Here, we apply our framework to a parameter calibration problem in the financial option pricing problem.
A derivative option is a financial asset whose future payoffs are determined by underlying assets, such as a stock price $X$.
The present value of an option $V$ is evaluated through mathematical models, and their model parameters must be learned from observed prices in financial markets.
We demonstrate parameter calibration from a call option with the Black-Scholes model~\cite{hull2016options}, which can be described by a linear parabolic PDE, formulated as
\begin{align}
    \diff{V}{\tau} + rX \diff{V}{X} + \frac{\sigma^2 X^2}{2} \ddiff{V}{X} = rV,
\end{align}
where $\tau$ is time, $r$ is a risk-free rate, and $\sigma$ is the volatility, the standard deviation of the stock's returns.
The final and boundary conditions are determined as
\begin{align}
    V(X, T) &= \max(X - K, 0) \\
    V(0, \tau) &= 0 \\
    \diff{V(X, \tau)}{X} &= 1 \text{ as } X \rightarrow \infty,
\end{align}
where $K$ is a strike price, and $T$ is an expiration time of the option.
By applying a change of variables $X = e^x$, $u = V - e^x$, $t = T - \tau$, we transform the Black-Scholes equation into the backward PDE with the initial and boundary conditions as follows:
\begin{align}
    \diff{u}{t} &= \left( r - \frac{\sigma^2}{2} \right) \diff{u}{x} + \frac{\sigma^2}{2} \ddiff{u}{x} - ru \label{eq:BSE} \\
    u(x, 0) &= \max(e^x-K, 0) - e^x \label{eq:BSE_init} \\
    u(x, t) &= 0 \text{ as } x \rightarrow -\infty \label{eq:BSE_left} \\
    \diff{u(x, t)}{x} &= 0 \text{ as } x \rightarrow \infty, \label{eq:BSE_right}
\end{align}
which is a constant-coefficient PDE where $t$ is the time until expiration.
Here, we use the notation $\tau$ for a time and $t$ for the time until expiration to make Eq.~\eqref{eq:BSE} consistent with our formulation in Eq.~\eqref{eq:ODECO}, although the symbols $t$ and $\tau$ are conventionally defined in the opposite manner.

In this study, we consider an inverse problem to estimate the volatility $\sigma$
when the call option price is observed in the financial market, i.e., the market price $u_{\mathrm{ref}} \approx u(x, T)$ is observed.
To make the notation clear, we denote the model price by $u(x, T; \xi)$ instead of $u(x, T)$.
The calibration problem is formulated as follows.
\begin{equation}
\begin{aligned}
        & \min_{\xi \in [0, 1]} &&\mathcal{F}_\mathrm{err}(\xi) =  \| u(x, T; \xi) - u_{\mathrm{ref}} \|^2  \\ 
        & \text{ subject to } && \text{Eqs.~\eqref{eq:BSE}---\eqref{eq:BSE_right}} \\
        & && \sigma^2 = (\sigma_\mathrm{max}^2 - \sigma_\mathrm{min}^2) \xi + \sigma_\mathrm{min}^2,
\end{aligned} \label{eq:BSECO}
\end{equation}
where $u_{\mathrm{ref}}$ is the data point at $x = \log(X)$, and $\sigma_\mathrm{min}$ and $\sigma_\mathrm{max}$ are the lower and upper bounds of volatility, respectively.
This forward problem is a simple yet well-studied example using quantum simulation~\cite{gonzalez2023efficient, jin2026quantum}, while the analytical solution exists.
We demonstrate the inverse problem having one variable $\sigma$
since the other parameter $r$ can typically be calibrated from other datasets independently.
Also, $(K, x, T)$ are not model parameters but variables to specify the date points.
To solve this problem, we first truncate the domain to the interval $x \in [ x_\mathrm{min}, x_\mathrm{max} ]$ where $x_\mathrm{min}$ is sufficiently small and $x_\mathrm{max}$ is sufficiently large, and discretize it using a uniform grid with $N$ nodes.
Applying the finite difference method, the Black-Scholes equation reduces to an ODE, as follows.
\begin{align}
    \difft{u} = \left( \left( r - \frac{\sigma^2}{2} \right) D^{\pm}_{\mathrm{D}, N} + \frac{\sigma^2}{2} L_{\mathrm{D}, N} - rI \right) u,
\end{align}
where $D^{\pm}_{\mathrm{D}, N}$ and $L_{\mathrm{D}, N}$ are central difference and discrete Laplace operators, respectively, both subject to Dirichlet boundary conditions on the left endpoint and Neumann conditions on the right endpoint.
The coefficient matrix of this ODE is rearranged to the form in Eq.~\eqref{eq:A_xi} as
\begin{align}
    A &= -\left( r - \frac{\sigma^2}{2} \right) D^{\pm}_{\mathrm{D}, N} - \frac{\sigma^2}{2} L_{\mathrm{D}, N} + rI \nonumber \\
    &= \underbrace{-\left( r - \frac{\sigma_\mathrm{min}^2}{2} \right) D^{\pm}_{\mathrm{D}, N} - \frac{\sigma_\mathrm{min}^2}{2} L_{\mathrm{D}, N} + rI}_{=A_0} \nonumber \\
    & + \underbrace{\frac{\sigma_\mathrm{max}^2 - \sigma_\mathrm{min}^2}{2} (D^{\pm}_{\mathrm{D}, N} - L_{\mathrm{D}, N} )}_{A_1} \xi.
\end{align}
Thus, we need to implement the block-encoding of $A_0$ and $A_1$ to apply our framework.
Based on the block-encoding of finite difference operators given in Appendix~\ref{sec:BE_PDE}, we can construct the block-encodings of $A_0$ and $A_1$ by $\mathcal{O}((\log N)^2)$ single and two-qubit gates.
Alternatively, access to Hamiltonian simulation oracles $e^{-i H_0 \tau}$, $e^{-i L_0 \tau}$, $e^{-i H_1 \tau}$ and $e^{-i L_1 \tau}$ also suffices to apply our method, which we will employ in the following numerical experiments as the simplest implementation.

\subsubsection{Numerical experiments}
Here, we provide the results of numerical experiments to demonstrate the validity of our method.
As parameters of the Black-Scholes equation, we set time to maturity $T = 1$, risk-free rate $r = 0.02$, and strike price $K = 30$. 
For truncating the domain, we set $x_\mathrm{min} = \ln(10^{-4})$ and $x_\mathrm{max} = \ln (10K)$, and the domain is discretized into a uniform grid of 64 nodes using 6 qubits.
To implement the LCHS for forward simulation, we use Trotterization and the kernel $f(k) = 1 / (\pi (1 + ik))$ in Eq.~\eqref{eq:lchs} as the simplest implementation, where we use 8 qubits to discretize the truncated $k$-space from $-64$ ($-2^6$) to $63.5$ ($2^6 - 2^{-1}$) with interval 0.5.
We set the time increment of Trotterization to $0.02$.
For optimization, we discretize a parameter $\xi$ using 4 qubits, and set the range of volatility as $\sigma_\mathrm{min} = 10^{-2}$ and $\sigma_\mathrm{max} = 0.5$.
To prepare the reference state, i.e., the data points in Eq.~\eqref{eq:BSECO}, we use the analytical expression of the European call option price, which is given in the form~\cite{hull2016options,jin2026quantum}:
\begin{align}
    V^\ast (X, \tau) = X \Phi(y_1) - K e^{-r (T - \tau)} \Phi(y_2),
\end{align}
where 
\begin{align}
    y_1 &= \frac{\ln \frac{X}{K} + (r + \frac{\sigma^2}{2})( T - \tau)}{\sigma \sqrt{T - \tau}} \\
    y_2 &= y_1 - \sigma \sqrt{T - \tau},
\end{align}
and $\Phi(y)$ is the cumulative distribution function of the standard normal distribution.
Selecting a point in $x$-space, denoted by $x_\mathrm{ref}$, which is the closest point to $\ln (Ke^{-r})$, we prepare a data point as
\begin{align}
    u_\mathrm{ref} = V^\ast (e^{x_\mathrm{ref}}, 0) - e^{x_\mathrm{ref}},
\end{align}
where we choose $\sigma=0.3$.
The number of qubits required for block-encoding is 19, which consists of $n=6$, $a_\mathrm{for}=8$, $M=1$, $m=4$, and one additional ancillary qubit for LCU in Eq.~\eqref{eq:LCU}.

\begin{figure}[t]
    \centering
    \includegraphics[width=\columnwidth]{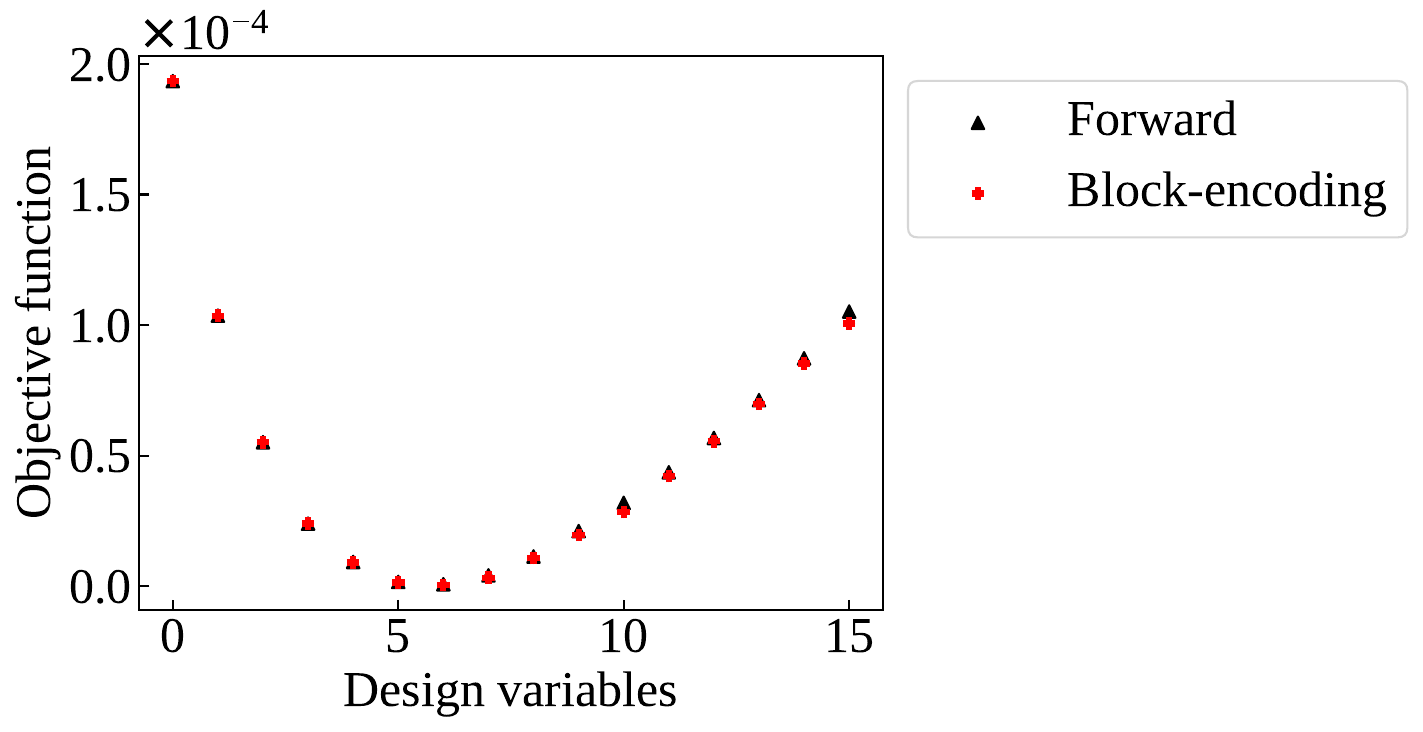}
    \caption{The objective function values computed by the block-encoding and the forward simulation. The horizontal axis represents the index of the discretized design variables, i.e., decimal value of the basis $\ket{\xi}$.} \label{fig:block-encoding_vs_forward_bs1d}
\end{figure}

Figure~\ref{fig:block-encoding_vs_forward_bs1d} illustrates the objective function values computed by the block-encoding and the forward simulation.
In the case of forward simulation, we performed LCHS for each value of the design variable, and evaluated the objective function by the expectation value of the projector $P$.
We confirmed that the objective function values computed by the block-encoding agree well with those by forward simulation, which demonstrates the validity of the proposed block-encoding.
The optimal solution is $\ket{\xi} = \ket{6}$, which corresponds to $\xi = 6 / 15$.

\begin{figure}[t]
    \centering
    {\includegraphics[width=\columnwidth]{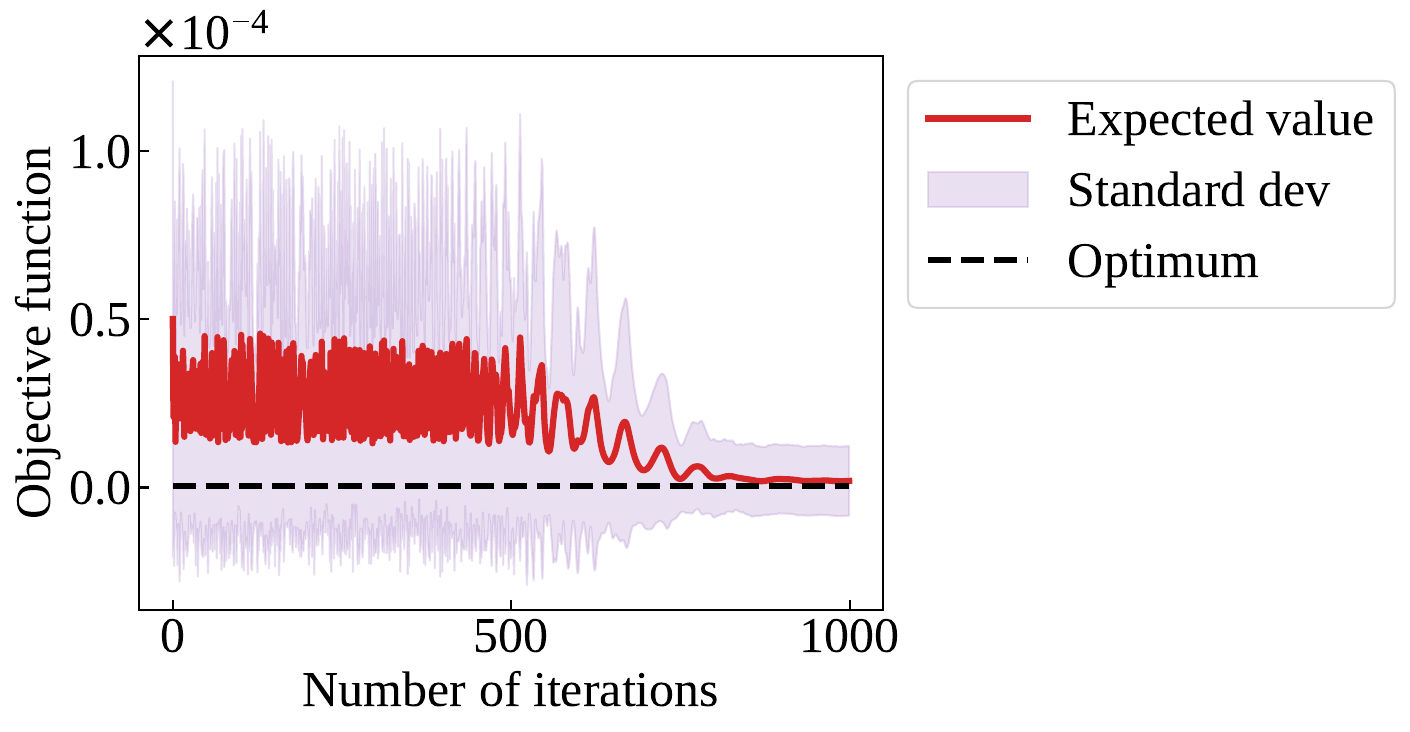}}
    \caption{The history of the expected value of the objective function (red line) with its standard deviation (purple region). The horizontal axis represents the number of iterations of the product formula. The black dashed lines represent the optimal value $\mathcal{F}(6 / 15)$. } \label{fig:history_expectation_bs1d}
\end{figure}

\begin{figure*}[t]
    \centering
    \includegraphics[width=0.9\textwidth]{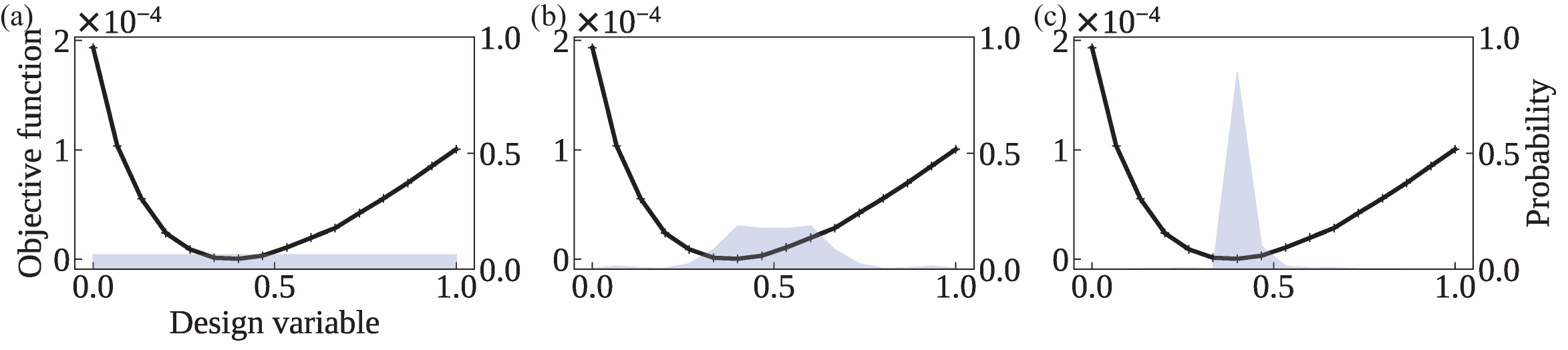}
    \caption{The probability distribution of all possible design variables at (a) initial, (b) intermediate (100th), and (c) final steps of the optimization. The black line shows the landscape of the objective function, and the blue region represents the probability distribution of the design variable.} \label{fig:history_probability_bs1d}
\end{figure*}

Next, we performed optimization by the time-dependent Hamiltonian simulation to examine the validity of the proposed method.
We empirically set the time increment $\Delta s = 0.01 / \sqrt{\bar{\mathcal{F}}}$, and the total time $S = 10 / \sqrt{\bar{\mathcal{F}}}$, which yields $N_\mathrm{step} = 1000$, where $\bar{\mathcal{F}}$ is the mean value of the objective function for all parameters, which can be evaluated by amplitude estimation (See Appendix~\ref{sec:friction}).
Figure~\ref{fig:history_expectation_bs1d} shows the history of the expected value of the objective function $\bra{\psi(t)} \hat{\mathcal{F}} \ket{\psi(t)}$ with its standard deviation $\sqrt{\bra{\psi(t)} \hat{\mathcal{F}}^2 \ket{\psi(t)} - \bra{\psi(t)} \hat{\mathcal{F}} \ket{\psi(t)}^2}$.
We observed that the expected value of the objective function converged to the optimum, while its variance decreased progressively.
Figure~\ref{fig:history_probability_bs1d} shows the distributions of all possible values of the design variable at the initial, intermediate, and final steps of the optimization.
This figure clearly shows that the probability (blue distribution) concentrates on the optimal solution $\xi = 6/15$, which has the minimum value of the black plot.

\begin{figure*}[t]
    \centering
    \includegraphics[width=0.9\textwidth]{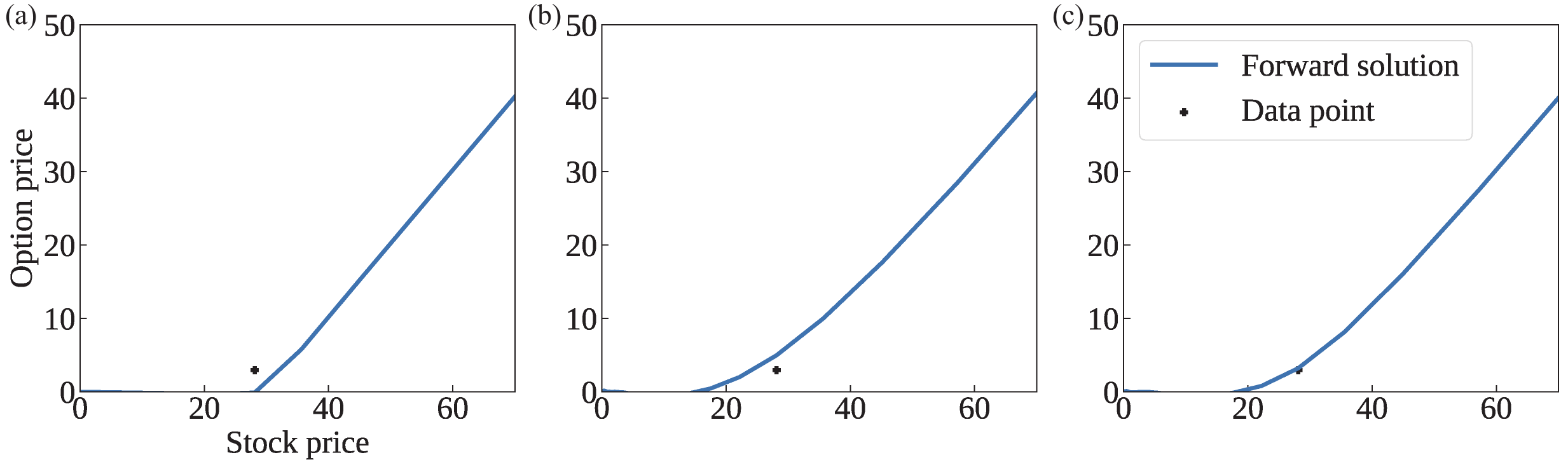}
    \caption{The forward simulation results for design variables (a) $\xi = 0$, (b) $\xi = 1$, and (c) $\xi = 6/15$. Blue lines illustrate the plots of the option price vs stock price obtained by forward simulation. The black points are data points used to define the reference state.} \label{fig:forward_result_bs1d}
\end{figure*}

We also examined the validity of the optimal solution.
Figure~\ref{fig:forward_result_bs1d} shows the forward simulation results for the optimal solution and the other reference solutions.
The blue lines indicate option prices with respect to the stock price calculated by forward simulation. 
The black points represent data points used to define the reference state.
In Figs.~\ref{fig:forward_result_bs1d}(a) and (b), the model fails to fit the data, whereas the optimized model in Fig.~\ref{fig:forward_result_bs1d}(c) shows good agreement with data points.

\subsection{Material parameter design in wave equation}
\subsubsection{Motivation and formulation}
We also apply our framework to a material parameter design problem of the one-dimensional wave equation.
The wave equation describes the dynamics of waves such as vibrations of a string, sound, and light, appearing in various fields, including mechanics, acoustics, and electromagnetics.
In wave-manipulating devices such as optical and acoustic devices, careful design for controlling reflection, transmission, and propagation velocity plays a critical role in determining device performance.
Hence, we herein consider a material parameter design problem to concentrate the energy of waves in a certain region.

\begin{figure}[t]
    \centering
    \includegraphics[width=\columnwidth]{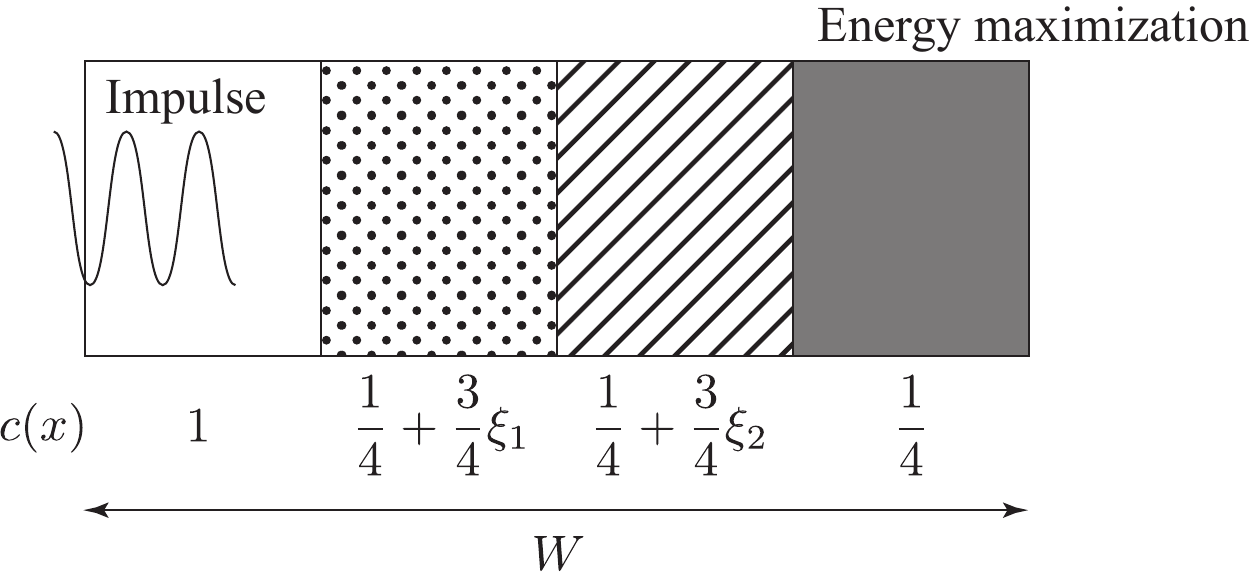}
    \caption{Problem setting of the material parameter design in the wave equation.} \label{fig:problem_wave1d}
\end{figure}

The one-dimensional wave equation is a linear hyperbolic PDE described as 
\begin{align}
    \ddiff{w}{t} = c(x)^2 \ddiff{w}{x}, \label{eq:WE}
\end{align}
where $w$ is a state variable, and $c$ is the speed of sound.
We focus on the equation on a segment $\Omega := (0, W)$ with the initial and boundary conditions
\begin{align}
    & w(x, 0) = w_0, ~ \diff{w(x, 0)}{t} = v_0 \label{eq:WE_init} \\
    & \diff{w(0, t)}{x} = 0, ~ w(W, t) = 0 \label{eq:WE_bnd_x},
\end{align}
where $w_0$ and $v_0$ are the initial state and its time derivative, respectively.
Since the speed of the wave depends on a material parameter in which the wave propagates, we parameterize it using design variables as follows.
\begin{align}
    c(x) = \begin{cases}
    c_1 & \text{ if } x \leq \frac{W}{4} \\
    c_2 + (c_1 - c_2) \xi_1 & \text{ if } \frac{W}{4} < x \leq \frac{W}{2} \\
    c_2 + (c_1 - c_2) \xi_2 & \text{ if } \frac{W}{2} < x \leq \frac{3W}{4} \\
    c_2 & \text{otherwise},
    \end{cases} \label{eq:c_param}
\end{align}
where $c_1$ is the speed of the wave in a material and $c_2$ is that in the other material.
Then, we formulate the material parameter design problem as
\begin{equation}
    \begin{aligned}
        & \max_{\xi \in [0, 1]^2} && \int_\frac{3W}{4}^1 \left( c(x)^2 \left( \frac{\partial w(x, T)}{\partial x} \right)^2 + \left( \diff{w(x, T)}{t} \right)^2 \right) \mathrm{d}x \\ 
        & \text{ subject to } && \text{Eqs.~\eqref{eq:WE}---\eqref{eq:c_param}}.
    \end{aligned} \label{eq:WECO}
\end{equation}
Here, we focus on the maximization problem, which can be converted into the minimization problem by adding the minus sign to the objective function.
This corresponds to the replacement of the reflection operator $R_P$ into $-R_P$ in constructing the block-encoding of the objective function.
Figure~\ref{fig:problem_wave1d} illustrates the problem setting of the material parameter design. 
The impulse wave is provided at the left boundary and the optimization aims to maximize the wave energy at the gray region by designing speed of wave (material parameters) of the intermediate domains.
To solve this problem, we discretize the wave equation in the direction of space and rearrange it to the following form:
\begin{align}
    \difft{} \underbrace{\begin{pmatrix}
    w^\mathrm{t} \\
    iw^\mathrm{x}
    \end{pmatrix}}_{u} = - \underbrace{i \begin{pmatrix}
        0 & C D^{-}_\mathrm{D} \\
        -D^{+}_\mathrm{D} C & 0
    \end{pmatrix}}_{A} \underbrace{\begin{pmatrix}
        w^\mathrm{t} \\
        iw^\mathrm{x}
    \end{pmatrix}}_{u}, \label{eq:ODE_wave}
\end{align}
where $w^\mathrm{t}$ and $w^\mathrm{x}$ are discretized state vectors of $(1 / c(x)) \partial w / \partial t$ and $\partial w / \partial x$, respectively, $D^{+}_\mathrm{D}$ and $D^{-}_\mathrm{D}$ are forward and backward difference operators with the Dirichlet boundary conditons, respectively.
The matrix $C$ is a diagonal matrix whose $j$-th entry corresponds to the speed of wave $c(x)$ at the $j$-th node.
That is, based on the parameterization in Eq.~\eqref{eq:c_param}, it can be decomposed into the following form:
\begin{align}
    C = c_2 I + C_0 + C_1 \xi_1 + C_2 \xi_2,
\end{align}
where $C_0$, $C_1$, and $C_2$ are diagonal matrices whose entries are $c_1 - c_2$ for nodes satisfying $x \leq W/4$, $W/4 < x \leq W/2$, and $W/2 < x \leq 3W/4$, respectively, and $0$ otherwise.
The coefficient matrix of this ODE is then rearranged to the form in Eq.~\eqref{eq:A_xi} as
\begin{align}
    A &= i \begin{pmatrix}
    0 & C D^{-}_\mathrm{D} \\
    -D^{+}_\mathrm{D} C & 0
    \end{pmatrix} \nonumber \\
    &= \underbrace{i \begin{pmatrix}
    0 & \left( c_2 I + C_0 \right) D^{-}_\mathrm{D} \\
    -D^{+}_\mathrm{D} \left( c_2 I + C_0 \right) & 0
    \end{pmatrix}}_{A_0} \nonumber \\
    &\quad + \underbrace{i \begin{pmatrix}
    0 & C_1 D^{-}_\mathrm{D} \\
    -D^{+}_\mathrm{D} C_1 & 0
    \end{pmatrix}}_{A_1} \xi_1 + \underbrace{i \begin{pmatrix}
    0 & C_2 D^{-}_\mathrm{D} \\
    -D^{+}_\mathrm{D} C_2 & 0
    \end{pmatrix}}_{A_2} \xi_2. 
    \end{align}
Thus, we need to implement the block-encodings of $A_0$, $A_1$, and $A_2$ to apply our framework.
Alternatively, it is also sufficient to implement Hamiltonian simulation oracles $e^{-i H_0 \tau}$, $e^{-i H_1 \tau}$, and $e^{-i H_2 \tau}$, which we will use in the subsequent experiments as the simplest implementation.
Note that since $A_0$, $A_1$, and $A_2$ are anti-Hermitian, we perform simple Hamiltonian simulation, rather than LCHS.

\begin{figure}[t]
    \centering
    \includegraphics[width=\columnwidth]{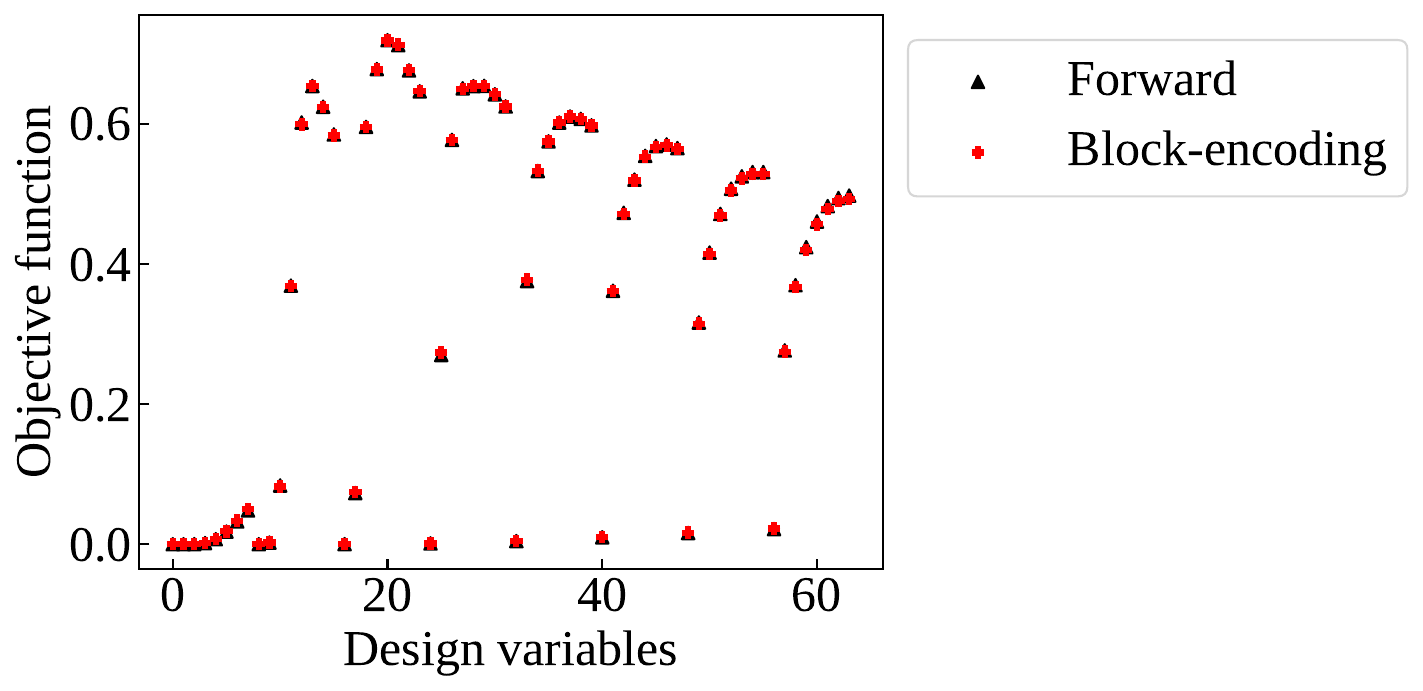}
    \caption{The objective function values evaluated by the block-encoding and the forward simulation. The horizontal axis represents the index of the discretized design variables, i.e., decimal value of the basis $\ket{\xi_2, \xi_1}$.} \label{fig:block-encoding_vs_forward_wave1d}
\end{figure}

\begin{figure}[t]
    \centering
    \includegraphics[width=0.75\columnwidth]{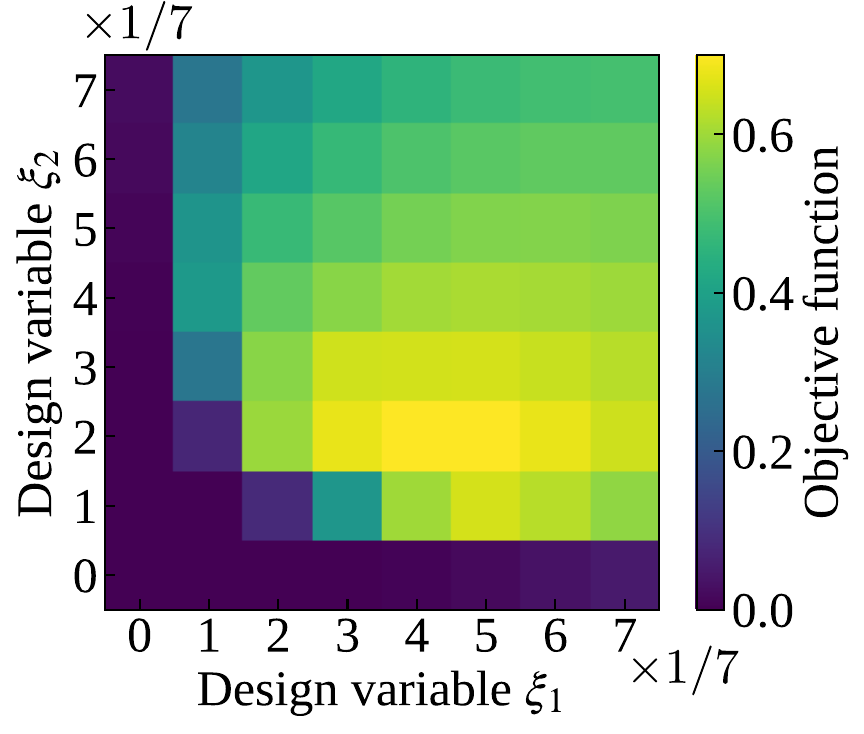}
    \caption{The objective function values evaluated by the block-encoding. Each axis represents each index of the discretized design variables, i.e., decimal value of the basis $\ket{\xi_1}$ and $\ket{\xi_2}$, respectively.} \label{fig:block-encoding_objective_wave1d}
\end{figure}

\begin{figure}[t]
    \centering
    {\includegraphics[width=\columnwidth]{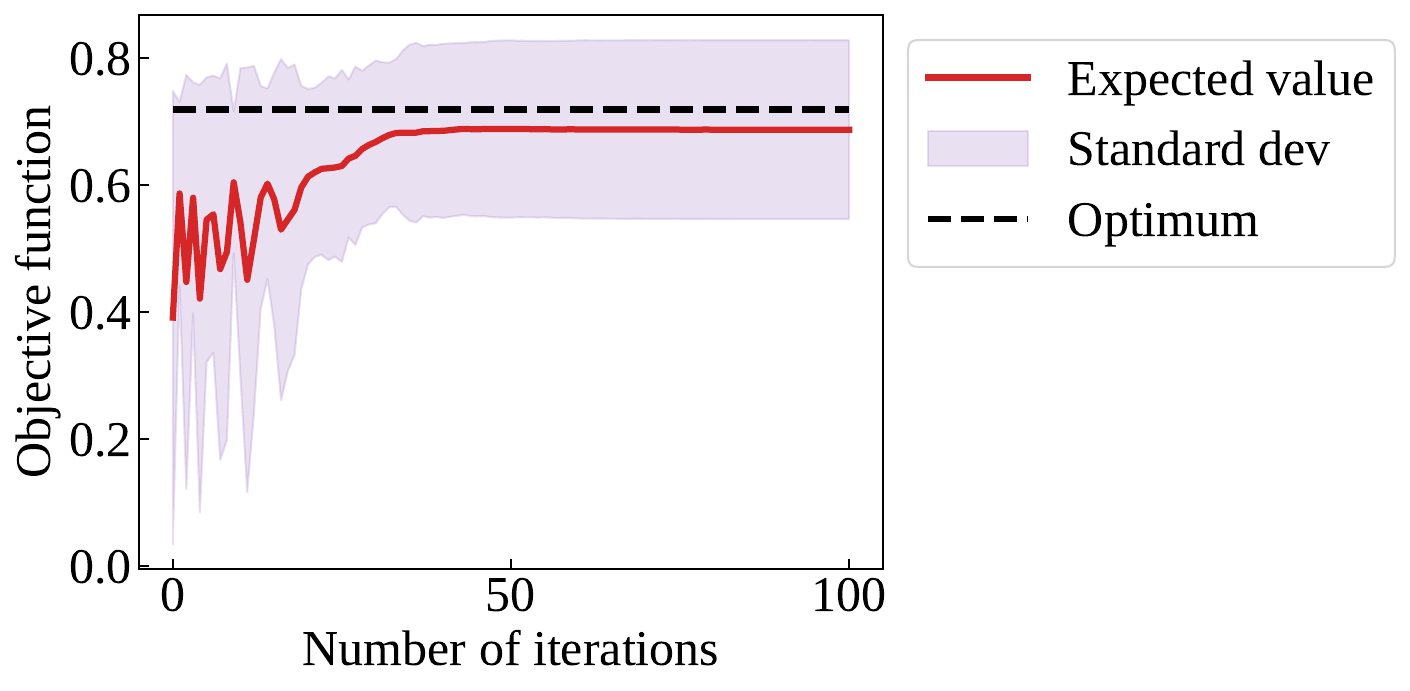}}
    \caption{The history of the expected value of the objective function (red line) with the purple region indicating the standard deviation of the objective function. The horizontal axis represents the number of iterations of the product formula. The red lines illustrate the expected value of the objective function, while the black dashed lines represent the optimal value $\mathcal{F}(4 / 7, 2 / 7)$. } \label{fig:history_expectation_wave1d}
\end{figure}

\subsubsection{Numerical experiments}
Here, we provide numerical examples to demonstrate the validity of our method.
As parameters of the wave equation, we set the domain length $W=2^5 - 1$, simulation time $T=48$, and the speed of the wave $c_1 = 1$ and $c_2 = 1/4$.
We set the initial conditions as
\begin{align}
    w_0 &= 0 \\
    v_0 &= \begin{cases}
        e^{-\frac{x^2}{2 \cdot 2^2}} & x < 4 \\
        0 & x \geq 4,
    \end{cases}
\end{align}
which is sparse, and thus can be prepared efficiently.
We implement the state preparation oracle by amplitude encoding~\cite{iten2016quantum} through Qiskit.
We discretize the segment into a uniform grid using 5 qubits with one additional ancillary qubit to encode the wave equation into an ODE in Eq.~\eqref{eq:ODE_wave}, which corresponds to $n=6$.
For forward simulation, we use Trotterization with the time increment $0.5$.
For optimization, we discretize each parameter $\xi_1$ $\xi_2$ using 3 qubits.
The number of qubits required for block-encoding is 12, which consists of $n=6$, $a_\mathrm{for}=0$, $M=2$, $m=3$.

Figure~\ref{fig:block-encoding_vs_forward_wave1d} illustrates the objective function values evaluated by the block-encoding and the forward simulation.
In forward simulation, we evaluated the objective function for each value of the design variable by performing Hamiltonian simulation, followed by computing the expectation value of the projector $P$.
As shown in Fig.~\ref{fig:block-encoding_vs_forward_wave1d}, the objective function values computed by the block-encoding agree well with those obtained by forward simulation.
Figure~\ref{fig:block-encoding_objective_wave1d} shows the landscape of the objective function computed by its block-encoding.
Since we focus on the maximization problem, the optimal solution is $\ket{\xi_2, \xi_1} = \ket{2, 4}$, which corresponds to $\xi_1 = 4 / 7$ and $\xi_2 = 2 / 7$.

\begin{figure*}[t]
    \centering
    \includegraphics[width=0.9\textwidth]{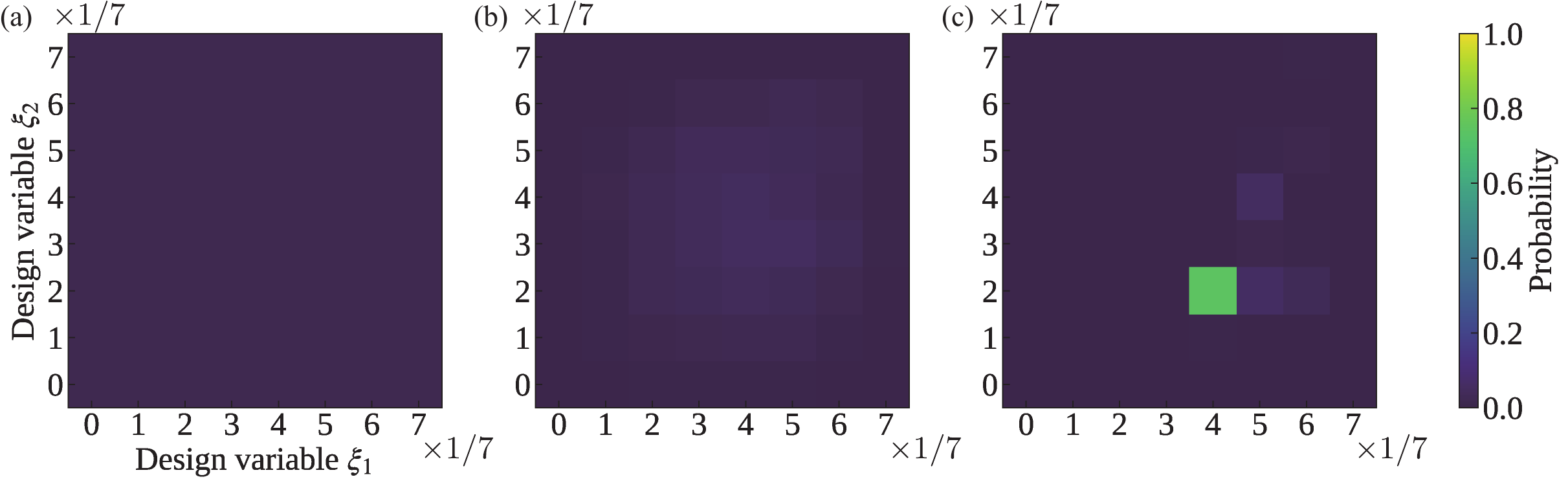}
    \caption{The probability distribution of all possible design variables at (a) initial, (b) intermediate (10th), and (c) final steps of the optimization.} \label{fig:history_probability_wave1d}
\end{figure*}

\begin{figure*}[t]
    \centering
    \includegraphics[width=\textwidth]{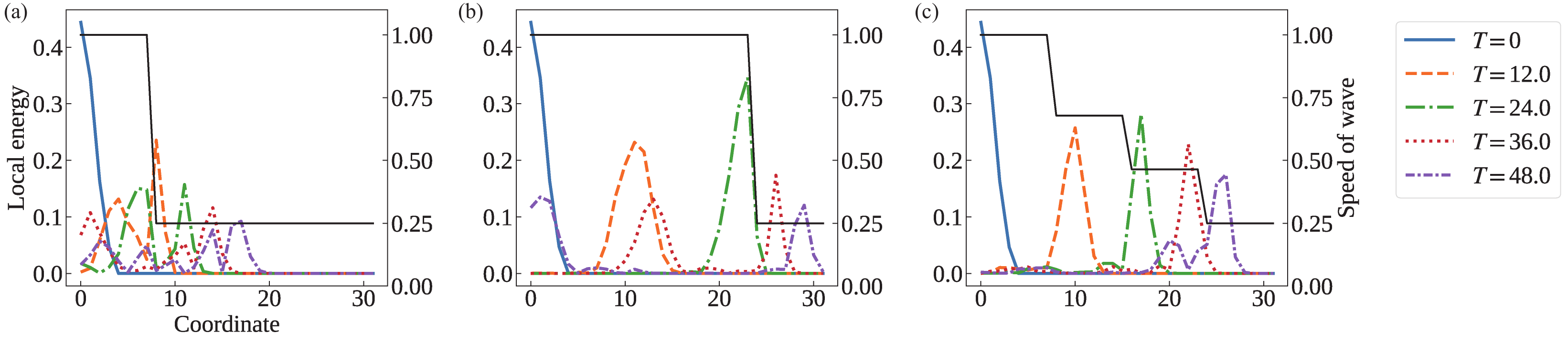}
    \caption{The forward simulation results for design variables (a) $(\xi_1, \xi_2) = (0, 0)$, (b) $(\xi_1, \xi_2) = (1, 1)$, and (c) $(\xi_1, \xi_2) = (4/7, 2/7)$. Each color line illustrates the local energy of the wave field at each time slice. The black line shows the speed of wave at each spatial coordinate, which depends on the design variables.} \label{fig:forward_result_wave1d}
\end{figure*}

Next, we provide numerical experiments of optimization by the time-dependent Hamiltonian simulation to verify whether our method can find the optimal solution.
We set the time increment $\Delta s = 0.1 / \sqrt{\bar{\mathcal{F}}}$, and the total time $S = 10 / \sqrt{\bar{\mathcal{F}}}$, which yields $N_\mathrm{step} = 100$.
Figure~\ref{fig:history_expectation_wave1d} shows the history of the expected value of the objective function $\bra{\psi(t)} \hat{\mathcal{F}} \ket{\psi(t)}$ with its standard deviation $\sqrt{\bra{\psi(t)} \hat{\mathcal{F}}^2 \ket{\psi(t)} - \bra{\psi(t)} \hat{\mathcal{F}} \ket{\psi(t)}^2}$, where $\hat{\mathcal{F}} := (U_\mathrm{obj} + I) / 2$ is an objective function operator.
The expected value of the objective function converged near the optimal value with a gradual decrease in the standard deviation.
As the standard deviation appears to remain moderately large, we next examined the probability distributions of the design variables.
Figure~\ref{fig:history_probability_wave1d} shows the distributions of the design variables at the initial, intermediate, and final steps of the optimization.
This figure clearly shows that the probability concentrates on the optimal solution $(\xi_1, \xi_2) = (4/7, 2/7)$.
Note that the expectation value in Fig.~\ref{fig:history_expectation_wave1d} does not converge exactly to the optimal value, and the success probability in Fig.~\ref{fig:history_probability_wave1d} remains below unity.
This is most likely caused by discretization effects in the present simulation.

Finally, we examined the validity of the optimal solution.
Figure~\ref{fig:forward_result_wave1d} illustrates the forward simulation results for the optimal solution and the other reference solutions.
Each color line illustrates the local energy of the wave field at each time slice, while the black line shows the speed of wave at each spatial coordinate, depending on the design variables.
In Figs.~\ref{fig:forward_result_wave1d}(a) and (b), the wave reflects at the boundary of two phases (areas with $c(x)=1$ and $c(x)=1/4$), which prevents the wave from propagating smoothly into the forward direction.
In the optimal solution in Fig.~\ref{fig:forward_result_wave1d}, on the other hand, the speed of wave changes in a stepwise manner, which suppresses the reflection at the boundary of adjacent phases.
This resulting structure has typical characteristics of antireflection (AR) designs; its discovery through optimization validates the proposed method.

\section{Conclusions} \label{sec:conlusion}
This study proposed a quantum algorithm for PDE-constrained optimization with continuous design variables.
Specifically, we constructed an explicit block-encoding of the objective function that coherently bridges a quantum PDE solver and a quantum optimizer, thereby avoiding any classical readout of the output of the quantum PDE solver, which could negate the potential quantum speedups. 
We then derived the query and gate complexities of the proposed method.
Under the assumption that the optimization landscape is strongly convex, we showed that, compared with conventional classical approaches, our algorithm can provide a polynomial speedup for the system size of PDEs $N_x$, and an exponential speedup for the spatial dimension of PDEs, $d$.
We note that the block-encoding construction and the coherent
interface between the PDE solver and the optimizer do not rely on strong convexity.
Strong convexity and smoothness are assumed only to obtain the rigorous convergence guarantee for QHD and the corresponding
end-to-end complexity bound.
Our algorithm also exhibited the exponential space advantage over classical approaches with respect to the system size of PDEs.
These results suggest that quantum optimization can inherit quantum speedup from the underlying quantum PDE solver.
As numerical experiments, we demonstrated two applications: a parameter calibration problem in the Black-Scholes equation and a material parameter design problem in the wave equation, which support the effectiveness of the proposed method.

Beyond the use of QHD as an optimizer, the present oracle-construction framework may also serve as a coherent access primitive for broader continuous-optimization schemes. In particular, Appendix~\ref{sec:gradient_block_encoding} shows that our framework can be extended to gradient block-encodings, suggesting a possible interface between quantum PDE solvers and more general coherent local/global optimization frameworks that assume objective and derivative access~\cite{catli2026efficient}. Exploring such connections, including possible interfaces with thermal-state-based global optimization methods, is an interesting direction for future work.

More broadly, this work presents the concept of composing quantum subroutines so that the prohibitive readout overhead of one can be avoided by using it coherently as input to another, paving the way to exploiting quantum speedups in practical scenarios.

\section*{Acknowledgement}
This work is supported by MEXT Quantum Leap Flagship Program Grant Number JPMXS0118067285 and JPMXS0120319794, and JSPS KAKENHI Grant Number 20H05966. 
J.K. acknowledges support by SIP Grant Number JPJ012367.
\appendix

\section{Block-encoding of finite difference operator} \label{sec:BE_PDE}

Let us consider a one-dimensional domain $\Omega = (0, 1)$ in which a scalar field $u(x)$ is defined with $x$ the spatial coordinate.
For finite difference approximation, we discretize the domain $\Omega$ into a grid with $N_x$ nodes, and represent the scalar field by its value at each node, i.e., $u :=[u_1, \dots, u_{N_x}]^\top$ where $u_j$ is the value at the $j$-th node.
The forward, central, and backward difference schemes, denoted by $D^{+}$, $D^{-}$, and $D^{\pm}$ respectively, approximate the first-order spatial derivative as follows:
\begin{align} \label{eq:diff_operator}
    (D^{+}u)_j &= \frac{u_{j+1} - u_j}{ h } + O(h), \nonumber \\
    (D^{-}u)_j &= \frac{u_j - u_{j-1}}{ h } + O(h), \nonumber \\
    (D^{\pm}u)_j &= \frac{u_{j+1} - u_{j-1}}{ 2h } + O(h^2), 
\end{align}
where $h$ is the interval of adjacent nodes.
Similarly, the Laplace operator by the central difference scheme is given as
\begin{align}
    (\Lambda u)_j = \frac{u_{j+1} - 2u_j + u_{j-1}}{ h^2 } + O(h^2).
\end{align}
With the Dirichlet boundary condition, these finite difference operators can be represented by using the identity and shift operators as follows:
\begin{align}
    D^{+} &= \frac{1}{h} \left( \texttt{SHIFT}^{-} - I - \sigma_{10}^{\otimes n_x} \right) \nonumber \\
    D^{-} &= \frac{1}{h} \left( I - \texttt{SHIFT}^{+} + \sigma_{01}^{\otimes n_x} \right) \nonumber \\
    D^{\pm} &= \frac{1}{2h} \left( \texttt{SHIFT}^{-} - \texttt{SHIFT}^{+} + \sigma_{01}^{\otimes n_x} - \sigma_{10}^{\otimes n_x} \right) \nonumber \\
    \Lambda &= \frac{1}{h^2} \left( \texttt{SHIFT}^{-} + \texttt{SHIFT}^{+} - 2 I - \sigma_{01}^{\otimes n_x} - \sigma_{10}^{\otimes n_x} \right)
\end{align}
where $n_x := \log_2 N_x$, $\texttt{SHIFT}^{+} := \sum_{j=0}^{2^{n_x}-1} \ket{j+1} \! \bra{j}$, $\texttt{SHIFT}^{-} := \sum_{j=0}^{2^{n_x}-1} \ket{j} \! \bra{j+1}$, $\sigma_{01} := \ket{0} \! \bra{1}$, and $\sigma_{10} := \ket{1} \! \bra{0}$.
The terms including ladder operators $\sigma_{01}$ and $\sigma_{10}$ are for the Dirichlet boundary conditions.
Here, we construct a block-encoding of the following Hermitian operator $B(\alpha_1, \alpha_2, \phi)$:
\begin{align}
    B(\alpha_1, \alpha_2, \phi) &:= \alpha_1 e^{i \phi} (\texttt{SHIFT}^{-} - \sigma_{10}^{\otimes n_x}) \nonumber \\
    &\quad + \alpha_1 e^{-i \phi} (\texttt{SHIFT}^{+} - \sigma_{01}^{\otimes n_x}) + \alpha_2 I,
\end{align}
where $\alpha_1$, $\alpha_2$ are real coefficients, and $\lambda$ is a phase parameter.
For example, $B(1/h^2, -2/h^2, 0)$ corresponds to the Laplace operator while $B(1/2h, 0, \pi/2)$ to the central difference operator up to global phase $e^{i \pi/2}$.
Since the terms of ladder operators are not unitary, we rearrange them as follows.
\begin{align}
    & e^{-i \phi} \sigma_{01}^{\otimes n_x} + e^{i \phi} \sigma_{10}^{\otimes n_x} \nonumber \\
    &= \frac{\ket{0}^{\otimes n_x} + e^{i \phi} \ket{1}^{\otimes n_x} }{\sqrt{2}} \frac{\bra{0}^{\otimes n_x} + e^{- i \phi} \bra{1}^{\otimes n_x} }{\sqrt{2}} \nonumber \\
    &\quad - \frac{\ket{0}^{\otimes n_x} - e^{i \phi} \ket{1}^{\otimes n_x} }{\sqrt{2}} \frac{\bra{0}^{\otimes n_x} - e^{-i \phi} \bra{1}^{\otimes n_x} }{\sqrt{2}} \nonumber \\
    & = U_\mathrm{GHZ}(\phi) Z \otimes \ket{0} \! \bra{0}^{\otimes n_x - 1} U_\mathrm{GHZ} (\phi)^\dagger, \nonumber \\
    & = \frac{1}{2} U_\mathrm{GHZ}(\phi) (Z \otimes R_0) U_\mathrm{GHZ} (\phi)^\dagger \nonumber \\
    &\quad + \frac{1}{2} U_\mathrm{GHZ}(\phi) (Z \otimes I) U_\mathrm{GHZ} (\phi)^\dagger,
\end{align}
where $Z$ is a Pauli Z operator, $R_0 := 2 \ket{0} \! \bra{0}^{\otimes n_x - 1} - I$ is a reflection operator, and $U_\mathrm{GHZ}(\phi)$ is a unitary that prepares $U_\mathrm{GHZ} \ket{0}^{\otimes n_x} = (\ket{0}^{\otimes n_x} + e^{i \phi} \ket{1}^{\otimes n_x} ) / \sqrt{2}$ and $U_\mathrm{GHZ} \ket{1} \! \ket{0}^{\otimes n_x-1} = (\ket{0}^{\otimes n_x} - e^{i \phi} \ket{1}^{\otimes n_x} ) / \sqrt{2}$.
Then, the Hermitian operator $B$ is written as a linear combination of unitaries with five terms, as follows:
\begin{align}
    B(\alpha_1, \alpha_2, \phi) &= \alpha_1 e^{i \phi} \texttt{SHIFT}^{-} + \alpha_1 e^{-i \phi} \texttt{SHIFT}^{+} + \alpha_2 I \nonumber \\
    &\quad - \frac{\alpha_1}{2} U_\mathrm{GHZ}(\phi) (Z \otimes R_0) U_\mathrm{GHZ} (\phi)^\dagger \nonumber \\
    &\quad - \frac{\alpha_1}{2} U_\mathrm{GHZ}(\phi) (Z \otimes I) U_\mathrm{GHZ} (\phi)^\dagger,
\end{align}
which means that LCU yields $(3 | \alpha_1 | + | \alpha_2 |, 3, 0)$-block-encoding of $B$.
The shift operators include $j$-controlled Toffoli gates for $j=0, 1, \dots, n_x - 1$, which requires $\mathcal{O}(j)$ single and two-qubit gates~\cite{barenco1995elementary}.
Thus, the shift operators require $\mathcal{O}(n_x^2)$ single and two-qubit gates.
On the other hand, the fourth and fifth terms yield $\mathcal{O}(n_x)$ single and two-qubit gates.
Consequently, the number of single and two-qubit gates for implementing the block-encoding of $B$ is $\mathcal{O}(n_x^2)$. 

For $d$-dimensional cases, the Hermitian operator is extended to
\begin{align}
    &B^{(d)} \left( \{ \alpha_1^{(\mu)} \}_{\mu=1}^d, \{ \alpha_2^{(\mu)} \}_{\mu=1}^d, \{ \phi^{(\mu)} \}_{\mu=1}^d \right) \nonumber \\
    &= \sum_{\mu=1}^{d} I^{\otimes (d - \mu)} \otimes B(\alpha_1^{(\mu)}, \alpha_2^{(\mu)}, \phi^{(\mu)}) \otimes I^{\otimes (\mu - 1)},
\end{align}
which is also the form of a linear combination of unitaries with $5d$ terms.
This means that LCU yields $(3\sum_{\mu=1}^d | \alpha_1^{(\mu)} | + \sum_{\mu=1}^d | \alpha_2^{(\mu)} |, \lceil \log (5d) \rceil, 0)$-block-encoding of $B^{(d)}$, which requires $\mathcal{O}(d n_x^2) = \mathcal{O}(n^2 / d)$ single and two-qubit gates where we assume that we assign $n/d$ qubits for each dimension.
For example, an operator $\alpha_\mathrm{PDE} \nabla^\eta$ in the $d$-dimensional space  with the order of spatial derivative $\eta$ can be discretized by $B^{(d)}$ by substituting $\alpha_1^{(\mu)} = \alpha_\mathrm{PDE}/h_x^\eta$, $\alpha_2^{(\mu)} = -2\alpha_\mathrm{PDE}/h_x^\eta$, and $\phi^{(\mu)} = 0$.

Since $B^{(d)} \left( \{ \alpha_1^{(\mu)} \}_{\mu=1}^d, \{ \alpha_2^{(\mu)} \}_{\mu=1}^d, \{ \phi^{(\mu)} \}_{\mu=1}^d \right)$ corresponds to the discrete Laplace operator $\Lambda^{(M)}$ when $\alpha_1^{(\mu)} = 1/h_\xi^2$, $\alpha_2^{(\mu)} = -2 /h_\xi^2$, $\phi^{(\mu)} = 0$, and $d=M$, $n=mM$, we can construct the $(5M/h_\xi^2, \lceil \log(5M) \rceil, 0)$-block-encoding of $\Lambda^{(M)}$ by using $\mathcal{O}(Mm^2)$ single and two-qubit gates.

\section{Block-encoding of spatially varying coefficients of PDEs} \label{sec:BE_C}
Here, we construct a block-encoding of the diagonal matrix derived from the spatially varying coefficients of PDEs.
Let $c(x)$ denote a spatially varying coefficient of a PDE such as speed of sound for acoustic simulation.
Discretizing $c(x)$ by the finite difference method yields a diagonal matrix $C$ whose $(j, j)$-entry corresponds to the value of $c(x)$ at the $j$-th node of the spatial grid.
In this study, we assume that $c(x)$ is a piece-wise constant function and the corresponding diagonal matrix $C$ can be represented using projectors $\sigma_{00} := \ket{0} \! \bra{0}$ and $\sigma_{11} := \ket{1} \! \bra{1}$ for a constant value $N_C$, as follows:
\begin{align}
    C = \sum_{k=1}^{N_\mathrm{C}} \alpha_\mathrm{PDE}^{(k)} C_n^{(k)} \otimes \cdots \otimes C_1^{(k)},
\end{align}
where $C_j^{(k)} \in \{\sigma_{00}, \sigma_{11}, I \}$, and $\alpha_\mathrm{PDE}^{(k)}$ is a coefficient.
Introducing a reflection operator $R_C^{(k)} :=2C_n^{(k)} \otimes \cdots \otimes C_1^{(k)} - I$, we can represent the diagonal matrix as
\begin{align}
    C = \sum_{k=1}^{N_\mathrm{C}} \alpha_\mathrm{PDE}^{(k)} \frac{R_C^{(k)}+I}{2} = \sum_{k=1}^{N_\mathrm{C}} \frac{\alpha_\mathrm{PDE}^{(k)}}{2} R_C^{(k)} + \frac{I}{2} \sum_{k=1}^{N_\mathrm{C}} \alpha_\mathrm{PDE}^{(k)},
\end{align}
which is LCU requiring $\mathcal{O}(nN_c)$ single and two-qubit gates since each reflection operator requires $Z$ gates controlled by at most $n-1$-qubits.
The operator norm of $C$ is bounded by $\alpha_\mathrm{PDE} := N_C \max_k \alpha_\mathrm{PDE}^{(k)}$.
Thus, we obtain the $(\sum_k |\alpha_{\mathrm{PDE}}^{(k)} |, \lceil \log (N_C) \rceil , 0 )$-block-encoding of $C$. 
Note that we assume $N_c \in \mathcal{O}(1)$ in the complexity analysis in Section~\ref{sec:complexity_obj}.

\section{Block-encoding of the time-dependent Hamiltonian} \label{sec:BE_HAM-T}
Here, we derive the block-encoding of the time-dependent Hamiltonian in the interaction picture, $\mathcal{H}_\mathrm{I}(t)$, in Eq.~\eqref{eq:QHD_ham_I} to simulate the time evolution
\begin{align}
    U_\mathrm{TDS} (s_{j+1}, s_j) := \mathcal{T} e^{-i \int_{s_j}^{s_{j+1}} \mathcal{H}_\mathrm{I}(\tau) \mathrm{d} \tau}.
\end{align}
Specifically, based on Ref.~\cite{catli2026efficient}, we implement the block-encoding $\texttt{HAM-T}_j$ such that
\begin{widetext}
\begin{align}
    &(\bra{0}^{\otimes a_\mathrm{HT}} \otimes I^{\otimes (Mm + \lceil \log(Q) \rceil)}) \texttt{HAM-T}_j (\ket{0}^{\otimes a_\mathrm{HT}} \otimes I^{\otimes (Mm + \lceil \log(Q) \rceil)}) = \sum_{q=0}^{Q-1} \ket{q} \!\bra{q} \otimes \frac{1}{\alpha_\Lambda} \mathcal{H}_\mathrm{I}\left(s_j + \frac{q\Delta s}{Q} \right), \label{eq:HAM-T}
\end{align}
\end{widetext}
where $Q \in \mathcal{O}(\frac{S}{\epsilon} (\alpha_\Lambda + \alpha_\mathrm{obj} e^{\nu S} + \nu))$ is the number of quadrature points for time integration from $s_j (= j\Delta s)$ to $s_{j+1} (= (j+1)\Delta s)$, $a_\mathrm{HT}$ is the number of ancillary qubits, and $\alpha_\Lambda$ is a normalization constant for block-encoding.
According to Eq.~\eqref{eq:QHD_ham_I}, the above equation \eqref{eq:HAM-T} can be further decomposed into:
\begin{widetext}
\begin{align}
    \sum_{q=0}^{Q-1} \ket{q} \!\bra{q} \otimes \frac{1}{\alpha_\Lambda} \mathcal{H}_\mathrm{I} ( t_{j,q} ) &= \underbrace{\left( \sum_{q=0}^{Q-1} \ket{q} \!\bra{q} \otimes e^{i \frac{1}{\nu} ( e^{\nu t_{j,q}} - 1) \hat{\mathcal{F}}} \right)}_{=\widetilde{\mathcal{F}} (t_{j, q})^\dagger} \underbrace{\left( \sum_{q=0}^{Q-1} \ket{q} \!\bra{q} \otimes e^{-\nu t_{j, q}} \frac{\Lambda^{(M)}}{2 \alpha_\Lambda} \right)}_{=: \widetilde{\Lambda} (t_{j, q})} \underbrace{\left( \sum_{q=0}^{Q-1} \ket{q} \!\bra{q} \otimes e^{-i \frac{1}{\nu} (e^{\nu t_{j,q}} - 1) \hat{\mathcal{F}}} \right)}_{=:\widetilde{\mathcal{F}} (t_{j, q})}
\end{align}
\end{widetext}
where $t_{j, q} := s_j + \frac{q\Delta s}{Q}$.
Thus, we focus on the implementation of the block-encoding of $\widetilde{\mathcal{F}}(t_{j,q})$ and $\widetilde{\Lambda}(t_{j,q})$.

The implementation of the block-encoding of $\widetilde{\mathcal{F}}(t_{j,q})$ consists of the following steps:
\begin{enumerate}
    \item Encode binary representation of $\frac{1}{\nu} (e^{\nu t_{j,q}} - 1)$ into $\lceil \log (  e^{\nu S} / (\nu \epsilon') ) \rceil$ ancillary qubits within an error $\epsilon'$, i.e.,
    \begin{align}
        &U_\mathrm{EN} \ket{q} \ket{0}^{\otimes \lceil \log ( e^{\nu S} / (\nu \epsilon') ) \rceil} \nonumber \\
        &= \ket{q} \ket{\frac{1}{\nu} (e^{\nu t_{j,q}} - 1)},
    \end{align}
    which can be implemented efficiently using $\mathcal{O}(\lceil \log (  e^{\nu S} / (\nu \epsilon') ) \rceil^2)$ Toffoli gates~\cite{haner2018optimizing}.

    \item Perform the time evolution of the objective function controlled by each ancillary qubit as
    \begin{align}
        \prod_{b=1}^{\lceil \log ( e^{\nu S} / (\nu \epsilon') ) \rceil} C_b \left( e^{-i 2^{b-1} \epsilon' \hat{\mathcal{F}}} \right), \label{eq:C_b}
    \end{align}
    where $C_b (e^{-i 2^{b-1} \epsilon' \hat{\mathcal{F}}} )$ represents the operation $e^{-i 2^{b-1} \epsilon' \hat{\mathcal{F}}}$ controlled by the $b$-th ancillary qubit.
    Combining QSVT with OAA~\cite[Corollary~62]{gilyen2019quantum}, we can implement $(1, a_\mathrm{obj}+2, \epsilon'')$-block-encoding of $C_b \left( e^{-i 2^{b-1} \epsilon' \hat{\mathcal{F}}} \right)$ using $6 \alpha_\mathrm{obj} 2^{b-1} \epsilon'' + 9 \log (12 / \epsilon'')$ queries to $(\alpha_\mathrm{obj}, a_\mathrm{obj}, \epsilon'' / (2^{b-1} \epsilon')$-block-encoding of the objective function $\hat{\mathcal{F}}$. 

\end{enumerate}
By setting $\epsilon'' < \epsilon' / \lceil \log ( e^{\nu S} / (\nu \epsilon') ) \rceil$ so that the error of implementing Eq.~\eqref{eq:C_b} could be within $\epsilon'$, we can construct $(1, a_\mathrm{obj} + 2 + \lceil \log ( e^{\nu S} / (\nu \epsilon') ) \rceil, 2\epsilon')$-block-encoding of $\widetilde{\mathcal{F}}(t_{j,q})$ using $\mathcal{O}(\alpha_\mathrm{obj} e^{\nu S} / \nu + \log(e^{\nu S} / \nu \epsilon') \log ( \log(e^{\nu S} / \nu \epsilon') / \epsilon'))$ queries to $(\alpha_\mathrm{obj}, a_\mathrm{obj}, \nu \epsilon' / (e^{\nu S} \log(e^{\nu S} / \nu \epsilon')) )$-block-encoding of the objective function $\hat{\mathcal{F}}$ or its inverse.
Similarly, we can implement the block-encoding of $\widetilde{\mathcal{F}}(t_{j,q})^\dagger$.
Given that the gate complexity of the $(\alpha_\mathrm{obj}, a_\mathrm{obj}, \epsilon''')$-block-encoding of the objective function $\hat{\mathcal{F}}$ for an error $\epsilon'''$ is $\tilde{\mathcal{O}}\left ( \gamma_u^{-1} \alpha_\mathrm{PDE} d N^\frac{\eta}{d} M^2 m T \log (1/\epsilon''')^{1+1/\beta} \right)$ based on the discussion in Section~\ref{sec:complexity_obj} around Eq.~\eqref{eq:alpha-obj}, the gate complexity of $(1, a_\mathrm{obj} + 2 + \lceil \log ( e^{\nu S} / (\nu \epsilon') ) \rceil, 2\epsilon')$-block-encoding of $\widetilde{\mathcal{F}}(t_{j,q})$ results in
\begin{align}
    \tilde{\mathcal{O}}\left ( \frac{1}{\nu} e^{\nu S} \gamma_u^{-1} \alpha_\mathrm{PDE} d N^\frac{\eta}{d} M^2 T \log \left( \frac{ e^{\nu S}} {\nu \epsilon'} \right)^{3+\frac{1}{\beta}} \right),
\end{align}
where we use $\alpha_\mathrm{obj} = (1 - \delta)^{-2} / 2 \in \mathcal{O}(1)$.

To implement the block-encoding of $\widetilde{\Lambda}(t_{j,q})$, we rearrange the terms, as follows:
\begin{align}
    \widetilde{\Lambda}(t_{j,q}) &= \sum_{q=0}^{Q-1} \ket{q} \!\bra{q} \otimes e^{-\nu t_{j, q}} \frac{\Lambda^{(M)}}{2 \alpha_\Lambda} \nonumber \\
    &= \left( \sum_{q=0}^{Q-1} e^{-\nu t_{j, q}} \ket{q} \!\bra{q} \right) \otimes \frac{\Lambda^{(M)}}{2 \alpha_\Lambda},
\end{align}
which means that $\widetilde{\Lambda}(t_{j,q})$ is separable.
Since the discussion in Appendix~\ref{sec:BE_PDE} provides a block-encoding of $\Lambda^{(M)}$, it suffices to consider the implementation of the block-encoding of $\sum_{q=0}^{Q-1} e^{-\nu t_{j, q}} \ket{q} \!\bra{q}$.
To this end, we further rearrange the term, as follows:
\begin{widetext}
\begin{align}
    \sum_{q=0}^{Q-1} e^{-\nu t_{j, q}} \ket{q} \!\bra{q} &= e^{-\nu j\Delta s } \sum_{q=0}^{Q-1}  (e^{-\nu \frac{\Delta s}{Q} })^q  \ket{q} \!\bra{q} \nonumber \\
    &= e^{-\nu j\Delta s } \sum_{q_{\lceil \log Q \rceil}=0}^{1} \cdots \sum_{q_1 = 0}^{1} (e^{-\nu \frac{\Delta s}{Q} })^{\sum_{b=1}^{{\lceil \log Q \rceil} } 2^{b-1} q_{b} } \ket{q_{\lceil \log Q \rceil}, \dots, q_1} \!\bra{ q_{\lceil \log Q \rceil}, \dots, q_1 } \nonumber \\
    &= e^{-\nu j\Delta s } \sum_{q_{\lceil \log Q \rceil}=0}^{1} (e^{-\nu \frac{\Delta s}{Q} })^{ 2^{{\lceil \log Q \rceil} - 1} q_{1} }  \ket{q_{\lceil \log Q \rceil}} \!\bra{ q_{\lceil \log Q \rceil} } \otimes \cdots \otimes \sum_{q_1 = 0}^{1}  (e^{-\nu \frac{\Delta s}{Q} })^{ q_{1} }  \ket{q_1} \!\bra{ q_1 },
\end{align}
\end{widetext}
which means that this term is also separable.
Each local operator can be represented as LCU, as follows.
\begin{align}
    &\sum_{q_b = 0}^{1}  (e^{-\nu \frac{\Delta s}{Q} })^{ 2^{b-1} q_{b} }  \ket{q_b} \!\bra{ q_b } \nonumber \\
    &= (e^{-\nu \frac{\Delta s}{Q} })^{ 2^{b-1} }  \ket{1} \!\bra{1} + \ket{0} \!\bra{0} \nonumber \\
    &= \frac{1 + (e^{-\nu \frac{\Delta s}{Q} })^{ 2^{b-1} } }{2}  I + \frac{1 - (e^{-\nu \frac{\Delta s}{Q} })^{ 2^{b-1} }  }{2} Z,
\end{align}
where the $\ell_1$-norm of the coefficients is one.
That is, LCU yields the $(1, 1, 0)$-block-encoding of each local operator, which means that we can implement the $(1, \lceil \log (Q) \rceil, 0)$-block-encoding of $\sum_{q=0}^{Q-1} e^{-\nu t_{j, q}} \ket{q} \!\bra{q}$.
Combined with $(5M/h_\xi^2, \lceil \log(5M) \rceil, 0)$-block-encoding of $\Lambda^{(M)}$, we obtain $(5M/h_\xi^2, \lceil \log(5M) \rceil + \lceil \log (Q) \rceil, 0)$-block-encoding of $\widetilde{\Lambda}(t_{j,q})$, which uses
\begin{align}
    \mathcal{O}(Mm^2 + \lceil \log Q \rceil)
\end{align}
single and two-qubit gates.

Since we now have both block-encodings of $\widetilde{\mathcal{F}}(t_{j,q})$ and $\widetilde{\Lambda}(t_{j,q})$, we can implement $\texttt{HAM-T}_j$ with $\alpha_\Lambda = 5M/(2h_\xi^2)$, which is the $(\alpha_\Lambda, a_\mathrm{HT}, 4 \alpha_\Lambda \epsilon')$-block-encoding where $a_\mathrm{HT}=\lceil \log(5M) \rceil + \lceil \log(Q) \rceil + 2a_\mathrm{obj} + 2\lceil \log ((1 / \nu) (e^{\nu S}) / \epsilon' ) \rceil + 4$ and requires
\begin{align}
    \tilde{\mathcal{O}}\left ( \frac{1}{\nu} e^{\nu S} \gamma_u^{-1} \alpha_\mathrm{PDE} d N^\frac{\eta}{d} M^2 T \log \left( \frac{ e^{\nu S}} {\nu \epsilon'} \right)^{3+\frac{1}{\beta}} \right. \nonumber \\
     + Mm^2 + \lceil \log Q \rceil \Bigg),
\end{align}
single and two-qubit gates.
The error $4 \alpha_\Lambda \epsilon'$ comes from the product of block-encodings~\cite[Lemma~53]{gilyen2019quantum}.
By setting $\epsilon' \leq \epsilon / (16 \alpha_\Lambda N_\mathrm{step})$, the error of the block-encoding is bounded by $\epsilon / (4 N_\mathrm{step})$, which satisfies the requirement for performing interaction picture simulation within error $\epsilon$~\cite[Lemma~13]{catli2026efficient}.

\section{Classical complexity analysis via adjoint variable method} \label{sec:classical_adjoint}
In this study, we consider the following optimization problem:
\begin{equation}
\begin{aligned}
    &\min_{\xi \in [0, 1]^M} && \mathcal{F}(u (T; \xi)) \\
    &\text{ subject to: } && \difft{u(t; \xi)} = -A(\xi) u(t; \xi), ~ u(0; \xi) = u_0.
\end{aligned}
\end{equation}
The constraint is the governing equation of the system of interest. It is an ODE obtained by discretizing a PDE, where $u(t) \in \mathbb{C}^N$ is the vector of discretized state variables, $u_0 \in \mathbb{C}^N$ is the initial state, and $A(\xi) \in \mathbb{C}^{N \times N}$ is a design-dependent coefficient matrix whose real part $\Re(A(\xi))$ is assumed to be positive semidefinite.

Here, we solve this problem using a gradient-based optimizer on a classical computer, and derive its computational complexity.
To compute the gradient of the objective function, we employ the adjoint variable method. 
We first formulate the Lagrangian as
\begin{align}
    \mathcal{L} = \mathcal{F}(u (T; \xi)) &+ \int_0^T v(t)^\dagger \left( \difft{u(t; \xi)} + A(\xi) u(t; \xi) \right) \mathrm{d} t \nonumber \\
    &+ v_0^\dagger \left( u(0; \xi) - u_0 \right) \nonumber \\
    &+ \int_0^T \left( \difft{u(t; \xi)} + A(\xi) u(t; \xi) \right)^\dagger v(t)  \mathrm{d} t \nonumber \\
    &+ \left( u(0; \xi) - u_0 \right)^\dagger v_0,
\end{align}
where $v(t) \in \mathbb{C}^N$ and $v_0 \in \mathbb{C}^N$ are Lagrange multipliers, also called adjoint variables.

We then consider the optimality conditions for this Lagrangian.
The first-order optimality condition requires that the partial derivative of $\mathcal{L}$ with respect to $u(t)$ in the direction $u_\delta(t)$ vanish, which yields
\begin{align}
    0 &= \left\langle \frac{\partial \mathcal{L}}{\partial u}, u_\delta \right\rangle \nonumber \\
    &= \left( \frac{\partial \mathcal{F}}{\partial u(T)} \right)^\dagger u_\delta(T) \nonumber \\
    &\quad + \int_0^T \left( - \difft{v(t)^\dagger} + v(t)^\dagger A(\xi) \right) u_\delta(t) \mathrm{d} t \nonumber \\
    &\quad + \left( v(T)^\dagger u_\delta(T) - v(0)^\dagger u_\delta(0) \right) + v_0^\dagger u_\delta(0).
\end{align}
Requiring this condition to hold for all admissible directions $u_\delta(t) \in \mathbb{C}^N$ for $t \in [0, T]$ gives the adjoint equation:
\begin{align}
    - \difft{v(t)} + A(\xi)^\dagger v(t) = 0, \label{eq:adjoint}
\end{align}
which is solved backward in time with the terminal condition
\begin{align}
    v(T) = - \frac{\partial \mathcal{F}}{\partial u(T)} = \begin{cases}
        -2 P u(T) & \text{ if } \mathcal{F} = \mathcal{F}^\mathrm{base}_\mathrm{quad} \\
        -2 P (u(T) - u_\mathrm{ref}) & \text{ if } \mathcal{F} = \mathcal{F}^\mathrm{base}_\mathrm{err}
    \end{cases}. \label{eq:final_condition}
\end{align}

Then, the partial derivative of $\mathcal{L}$ with respect to the design variables $\xi$ yields the gradient of the objective function:
\begin{align}
    \frac{\partial \mathcal{F}}{\partial \xi_\mu} = 2 \mathrm{Re} \left[ \int_0^T v^\dagger(t) A_\mu u(t) \mathrm{d} t \right], \label{eq:adjoint_grad}
\end{align}
where $u(t)$ and $v(t)$ satisfy the governing and adjoint equations, respectively.

We next determine the accuracy required for evaluating the gradient in
Eq.~\eqref{eq:adjoint_grad}.  We use the same objective-value accuracy scale
as in the main text, where an $\epsilon$-accurate solution for the normalized
objective corresponds to a $\gamma_u^2\epsilon$-accurate solution for the
physical objective $\mathcal{F}$ considered here.
Under the assumption that the objective function is smooth and strongly convex, the required gradient accuracy in the
Euclidean norm and the number of iterations sufficient to obtain a
$\gamma_u^2\epsilon$-accurate solution are
$\mathcal{O}(\gamma_u\sqrt{\epsilon})$ and
$\mathcal{O}(\log(1/\epsilon))$, respectively~\cite{friedlander2012hybrid, devolder2013first}.
Thus, if each gradient component in Eq.~\eqref{eq:adjoint_grad} is evaluated within additive error
$\epsilon_g$, then it suffices to take
\begin{align}
    \epsilon_g \in \mathcal{O}\left(
        \gamma_u
        \sqrt{\frac{\epsilon}{M}}
    \right).
\end{align}

We next translate this accuracy requirement into a tolerance for the forward and adjoint simulation used in the evaluation of Eq.~\eqref{eq:adjoint_grad}. 
Let
$e_u$ and $e_v$ denote the errors in the forward and adjoint states, respectively. 
The error arising in evaluating Eq.~\eqref{eq:adjoint_grad} is formulated as
\begin{align}
    \mathcal{E} &:= \left| \int_0^T f_\mu(t) \mathrm{d} t -  \sum_{j=1}^{N_t} w_j \hat{f}_\mu(t_j) \right| \nonumber \\
    & \leq \left| \int_0^T f_\mu(t) \mathrm{d} t -  \sum_{j=1}^{N_t} w_j f_\mu(t_j) \mathrm{d} t \right| \nonumber \\
    & \quad + \left| \sum_{j=1}^{N_t} w_j (f_\mu(t_j) -  \hat{f}_\mu(t_j) ) \right| \nonumber \\
    & \leq \left| \int_0^T f_\mu(t) \mathrm{d} t -  \sum_{j=1}^{N_t} w_j f_\mu(t_j) \mathrm{d} t \right| \nonumber \\
    & \quad + T \max_j \left| (f_\mu(t_j) -  \hat{f}_\mu(t_j) ) \right|
\end{align}
where $f_\mu(t) := v^\dagger(t) A_\mu u(t)$, $\hat{f}_\mu(t) := (v(t) + e_v(t) )^\dagger A_\mu ( u(t) + e_u(t) )$, $N_t$ is the number of time steps, and $w_j$ is the weight for numerical integration satisfying $\sum_j w_j = T$.
The first term is the numerical integration error, while the second term is errors arising from inexact solutions.
The numerical integration error is bounded by $\mathcal{O}( \| f_\mu^{(q+1)} \|_\infty T^{q+2} / N_t^{q+1})$ where $q$ is the order of numerical integration (e.g., $q=1$ for trapezoidal rule, and $q=3$ for Simpson's rule).
Since $u(t)$ and $v(t)$ are the solution of ODEs with coefficient matrices $A$ and $A^\dagger$, respectively, we obtain the bound $\| f_\mu^{(q)} \|_\infty \le 2^q \| A \|^{q+1} U V$ where $V := \sup_{t\in[0,T]} \|v(t)\|$ and $U := \sup_{t\in[0,T]} \|u(t)\|$.
We have that $U, V \in \mathcal{O}(1)$ based on the definition of $\| u(0) \| = 1$ and Eq.~\eqref{eq:final_condition}. 
Consequently, the first term of the error is asymptotically bounded by $\mathcal{O}(  2^{q+1} \|A \|^{q+2} T^{q+2} / N_t^{q+1})$.
The error arising from these inexact solutions can be
bounded as
\begin{align}
    T \max_j \left| f_\mu(t_j) - \hat{f}_\mu(t_j) \right|  &\le \|A \| T \left( V \epsilon_u + U \epsilon_v + \epsilon_u  \epsilon_v \right) \nonumber \\
    & \in \mathcal{O} \left( \|A \| T N_t \left( \frac{\| A \| T}{N_t} \right)^{p+1} \right)
\label{eq:forward_adjoint_error_bound}
\end{align}
where $p$ is the order of time integration for forward and adjoint simulations, $\epsilon_u := \sup_{t \in [0, T]} \| e_u(t) \|$ and $\epsilon_v := \sup_{t \in [0, T]} \| e_v(t) \|$.
Thus, the total numerical error arising in evaluating Eq.~\eqref{eq:adjoint_grad} is
\begin{align}
    &\mathcal{O} \left( \| A \|^{q+2} T^{q+2} / N_t^{q+1} + \| A \|^{p+2} T^{p+2} / N_t^{p} \right) \nonumber \\
    &= \mathcal{O} \left( \| A \|^{p+2} T^{p+2} / N_t^{p} \right)
\end{align}
where we employ $q = p$.
Therefore, it suffices to impose
\begin{equation}
N_t
\in \Theta \left(\frac{\| A \|^{1 + \frac{2}{p}} M^\frac{1}{2p} T^{1 + \frac{2}{p}}}{\gamma_u^\frac{1}{p} \epsilon^\frac{1}{2p}} \right).
\end{equation}
In each step of forward and adjoint simulations, we need to calculate the multiplication of $Au$.
If $A_\mu$ is $s$-sparse, the operator $A$ can be
$\mathcal{O}(Ms)$-sparse. Hence, $N_t$ times multiplication of $Au$ requires $\mathcal{O}(N_t M s)$ arithmetic operations.
Then, we evaluate the gradient by Eq.~\eqref{eq:adjoint_grad}, which requires $N_t$ times multiplication of $v^\dagger(t) A_\mu u(t)$ for $M$ components.
Consequently, one single iteration of optimization requires $\mathcal{O} \left(\frac{\| A \|^{1 + \frac{2}{p}} N M^{1+\frac{1}{2p}} T^{1 + \frac{2}{p}}}{\gamma_u^\frac{1}{p} \epsilon^{\frac{1}{2p}}} \right)$ where we assume $s \in \mathcal{O}(1)$.
Finally, using $\| A \| \leq (M+1) \max(\| A_0 \|, \max_\mu \| A_\mu \|) \leq (M+1) \alpha_A$ with $\alpha_A \geq \max (\| A_0 \|, \max_\mu \| A_\mu \|)$, $\alpha_A = \mathcal{O} (\alpha_{\mathrm{PDE}} d N^{n/d})$ as in the main text, and the required steps of optimization $\mathcal{O}(\log (1 / \epsilon))$, we obtain the classical computational cost as
\begin{equation}
    \tilde{\mathcal{O}} \left(\frac{\alpha_\mathrm{PDE}^{1 + \frac{2}{p}} d^{1 + \frac{2}{p}} N^{\frac{\eta}{d} (1 + \frac{2}{p} + \frac{d}{\eta})} M^{2+\frac{5}{2p}} T^{1 + \frac{2}{p}}}{\gamma_u^\frac{1}{p} \epsilon^{\frac{1}{2p}}} \right).
\label{eq:gradient_arithmetic_cost}
\end{equation}

Regarding the space complexity, the adjoint variable-based classical approach require $\mathcal{O}(N N_t)$ bits for the state vector and adjoint vector, $\mathcal{O}(MN)$ bits for coefficient matrices $A_\mu$, and $\mathcal{O}(M)$ bits for the design variables.
Consequently, the space complexity of the classical approach results in
\begin{equation}
    \mathcal{O} \left( \frac{\alpha_\mathrm{PDE}^{1 + \frac{2}{p}} d^{1 + \frac{2}{p}} N^{\frac{\eta}{d}(1 + \frac{2}{p} + \frac{d}{\eta})} M^{1 + \frac{5}{2p}} T^{1 + \frac{2}{p}}}{\gamma_u^\frac{1}{p} \epsilon^\frac{1}{2p}} \right).
\end{equation}

\section{Friction parameter setting by normalizing objective function} \label{sec:friction}
Quantum Hamiltonian descent employs the quantum counterpart of the classical dynamics governed by Newton's equation of motion
\begin{align}
    \ddot{\xi} + \nu \dot{\xi} + \nabla \mathcal{F} = 0,
\end{align}
which describes the frictional dynamics driven by a force $\nabla \mathcal{F}$.
Hence, the time to solution highly depends on the contrast of the friction parameter $\nu$ and the force $\nabla \mathcal{F}$, which makes it important to appropriately set the friction parameter based on the scale of $\mathcal{F}$.
When a lower curvature bound of the landscape of the objective function, such as a strong-convexity parameter, is available, a curvature-dependent choice of the friction parameter is theoretically preferable, as in Ref.~\cite{leng2025quantum, catli2026efficient}.
In this appendix, we describe a practical scale-normalization heuristic used in the numerical examples when such curvature information is not available a priori.

Here, we set the friction parameter by normalizing Newton's equation of motion using the characteristic objective function value $\bar{\mathcal{F}}$.
Dividing Newton's equation of motion by $\bar{\mathcal{F}}$, we obtain
\begin{align}
    \frac{1}{\bar{\mathcal{F}}} \frac{\mathrm{d}^2 \xi}{\mathrm{d} t^2} + \frac{\nu}{\bar{\mathcal{F}}} \frac{\mathrm{d} \xi}{\mathrm{d} t} + \frac{1}{\bar{\mathcal{F}}} \nabla \mathcal{F} = 0.
\end{align}
Then, we normalize time, the friction parameter, and the objective function as
\begin{align}
    \tilde{t} := \sqrt{\bar{\mathcal{F}}} t, ~
    \tilde{\nu} := \frac{\nu}{\sqrt{\mathcal{\bar{F}}}}, ~
    \tilde{\mathcal{F}} := \frac{\mathcal{F}}{\bar{\mathcal{F}}},
\end{align}
which yields the normalized Newton's equation of motion
\begin{align}
    \frac{\mathrm{d}^2 \xi}{\mathrm{d} \tilde{t}^2} + \tilde{\nu} \frac{\mathrm{d} \xi}{\mathrm{d} \tilde{t}} + \nabla \tilde{\mathcal{F}} = 0.
\end{align}
Since this equation no longer depends on the scale of the objective function, we can set the friction parameter $\tilde{\nu}$ independent of the problem of interest.
From empirical study, we employ $\tilde{\nu} = 1$, which therefore means that we set $\nu = \sqrt{\bar{\mathcal{F}}}$ for QHD. 
As the characteristic objective function value, we use the mean value of the normalized objective function for all possible parameters, that is,
\begin{align}
    \bar{\mathcal{F}} &= \frac{1}{2^{Mm}} \sum_{\xi} \mathcal{F}(\xi) \nonumber \\
    &= \alpha_\mathrm{obj} \left( \bra{\psi(0)} \! \bra{0}^{\otimes a_\mathrm{obj}} U_\mathrm{obj} \ket{\psi(0)} \! \ket{0}^{\otimes a_\mathrm{obj}} + 1 \right),
\end{align}
where $\ket{\psi(0)} = 2^{-Mm/2}\sum_\xi \ket{\xi}$ is the initial state of QHD in this study.
This value can be estimated by $\mathcal{O}(\varepsilon^{-1})$ queries to $U_\mathrm{obj}$ and $\mathcal{O}(\varepsilon^{-1})$ times state preparation to achieve precision $\varepsilon$~\cite{knill2007optimal}.

\section{Block-encoding of finite-difference gradient components of the objective function}
\label{sec:gradient_block_encoding}

In the main text, we construct a block-encoding of the objective operator of the form
\begin{align}
    &\left(I^{\otimes Mm} \otimes \bra{0}^{\otimes a_{\mathrm{obj}}}\right)
    U_{\mathrm{obj}}
    \left(I^{\otimes Mm} \otimes \ket{0}^{\otimes a_{\mathrm{obj}}}\right)
    \nonumber \\
    &=
    \frac{1}{\alpha_{\mathrm{obj}}}
    \left(
        \hat{\mathcal{F}} - \delta_{\mathrm{obj}} I^{\otimes Mm}
    \right)
    + E_{\mathrm{obj}},
\end{align}
where
\begin{align}
    \hat{\mathcal{F}}
    :=
    \sum_{\xi \in \Xi_h}
    \mathcal{F}(\xi)\ket{\xi}\!\bra{\xi},
    \qquad
    \|E_{\mathrm{obj}}\| \le \varepsilon_{\mathrm{obj}},
\end{align}
and $\Xi_h$ denotes the discretized design grid.
This construction can be extended in a straightforward manner to obtain coherent first-order access on the discretized design space.
Such access may be useful as an interface primitive for coherent optimization frameworks requiring derivative information.

For each design coordinate $\mu \in [M]$, we define the cyclic shift operators on the $\mu$-th $m$-qubit design register by
\begin{align}
    \texttt{SHIFT}^{+}
    &:=
    \sum_{j=0}^{2^m-1} \ket{j+1 \bmod 2^m}\!\bra{j},
    \\
    \texttt{SHIFT}^{-}
    &:=
    (\texttt{SHIFT}^{+})^\dagger
    =
    \sum_{j=0}^{2^m-1} \ket{j}\!\bra{j+1 \bmod 2^m},
\end{align}
and
\begin{align}
    \texttt{SHIFT}^{+}_{\mu}
    &:=
    I^{\otimes m(M-\mu)}
    \otimes \texttt{SHIFT}^{+}
    \otimes I^{\otimes m(\mu-1)},
    \\
    \texttt{SHIFT}^{-}_{\mu}
    &:=
    I^{\otimes m(M-\mu)}
    \otimes \texttt{SHIFT}^{-}
    \otimes I^{\otimes m(\mu-1)}.
\end{align}
Since each computational basis state $\ket{\xi}$ represents a discretized design vector with mesh size $h_\xi$,
these unitaries acts as
\begin{align}
    \texttt{SHIFT}^{\pm}_{\mu}\ket{\xi}
    =
    \ket{\xi \pm h_\xi e_\mu},
\end{align}
where $e_\mu \in \mathbb{R}^M$ is the $\mu$-th standard basis vector, and the addition is understood modulo the discrete grid.

We then define the shifted objective block-encodings
\begin{align}
    U_{\mu,+}
    &:=
    \left(I^{\otimes a_{\mathrm{obj}}} \otimes \texttt{SHIFT}^-_\mu\right)
    U_{\mathrm{obj}}
    \left(I^{\otimes a_{\mathrm{obj}}} \otimes \texttt{SHIFT}^+_\mu\right),
    \\
    U_{\mu,-}
    &:=
    \left(I^{\otimes a_{\mathrm{obj}}} \otimes \texttt{SHIFT}^+_\mu\right)
    U_{\mathrm{obj}}
    \left(I^{\otimes a_{\mathrm{obj}}} \otimes \texttt{SHIFT}^-_\mu\right).
\end{align}
These are again block-encodings with the same normalization, ancillary qubits, and error:
\begin{align}
    &
    \left(I^{\otimes Mm} \otimes \bra{0}^{\otimes a_{\mathrm{obj}}}\right)
    U_{\mu,\pm}
    \left(I^{\otimes Mm} \otimes \ket{0}^{\otimes a_{\mathrm{obj}}}\right)
    \nonumber\\
    &=
    \frac{1}{\alpha_{\mathrm{obj}}}
    \left(
        \hat{\mathcal{F}}_{\mu,\pm}
        -
        \delta_{\mathrm{obj}} I^{\otimes Mm}
    \right)
    +
    E_{\mu,\pm},
\end{align}
where $\|E_{\mu,\pm}\|\le \varepsilon_{\mathrm{obj}}$ and
\begin{align}
    \hat{\mathcal{F}}_{\mu,\pm}
    :=
    \sum_{\xi\in\Xi_h}
    \mathcal{F}(\xi \pm h_\xi e_\mu)\ket{\xi}\!\bra{\xi}.
\end{align}

Therefore, the offset term cancels in the finite difference, and the central-difference operator
\begin{align}
    \hat{\mathcal{F}'_\mu}
    &:=\frac{1}{2 h_\xi}(\hat{\mathcal{F}}_{\mu,+} - \hat{\mathcal{F}}_{\mu,-}) \notag \\
    &=  
    \sum_{\xi\in\Xi_h}
    \frac{
        \mathcal{F}(\xi + h_\xi e_\mu)
        -
        \mathcal{F}(\xi - h_\xi e_\mu)
    }{2h_\xi}
    \ket{\xi}\!\bra{\xi}
\end{align}
can be constructed as a linear combination of the two shifted block-encodings.
By the standard lemma for block-encoded matrices~\cite[Lemma~52]{gilyen2019quantum}, this yields an $( \alpha_{\mathrm{obj}} / h_\xi, a_{\mathrm{obj}} + 1, \alpha_\mathrm{obj} \varepsilon_{\mathrm{obj}} / h_\xi)$-block-encoding of $\hat{\mathcal{F}'_\mu}$.
If the objective function is sufficiently smooth with respect to $\xi_\mu$, then for interior grid points the above operator approximates the corresponding partial derivative with the second-order truncation error.

Note that the cyclic shifts above impose a periodic boundary condition on the boundary of the design grid.
If one instead wishes to use one-sided finite differences at the boundaries, one may replace the cyclic shifts by truncated shifts together with boundary-supported correction terms acting only on the basis states corresponding to $\xi_\mu=0$ and $\xi_\mu=1$.
The same idea also extends to higher-order finite-difference formulas by combining multiple shifted copies of the objective block-encoding with appropriate coefficients.

\bibliography{ref}

\begin{thebibliography}{61}%
\makeatletter
\providecommand \@ifxundefined [1]{%
 \@ifx{#1\undefined}
}%
\providecommand \@ifnum [1]{%
 \ifnum #1\expandafter \@firstoftwo
 \else \expandafter \@secondoftwo
 \fi
}%
\providecommand \@ifx [1]{%
 \ifx #1\expandafter \@firstoftwo
 \else \expandafter \@secondoftwo
 \fi
}%
\providecommand \natexlab [1]{#1}%
\providecommand \enquote  [1]{``#1''}%
\providecommand \bibnamefont  [1]{#1}%
\providecommand \bibfnamefont [1]{#1}%
\providecommand \citenamefont [1]{#1}%
\providecommand \href@noop [0]{\@secondoftwo}%
\providecommand \href [0]{\begingroup \@sanitize@url \@href}%
\providecommand \@href[1]{\@@startlink{#1}\@@href}%
\providecommand \@@href[1]{\endgroup#1\@@endlink}%
\providecommand \@sanitize@url [0]{\catcode `\\12\catcode `\$12\catcode `\&12\catcode `\#12\catcode `\^12\catcode `\_12\catcode `\%12\relax}%
\providecommand \@@startlink[1]{}%
\providecommand \@@endlink[0]{}%
\providecommand \url  [0]{\begingroup\@sanitize@url \@url }%
\providecommand \@url [1]{\endgroup\@href {#1}{\urlprefix }}%
\providecommand \urlprefix  [0]{URL }%
\providecommand \Eprint [0]{\href }%
\providecommand \doibase [0]{https://doi.org/}%
\providecommand \selectlanguage [0]{\@gobble}%
\providecommand \bibinfo  [0]{\@secondoftwo}%
\providecommand \bibfield  [0]{\@secondoftwo}%
\providecommand \translation [1]{[#1]}%
\providecommand \BibitemOpen [0]{}%
\providecommand \bibitemStop [0]{}%
\providecommand \bibitemNoStop [0]{.\EOS\space}%
\providecommand \EOS [0]{\spacefactor3000\relax}%
\providecommand \BibitemShut  [1]{\csname bibitem#1\endcsname}%
\let\auto@bib@innerbib\@empty
\bibitem [{\citenamefont {Hinze}\ \emph {et~al.}(2008)\citenamefont {Hinze}, \citenamefont {Pinnau}, \citenamefont {Ulbrich},\ and\ \citenamefont {Ulbrich}}]{hinze2008optimization}%
  \BibitemOpen
  \bibfield  {author} {\bibinfo {author} {\bibfnamefont {M.}~\bibnamefont {Hinze}}, \bibinfo {author} {\bibfnamefont {R.}~\bibnamefont {Pinnau}}, \bibinfo {author} {\bibfnamefont {M.}~\bibnamefont {Ulbrich}},\ and\ \bibinfo {author} {\bibfnamefont {S.}~\bibnamefont {Ulbrich}},\ }\href@noop {} {\emph {\bibinfo {title} {{Optimization with PDE constraints}}}},\ Vol.~\bibinfo {volume} {23}\ (\bibinfo  {publisher} {Springer Science \& Business Media},\ \bibinfo {year} {2008})\BibitemShut {NoStop}%
\bibitem [{\citenamefont {De~los Reyes}(2015)}]{de2015numerical}%
  \BibitemOpen
  \bibfield  {author} {\bibinfo {author} {\bibfnamefont {J.~C.}\ \bibnamefont {De~los Reyes}},\ }\href@noop {} {\emph {\bibinfo {title} {{Numerical PDE-constrained optimization}}}}\ (\bibinfo  {publisher} {Springer},\ \bibinfo {year} {2015})\BibitemShut {NoStop}%
\bibitem [{\citenamefont {Deb}(2012)}]{deb2012optimization}%
  \BibitemOpen
  \bibfield  {author} {\bibinfo {author} {\bibfnamefont {K.}~\bibnamefont {Deb}},\ }\href@noop {} {\emph {\bibinfo {title} {Optimization for engineering design: Algorithms and examples}}}\ (\bibinfo  {publisher} {PHI Learning Pvt. Ltd.},\ \bibinfo {year} {2012})\BibitemShut {NoStop}%
\bibitem [{\citenamefont {Martins}\ and\ \citenamefont {Ning}(2021)}]{martins2021engineering}%
  \BibitemOpen
  \bibfield  {author} {\bibinfo {author} {\bibfnamefont {J.~R.}\ \bibnamefont {Martins}}\ and\ \bibinfo {author} {\bibfnamefont {A.}~\bibnamefont {Ning}},\ }\href@noop {} {\emph {\bibinfo {title} {Engineering design optimization}}}\ (\bibinfo  {publisher} {Cambridge University Press},\ \bibinfo {year} {2021})\BibitemShut {NoStop}%
\bibitem [{\citenamefont {Sokolowski}\ and\ \citenamefont {Zol{\'e}sio}(1992)}]{sokolowski1992introduction}%
  \BibitemOpen
  \bibfield  {author} {\bibinfo {author} {\bibfnamefont {J.}~\bibnamefont {Sokolowski}}\ and\ \bibinfo {author} {\bibfnamefont {J.-P.}\ \bibnamefont {Zol{\'e}sio}},\ }\href@noop {} {\emph {\bibinfo {title} {Introduction to Shape Optimization: Shape Sensitivity Analysis}}}\ (\bibinfo  {publisher} {Springer},\ \bibinfo {year} {1992})\ pp.\ \bibinfo {pages} {5--12}\BibitemShut {NoStop}%
\bibitem [{\citenamefont {Haslinger}\ and\ \citenamefont {M{\"a}kinen}(2003)}]{haslinger2003introduction}%
  \BibitemOpen
  \bibfield  {author} {\bibinfo {author} {\bibfnamefont {J.}~\bibnamefont {Haslinger}}\ and\ \bibinfo {author} {\bibfnamefont {R.~A.}\ \bibnamefont {M{\"a}kinen}},\ }\href@noop {} {\emph {\bibinfo {title} {Introduction to shape optimization: theory, approximation, and computation}}}\ (\bibinfo  {publisher} {SIAM},\ \bibinfo {year} {2003})\BibitemShut {NoStop}%
\bibitem [{\citenamefont {Bends{\o}e}\ and\ \citenamefont {Kikuchi}(1988)}]{bendsoe1988generating}%
  \BibitemOpen
  \bibfield  {author} {\bibinfo {author} {\bibfnamefont {M.~P.}\ \bibnamefont {Bends{\o}e}}\ and\ \bibinfo {author} {\bibfnamefont {N.}~\bibnamefont {Kikuchi}},\ }\bibfield  {title} {\bibinfo {title} {Generating optimal topologies in structural design using a homogenization method},\ }\href@noop {} {\bibfield  {journal} {\bibinfo  {journal} {Computer methods in applied mechanics and engineering}\ }\textbf {\bibinfo {volume} {71}},\ \bibinfo {pages} {197} (\bibinfo {year} {1988})}\BibitemShut {NoStop}%
\bibitem [{\citenamefont {Tr{\"o}ltzsch}(2024)}]{troltzsch2024optimal}%
  \BibitemOpen
  \bibfield  {author} {\bibinfo {author} {\bibfnamefont {F.}~\bibnamefont {Tr{\"o}ltzsch}},\ }\href@noop {} {\emph {\bibinfo {title} {Optimal control of partial differential equations: theory, methods and applications}}},\ Vol.\ \bibinfo {volume} {112}\ (\bibinfo  {publisher} {American Mathematical Society},\ \bibinfo {year} {2024})\BibitemShut {NoStop}%
\bibitem [{\citenamefont {Harrow}\ \emph {et~al.}(2009)\citenamefont {Harrow}, \citenamefont {Hassidim},\ and\ \citenamefont {Lloyd}}]{harrow2009quantum}%
  \BibitemOpen
  \bibfield  {author} {\bibinfo {author} {\bibfnamefont {A.~W.}\ \bibnamefont {Harrow}}, \bibinfo {author} {\bibfnamefont {A.}~\bibnamefont {Hassidim}},\ and\ \bibinfo {author} {\bibfnamefont {S.}~\bibnamefont {Lloyd}},\ }\bibfield  {title} {\bibinfo {title} {Quantum algorithm for linear systems of equations},\ }\href@noop {} {\bibfield  {journal} {\bibinfo  {journal} {Physical Review Letters}\ }\textbf {\bibinfo {volume} {103}},\ \bibinfo {pages} {150502} (\bibinfo {year} {2009})}\BibitemShut {NoStop}%
\bibitem [{\citenamefont {Childs}\ \emph {et~al.}(2017)\citenamefont {Childs}, \citenamefont {Kothari},\ and\ \citenamefont {Somma}}]{childs2017quantum}%
  \BibitemOpen
  \bibfield  {author} {\bibinfo {author} {\bibfnamefont {A.~M.}\ \bibnamefont {Childs}}, \bibinfo {author} {\bibfnamefont {R.}~\bibnamefont {Kothari}},\ and\ \bibinfo {author} {\bibfnamefont {R.~D.}\ \bibnamefont {Somma}},\ }\bibfield  {title} {\bibinfo {title} {Quantum algorithm for systems of linear equations with exponentially improved dependence on precision},\ }\href@noop {} {\bibfield  {journal} {\bibinfo  {journal} {SIAM Journal on Computing}\ }\textbf {\bibinfo {volume} {46}},\ \bibinfo {pages} {1920} (\bibinfo {year} {2017})}\BibitemShut {NoStop}%
\bibitem [{\citenamefont {Martyn}\ \emph {et~al.}(2021)\citenamefont {Martyn}, \citenamefont {Rossi}, \citenamefont {Tan},\ and\ \citenamefont {Chuang}}]{martyn2021grand}%
  \BibitemOpen
  \bibfield  {author} {\bibinfo {author} {\bibfnamefont {J.~M.}\ \bibnamefont {Martyn}}, \bibinfo {author} {\bibfnamefont {Z.~M.}\ \bibnamefont {Rossi}}, \bibinfo {author} {\bibfnamefont {A.~K.}\ \bibnamefont {Tan}},\ and\ \bibinfo {author} {\bibfnamefont {I.~L.}\ \bibnamefont {Chuang}},\ }\bibfield  {title} {\bibinfo {title} {Grand unification of quantum algorithms},\ }\href@noop {} {\bibfield  {journal} {\bibinfo  {journal} {PRX quantum}\ }\textbf {\bibinfo {volume} {2}},\ \bibinfo {pages} {040203} (\bibinfo {year} {2021})}\BibitemShut {NoStop}%
\bibitem [{\citenamefont {Cao}\ \emph {et~al.}(2013)\citenamefont {Cao}, \citenamefont {Papageorgiou}, \citenamefont {Petras}, \citenamefont {Traub},\ and\ \citenamefont {Kais}}]{cao2013quantum}%
  \BibitemOpen
  \bibfield  {author} {\bibinfo {author} {\bibfnamefont {Y.}~\bibnamefont {Cao}}, \bibinfo {author} {\bibfnamefont {A.}~\bibnamefont {Papageorgiou}}, \bibinfo {author} {\bibfnamefont {I.}~\bibnamefont {Petras}}, \bibinfo {author} {\bibfnamefont {J.}~\bibnamefont {Traub}},\ and\ \bibinfo {author} {\bibfnamefont {S.}~\bibnamefont {Kais}},\ }\bibfield  {title} {\bibinfo {title} {Quantum algorithm and circuit design solving the poisson equation},\ }\href@noop {} {\bibfield  {journal} {\bibinfo  {journal} {New Journal of Physics}\ }\textbf {\bibinfo {volume} {15}},\ \bibinfo {pages} {013021} (\bibinfo {year} {2013})}\BibitemShut {NoStop}%
\bibitem [{\citenamefont {Montanaro}\ and\ \citenamefont {Pallister}(2016)}]{montanaro2016quantum}%
  \BibitemOpen
  \bibfield  {author} {\bibinfo {author} {\bibfnamefont {A.}~\bibnamefont {Montanaro}}\ and\ \bibinfo {author} {\bibfnamefont {S.}~\bibnamefont {Pallister}},\ }\bibfield  {title} {\bibinfo {title} {Quantum algorithms and the finite element method},\ }\href@noop {} {\bibfield  {journal} {\bibinfo  {journal} {Physical Review A}\ }\textbf {\bibinfo {volume} {93}},\ \bibinfo {pages} {032324} (\bibinfo {year} {2016})}\BibitemShut {NoStop}%
\bibitem [{\citenamefont {Berry}(2014)}]{berry2014high}%
  \BibitemOpen
  \bibfield  {author} {\bibinfo {author} {\bibfnamefont {D.~W.}\ \bibnamefont {Berry}},\ }\bibfield  {title} {\bibinfo {title} {High-order quantum algorithm for solving linear differential equations},\ }\href@noop {} {\bibfield  {journal} {\bibinfo  {journal} {Journal of Physics A: Mathematical and Theoretical}\ }\textbf {\bibinfo {volume} {47}},\ \bibinfo {pages} {105301} (\bibinfo {year} {2014})}\BibitemShut {NoStop}%
\bibitem [{\citenamefont {An}\ \emph {et~al.}(2023)\citenamefont {An}, \citenamefont {Liu},\ and\ \citenamefont {Lin}}]{an2023linear}%
  \BibitemOpen
  \bibfield  {author} {\bibinfo {author} {\bibfnamefont {D.}~\bibnamefont {An}}, \bibinfo {author} {\bibfnamefont {J.-P.}\ \bibnamefont {Liu}},\ and\ \bibinfo {author} {\bibfnamefont {L.}~\bibnamefont {Lin}},\ }\bibfield  {title} {\bibinfo {title} {{Linear combination of Hamiltonian simulation for nonunitary dynamics with optimal state preparation cost}},\ }\href@noop {} {\bibfield  {journal} {\bibinfo  {journal} {Physical Review Letters}\ }\textbf {\bibinfo {volume} {131}},\ \bibinfo {pages} {150603} (\bibinfo {year} {2023})}\BibitemShut {NoStop}%
\bibitem [{\citenamefont {An}\ \emph {et~al.}(2026)\citenamefont {An}, \citenamefont {Childs},\ and\ \citenamefont {Lin}}]{an2023quantum}%
  \BibitemOpen
  \bibfield  {author} {\bibinfo {author} {\bibfnamefont {D.}~\bibnamefont {An}}, \bibinfo {author} {\bibfnamefont {A.~M.}\ \bibnamefont {Childs}},\ and\ \bibinfo {author} {\bibfnamefont {L.}~\bibnamefont {Lin}},\ }\bibfield  {title} {\bibinfo {title} {Quantum algorithm for linear non-unitary dynamics with near-optimal dependence on all parameters},\ }\href@noop {} {\bibfield  {journal} {\bibinfo  {journal} {Communications in Mathematical Physics}\ }\textbf {\bibinfo {volume} {407}},\ \bibinfo {pages} {19} (\bibinfo {year} {2026})}\BibitemShut {NoStop}%
\bibitem [{\citenamefont {Pocrnic}\ \emph {et~al.}(2025)\citenamefont {Pocrnic}, \citenamefont {Johnson}, \citenamefont {Katabarwa},\ and\ \citenamefont {Wiebe}}]{pocrnic2025constant}%
  \BibitemOpen
  \bibfield  {author} {\bibinfo {author} {\bibfnamefont {M.}~\bibnamefont {Pocrnic}}, \bibinfo {author} {\bibfnamefont {P.~D.}\ \bibnamefont {Johnson}}, \bibinfo {author} {\bibfnamefont {A.}~\bibnamefont {Katabarwa}},\ and\ \bibinfo {author} {\bibfnamefont {N.}~\bibnamefont {Wiebe}},\ }\bibfield  {title} {\bibinfo {title} {Constant-factor improvements in quantum algorithms for linear differential equations},\ }\href@noop {} {\bibfield  {journal} {\bibinfo  {journal} {arXiv preprint arXiv:2506.20760}\ } (\bibinfo {year} {2025})}\BibitemShut {NoStop}%
\bibitem [{\citenamefont {Low}\ and\ \citenamefont {Somma}(2025)}]{low2025optimal}%
  \BibitemOpen
  \bibfield  {author} {\bibinfo {author} {\bibfnamefont {G.~H.}\ \bibnamefont {Low}}\ and\ \bibinfo {author} {\bibfnamefont {R.~D.}\ \bibnamefont {Somma}},\ }\bibfield  {title} {\bibinfo {title} {Optimal quantum simulation of linear non-unitary dynamics},\ }\href@noop {} {\bibfield  {journal} {\bibinfo  {journal} {arXiv preprint arXiv:2508.19238}\ } (\bibinfo {year} {2025})}\BibitemShut {NoStop}%
\bibitem [{\citenamefont {Jin}\ \emph {et~al.}(2023)\citenamefont {Jin}, \citenamefont {Liu},\ and\ \citenamefont {Yu}}]{jin2023aquantum}%
  \BibitemOpen
  \bibfield  {author} {\bibinfo {author} {\bibfnamefont {S.}~\bibnamefont {Jin}}, \bibinfo {author} {\bibfnamefont {N.}~\bibnamefont {Liu}},\ and\ \bibinfo {author} {\bibfnamefont {Y.}~\bibnamefont {Yu}},\ }\bibfield  {title} {\bibinfo {title} {{Quantum simulation of partial differential equations: Applications and detailed analysis}},\ }\href {https://doi.org/10.1103/PhysRevA.108.032603} {\bibfield  {journal} {\bibinfo  {journal} {Physical Review A}\ }\textbf {\bibinfo {volume} {108}},\ \bibinfo {pages} {032603} (\bibinfo {year} {2023})}\BibitemShut {NoStop}%
\bibitem [{\citenamefont {Jin}\ \emph {et~al.}(2024)\citenamefont {Jin}, \citenamefont {Liu},\ and\ \citenamefont {Yu}}]{jin2024quantum}%
  \BibitemOpen
  \bibfield  {author} {\bibinfo {author} {\bibfnamefont {S.}~\bibnamefont {Jin}}, \bibinfo {author} {\bibfnamefont {N.}~\bibnamefont {Liu}},\ and\ \bibinfo {author} {\bibfnamefont {Y.}~\bibnamefont {Yu}},\ }\bibfield  {title} {\bibinfo {title} {Quantum simulation of partial differential equations via schr{\"o}dingerization},\ }\href@noop {} {\bibfield  {journal} {\bibinfo  {journal} {Physical Review Letters}\ }\textbf {\bibinfo {volume} {133}},\ \bibinfo {pages} {230602} (\bibinfo {year} {2024})}\BibitemShut {NoStop}%
\bibitem [{\citenamefont {Jin}\ \emph {et~al.}(2026{\natexlab{a}})\citenamefont {Jin}, \citenamefont {Liu}, \citenamefont {Ma}, \citenamefont {Peng},\ and\ \citenamefont {Yu}}]{jin2026schrodingerization}%
  \BibitemOpen
  \bibfield  {author} {\bibinfo {author} {\bibfnamefont {S.}~\bibnamefont {Jin}}, \bibinfo {author} {\bibfnamefont {N.}~\bibnamefont {Liu}}, \bibinfo {author} {\bibfnamefont {C.}~\bibnamefont {Ma}}, \bibinfo {author} {\bibfnamefont {Y.}~\bibnamefont {Peng}},\ and\ \bibinfo {author} {\bibfnamefont {Y.}~\bibnamefont {Yu}},\ }\bibfield  {title} {\bibinfo {title} {On the schr{\"o}dingerization method for linear non-unitary dynamics with optimal dependence on matrix queries},\ }\href@noop {} {\bibfield  {journal} {\bibinfo  {journal} {Communications in Mathematical Sciences}\ }\textbf {\bibinfo {volume} {24}},\ \bibinfo {pages} {565} (\bibinfo {year} {2026}{\natexlab{a}})}\BibitemShut {NoStop}%
\bibitem [{\citenamefont {Shang}\ \emph {et~al.}(2025)\citenamefont {Shang}, \citenamefont {Guo}, \citenamefont {An},\ and\ \citenamefont {Zhao}}]{shang2025designing}%
  \BibitemOpen
  \bibfield  {author} {\bibinfo {author} {\bibfnamefont {Z.-X.}\ \bibnamefont {Shang}}, \bibinfo {author} {\bibfnamefont {N.}~\bibnamefont {Guo}}, \bibinfo {author} {\bibfnamefont {D.}~\bibnamefont {An}},\ and\ \bibinfo {author} {\bibfnamefont {Q.}~\bibnamefont {Zhao}},\ }\bibfield  {title} {\bibinfo {title} {Designing a nearly optimal quantum algorithm for linear differential equations via lindbladians},\ }\href@noop {} {\bibfield  {journal} {\bibinfo  {journal} {Physical Review Letters}\ }\textbf {\bibinfo {volume} {135}},\ \bibinfo {pages} {120604} (\bibinfo {year} {2025})}\BibitemShut {NoStop}%
\bibitem [{\citenamefont {Fang}\ \emph {et~al.}(2025)\citenamefont {Fang}, \citenamefont {George},\ and\ \citenamefont {Tong}}]{fang2025qubit}%
  \BibitemOpen
  \bibfield  {author} {\bibinfo {author} {\bibfnamefont {D.}~\bibnamefont {Fang}}, \bibinfo {author} {\bibfnamefont {D.~L.}\ \bibnamefont {George}},\ and\ \bibinfo {author} {\bibfnamefont {Y.}~\bibnamefont {Tong}},\ }\bibfield  {title} {\bibinfo {title} {Qubit-efficient quantum algorithm for linear differential equations},\ }\href@noop {} {\bibfield  {journal} {\bibinfo  {journal} {arXiv preprint arXiv:2507.16995}\ } (\bibinfo {year} {2025})}\BibitemShut {NoStop}%
\bibitem [{\citenamefont {Costa}\ \emph {et~al.}(2019)\citenamefont {Costa}, \citenamefont {Jordan},\ and\ \citenamefont {Ostrander}}]{costa2019quantum}%
  \BibitemOpen
  \bibfield  {author} {\bibinfo {author} {\bibfnamefont {P.~C.}\ \bibnamefont {Costa}}, \bibinfo {author} {\bibfnamefont {S.}~\bibnamefont {Jordan}},\ and\ \bibinfo {author} {\bibfnamefont {A.}~\bibnamefont {Ostrander}},\ }\bibfield  {title} {\bibinfo {title} {Quantum algorithm for simulating the wave equation},\ }\href@noop {} {\bibfield  {journal} {\bibinfo  {journal} {Physical Review A}\ }\textbf {\bibinfo {volume} {99}},\ \bibinfo {pages} {012323} (\bibinfo {year} {2019})}\BibitemShut {NoStop}%
\bibitem [{\citenamefont {Miyamoto}\ \emph {et~al.}(2024)\citenamefont {Miyamoto}, \citenamefont {Yamazaki}, \citenamefont {Uchida}, \citenamefont {Fujisawa},\ and\ \citenamefont {Yoshida}}]{miyamoto2024quantum}%
  \BibitemOpen
  \bibfield  {author} {\bibinfo {author} {\bibfnamefont {K.}~\bibnamefont {Miyamoto}}, \bibinfo {author} {\bibfnamefont {S.}~\bibnamefont {Yamazaki}}, \bibinfo {author} {\bibfnamefont {F.}~\bibnamefont {Uchida}}, \bibinfo {author} {\bibfnamefont {K.}~\bibnamefont {Fujisawa}},\ and\ \bibinfo {author} {\bibfnamefont {N.}~\bibnamefont {Yoshida}},\ }\bibfield  {title} {\bibinfo {title} {Quantum algorithm for the vlasov simulation of the large-scale structure formation with massive neutrinos},\ }\href@noop {} {\bibfield  {journal} {\bibinfo  {journal} {Physical Review Research}\ }\textbf {\bibinfo {volume} {6}},\ \bibinfo {pages} {013200} (\bibinfo {year} {2024})}\BibitemShut {NoStop}%
\bibitem [{\citenamefont {Brearley}\ and\ \citenamefont {Laizet}(2024)}]{brearley2024quantum}%
  \BibitemOpen
  \bibfield  {author} {\bibinfo {author} {\bibfnamefont {P.}~\bibnamefont {Brearley}}\ and\ \bibinfo {author} {\bibfnamefont {S.}~\bibnamefont {Laizet}},\ }\bibfield  {title} {\bibinfo {title} {{Quantum algorithm for solving the advection equation using Hamiltonian simulation}},\ }\href@noop {} {\bibfield  {journal} {\bibinfo  {journal} {Physical Review A}\ }\textbf {\bibinfo {volume} {110}},\ \bibinfo {pages} {012430} (\bibinfo {year} {2024})}\BibitemShut {NoStop}%
\bibitem [{\citenamefont {Sato}\ \emph {et~al.}(2024)\citenamefont {Sato}, \citenamefont {Kondo}, \citenamefont {Hamamura}, \citenamefont {Onodera},\ and\ \citenamefont {Yamamoto}}]{sato2024hamiltonian}%
  \BibitemOpen
  \bibfield  {author} {\bibinfo {author} {\bibfnamefont {Y.}~\bibnamefont {Sato}}, \bibinfo {author} {\bibfnamefont {R.}~\bibnamefont {Kondo}}, \bibinfo {author} {\bibfnamefont {I.}~\bibnamefont {Hamamura}}, \bibinfo {author} {\bibfnamefont {T.}~\bibnamefont {Onodera}},\ and\ \bibinfo {author} {\bibfnamefont {N.}~\bibnamefont {Yamamoto}},\ }\bibfield  {title} {\bibinfo {title} {Hamiltonian simulation for hyperbolic partial differential equations by scalable quantum circuits},\ }\href@noop {} {\bibfield  {journal} {\bibinfo  {journal} {Physical Review Research}\ }\textbf {\bibinfo {volume} {6}},\ \bibinfo {pages} {033246} (\bibinfo {year} {2024})}\BibitemShut {NoStop}%
\bibitem [{\citenamefont {Wright}\ \emph {et~al.}(2024)\citenamefont {Wright}, \citenamefont {Mc~Keever}, \citenamefont {First}, \citenamefont {Johnston}, \citenamefont {Tillay}, \citenamefont {Chaney}, \citenamefont {Rosenkranz},\ and\ \citenamefont {Lubasch}}]{wright2024noisy}%
  \BibitemOpen
  \bibfield  {author} {\bibinfo {author} {\bibfnamefont {L.}~\bibnamefont {Wright}}, \bibinfo {author} {\bibfnamefont {C.}~\bibnamefont {Mc~Keever}}, \bibinfo {author} {\bibfnamefont {J.~T.}\ \bibnamefont {First}}, \bibinfo {author} {\bibfnamefont {R.}~\bibnamefont {Johnston}}, \bibinfo {author} {\bibfnamefont {J.}~\bibnamefont {Tillay}}, \bibinfo {author} {\bibfnamefont {S.}~\bibnamefont {Chaney}}, \bibinfo {author} {\bibfnamefont {M.}~\bibnamefont {Rosenkranz}},\ and\ \bibinfo {author} {\bibfnamefont {M.}~\bibnamefont {Lubasch}},\ }\bibfield  {title} {\bibinfo {title} {Noisy intermediate-scale quantum simulation of the one-dimensional wave equation},\ }\href {https://doi.org/10.1103/PhysRevResearch.6.043169} {\bibfield  {journal} {\bibinfo  {journal} {Physical Review Research}\ }\textbf {\bibinfo {volume} {6}},\ \bibinfo {pages} {043169} (\bibinfo {year} {2024})}\BibitemShut {NoStop}%
\bibitem [{\citenamefont {Sato}\ \emph {et~al.}(2025)\citenamefont {Sato}, \citenamefont {Tezuka}, \citenamefont {Kondo},\ and\ \citenamefont {Yamamoto}}]{sato2025quantum}%
  \BibitemOpen
  \bibfield  {author} {\bibinfo {author} {\bibfnamefont {Y.}~\bibnamefont {Sato}}, \bibinfo {author} {\bibfnamefont {H.}~\bibnamefont {Tezuka}}, \bibinfo {author} {\bibfnamefont {R.}~\bibnamefont {Kondo}},\ and\ \bibinfo {author} {\bibfnamefont {N.}~\bibnamefont {Yamamoto}},\ }\bibfield  {title} {\bibinfo {title} {Quantum algorithm for partial differential equations of nonconservative systems with spatially varying parameters},\ }\href@noop {} {\bibfield  {journal} {\bibinfo  {journal} {Physical Review Applied}\ }\textbf {\bibinfo {volume} {23}},\ \bibinfo {pages} {014063} (\bibinfo {year} {2025})}\BibitemShut {NoStop}%
\bibitem [{\citenamefont {Hu}\ \emph {et~al.}(2024)\citenamefont {Hu}, \citenamefont {Jin}, \citenamefont {Liu},\ and\ \citenamefont {Zhang}}]{hu2024quantum}%
  \BibitemOpen
  \bibfield  {author} {\bibinfo {author} {\bibfnamefont {J.}~\bibnamefont {Hu}}, \bibinfo {author} {\bibfnamefont {S.}~\bibnamefont {Jin}}, \bibinfo {author} {\bibfnamefont {N.}~\bibnamefont {Liu}},\ and\ \bibinfo {author} {\bibfnamefont {L.}~\bibnamefont {Zhang}},\ }\bibfield  {title} {\bibinfo {title} {{Quantum circuits for partial differential equations via Schr{\"o}dingerisation}},\ }\href@noop {} {\bibfield  {journal} {\bibinfo  {journal} {Quantum}\ }\textbf {\bibinfo {volume} {8}},\ \bibinfo {pages} {1563} (\bibinfo {year} {2024})}\BibitemShut {NoStop}%
\bibitem [{\citenamefont {Aaronson}(2015)}]{aaronson2015read}%
  \BibitemOpen
  \bibfield  {author} {\bibinfo {author} {\bibfnamefont {S.}~\bibnamefont {Aaronson}},\ }\bibfield  {title} {\bibinfo {title} {Read the fine print},\ }\href@noop {} {\bibfield  {journal} {\bibinfo  {journal} {Nature Physics}\ }\textbf {\bibinfo {volume} {11}},\ \bibinfo {pages} {291} (\bibinfo {year} {2015})}\BibitemShut {NoStop}%
\bibitem [{\citenamefont {Williams}\ \emph {et~al.}(2025)\citenamefont {Williams}, \citenamefont {Paine}, \citenamefont {Gentile}, \citenamefont {Berger},\ and\ \citenamefont {Kyriienko}}]{williams2025vortex}%
  \BibitemOpen
  \bibfield  {author} {\bibinfo {author} {\bibfnamefont {C.~A.}\ \bibnamefont {Williams}}, \bibinfo {author} {\bibfnamefont {A.~E.}\ \bibnamefont {Paine}}, \bibinfo {author} {\bibfnamefont {A.~A.}\ \bibnamefont {Gentile}}, \bibinfo {author} {\bibfnamefont {D.}~\bibnamefont {Berger}},\ and\ \bibinfo {author} {\bibfnamefont {O.}~\bibnamefont {Kyriienko}},\ }\bibfield  {title} {\bibinfo {title} {Vortex detection from quantum data},\ }\href@noop {} {\bibfield  {journal} {\bibinfo  {journal} {Physical Review A}\ }\textbf {\bibinfo {volume} {112}},\ \bibinfo {pages} {062409} (\bibinfo {year} {2025})}\BibitemShut {NoStop}%
\bibitem [{\citenamefont {Williams}\ \emph {et~al.}(2024)\citenamefont {Williams}, \citenamefont {Scali}, \citenamefont {Gentile}, \citenamefont {Berger},\ and\ \citenamefont {Kyriienko}}]{williams2024addressing}%
  \BibitemOpen
  \bibfield  {author} {\bibinfo {author} {\bibfnamefont {C.~A.}\ \bibnamefont {Williams}}, \bibinfo {author} {\bibfnamefont {S.}~\bibnamefont {Scali}}, \bibinfo {author} {\bibfnamefont {A.~A.}\ \bibnamefont {Gentile}}, \bibinfo {author} {\bibfnamefont {D.}~\bibnamefont {Berger}},\ and\ \bibinfo {author} {\bibfnamefont {O.}~\bibnamefont {Kyriienko}},\ }\bibfield  {title} {\bibinfo {title} {Addressing the readout problem in quantum differential equation algorithms with quantum scientific machine learning},\ }\href@noop {} {\bibfield  {journal} {\bibinfo  {journal} {arXiv preprint arXiv:2411.14259}\ } (\bibinfo {year} {2024})}\BibitemShut {NoStop}%
\bibitem [{\citenamefont {Cerezo}\ \emph {et~al.}(2021{\natexlab{a}})\citenamefont {Cerezo}, \citenamefont {Arrasmith}, \citenamefont {Babbush}, \citenamefont {Benjamin}, \citenamefont {Endo}, \citenamefont {Fujii}, \citenamefont {McClean}, \citenamefont {Mitarai}, \citenamefont {Yuan}, \citenamefont {Cincio} \emph {et~al.}}]{cerezo2021variational}%
  \BibitemOpen
  \bibfield  {author} {\bibinfo {author} {\bibfnamefont {M.}~\bibnamefont {Cerezo}}, \bibinfo {author} {\bibfnamefont {A.}~\bibnamefont {Arrasmith}}, \bibinfo {author} {\bibfnamefont {R.}~\bibnamefont {Babbush}}, \bibinfo {author} {\bibfnamefont {S.~C.}\ \bibnamefont {Benjamin}}, \bibinfo {author} {\bibfnamefont {S.}~\bibnamefont {Endo}}, \bibinfo {author} {\bibfnamefont {K.}~\bibnamefont {Fujii}}, \bibinfo {author} {\bibfnamefont {J.~R.}\ \bibnamefont {McClean}}, \bibinfo {author} {\bibfnamefont {K.}~\bibnamefont {Mitarai}}, \bibinfo {author} {\bibfnamefont {X.}~\bibnamefont {Yuan}}, \bibinfo {author} {\bibfnamefont {L.}~\bibnamefont {Cincio}}, \emph {et~al.},\ }\bibfield  {title} {\bibinfo {title} {Variational quantum algorithms},\ }\href@noop {} {\bibfield  {journal} {\bibinfo  {journal} {Nature Reviews Physics}\ }\textbf {\bibinfo {volume} {3}},\ \bibinfo {pages} {625} (\bibinfo {year} {2021}{\natexlab{a}})}\BibitemShut {NoStop}%
\bibitem [{\citenamefont {Tilly}\ \emph {et~al.}(2022)\citenamefont {Tilly}, \citenamefont {Chen}, \citenamefont {Cao}, \citenamefont {Picozzi}, \citenamefont {Setia}, \citenamefont {Li}, \citenamefont {Grant}, \citenamefont {Wossnig}, \citenamefont {Rungger}, \citenamefont {Booth} \emph {et~al.}}]{tilly2022variational}%
  \BibitemOpen
  \bibfield  {author} {\bibinfo {author} {\bibfnamefont {J.}~\bibnamefont {Tilly}}, \bibinfo {author} {\bibfnamefont {H.}~\bibnamefont {Chen}}, \bibinfo {author} {\bibfnamefont {S.}~\bibnamefont {Cao}}, \bibinfo {author} {\bibfnamefont {D.}~\bibnamefont {Picozzi}}, \bibinfo {author} {\bibfnamefont {K.}~\bibnamefont {Setia}}, \bibinfo {author} {\bibfnamefont {Y.}~\bibnamefont {Li}}, \bibinfo {author} {\bibfnamefont {E.}~\bibnamefont {Grant}}, \bibinfo {author} {\bibfnamefont {L.}~\bibnamefont {Wossnig}}, \bibinfo {author} {\bibfnamefont {I.}~\bibnamefont {Rungger}}, \bibinfo {author} {\bibfnamefont {G.~H.}\ \bibnamefont {Booth}}, \emph {et~al.},\ }\bibfield  {title} {\bibinfo {title} {The variational quantum eigensolver: a review of methods and best practices},\ }\href@noop {} {\bibfield  {journal} {\bibinfo  {journal} {Physics Reports}\ }\textbf {\bibinfo {volume} {986}},\ \bibinfo {pages} {1} (\bibinfo {year} {2022})}\BibitemShut {NoStop}%
\bibitem [{\citenamefont {Farhi}\ \emph {et~al.}(2014)\citenamefont {Farhi}, \citenamefont {Goldstone},\ and\ \citenamefont {Gutmann}}]{farhi2014quantum}%
  \BibitemOpen
  \bibfield  {author} {\bibinfo {author} {\bibfnamefont {E.}~\bibnamefont {Farhi}}, \bibinfo {author} {\bibfnamefont {J.}~\bibnamefont {Goldstone}},\ and\ \bibinfo {author} {\bibfnamefont {S.}~\bibnamefont {Gutmann}},\ }\bibfield  {title} {\bibinfo {title} {A quantum approximate optimization algorithm},\ }\href@noop {} {\bibfield  {journal} {\bibinfo  {journal} {arXiv preprint arXiv:1411.4028}\ } (\bibinfo {year} {2014})}\BibitemShut {NoStop}%
\bibitem [{\citenamefont {Zhou}\ \emph {et~al.}(2020)\citenamefont {Zhou}, \citenamefont {Wang}, \citenamefont {Choi}, \citenamefont {Pichler},\ and\ \citenamefont {Lukin}}]{zhou2020quantum}%
  \BibitemOpen
  \bibfield  {author} {\bibinfo {author} {\bibfnamefont {L.}~\bibnamefont {Zhou}}, \bibinfo {author} {\bibfnamefont {S.-T.}\ \bibnamefont {Wang}}, \bibinfo {author} {\bibfnamefont {S.}~\bibnamefont {Choi}}, \bibinfo {author} {\bibfnamefont {H.}~\bibnamefont {Pichler}},\ and\ \bibinfo {author} {\bibfnamefont {M.~D.}\ \bibnamefont {Lukin}},\ }\bibfield  {title} {\bibinfo {title} {Quantum approximate optimization algorithm: Performance, mechanism, and implementation on near-term devices},\ }\href@noop {} {\bibfield  {journal} {\bibinfo  {journal} {Physical Review X}\ }\textbf {\bibinfo {volume} {10}},\ \bibinfo {pages} {021067} (\bibinfo {year} {2020})}\BibitemShut {NoStop}%
\bibitem [{\citenamefont {Willsch}\ \emph {et~al.}(2020)\citenamefont {Willsch}, \citenamefont {Willsch}, \citenamefont {Jin}, \citenamefont {De~Raedt},\ and\ \citenamefont {Michielsen}}]{willsch2020benchmarking}%
  \BibitemOpen
  \bibfield  {author} {\bibinfo {author} {\bibfnamefont {M.}~\bibnamefont {Willsch}}, \bibinfo {author} {\bibfnamefont {D.}~\bibnamefont {Willsch}}, \bibinfo {author} {\bibfnamefont {F.}~\bibnamefont {Jin}}, \bibinfo {author} {\bibfnamefont {H.}~\bibnamefont {De~Raedt}},\ and\ \bibinfo {author} {\bibfnamefont {K.}~\bibnamefont {Michielsen}},\ }\bibfield  {title} {\bibinfo {title} {Benchmarking the quantum approximate optimization algorithm},\ }\href@noop {} {\bibfield  {journal} {\bibinfo  {journal} {Quantum Information Processing}\ }\textbf {\bibinfo {volume} {19}},\ \bibinfo {pages} {197} (\bibinfo {year} {2020})}\BibitemShut {NoStop}%
\bibitem [{\citenamefont {Cerezo}\ \emph {et~al.}(2021{\natexlab{b}})\citenamefont {Cerezo}, \citenamefont {Sone}, \citenamefont {Volkoff}, \citenamefont {Cincio},\ and\ \citenamefont {Coles}}]{cerezo2021cost}%
  \BibitemOpen
  \bibfield  {author} {\bibinfo {author} {\bibfnamefont {M.}~\bibnamefont {Cerezo}}, \bibinfo {author} {\bibfnamefont {A.}~\bibnamefont {Sone}}, \bibinfo {author} {\bibfnamefont {T.}~\bibnamefont {Volkoff}}, \bibinfo {author} {\bibfnamefont {L.}~\bibnamefont {Cincio}},\ and\ \bibinfo {author} {\bibfnamefont {P.~J.}\ \bibnamefont {Coles}},\ }\bibfield  {title} {\bibinfo {title} {Cost function dependent barren plateaus in shallow parametrized quantum circuits},\ }\href@noop {} {\bibfield  {journal} {\bibinfo  {journal} {Nature communications}\ }\textbf {\bibinfo {volume} {12}},\ \bibinfo {pages} {1791} (\bibinfo {year} {2021}{\natexlab{b}})}\BibitemShut {NoStop}%
\bibitem [{\citenamefont {Arrasmith}\ \emph {et~al.}(2022)\citenamefont {Arrasmith}, \citenamefont {Holmes}, \citenamefont {Cerezo},\ and\ \citenamefont {Coles}}]{arrasmith2022equivalence}%
  \BibitemOpen
  \bibfield  {author} {\bibinfo {author} {\bibfnamefont {A.}~\bibnamefont {Arrasmith}}, \bibinfo {author} {\bibfnamefont {Z.}~\bibnamefont {Holmes}}, \bibinfo {author} {\bibfnamefont {M.}~\bibnamefont {Cerezo}},\ and\ \bibinfo {author} {\bibfnamefont {P.~J.}\ \bibnamefont {Coles}},\ }\bibfield  {title} {\bibinfo {title} {Equivalence of quantum barren plateaus to cost concentration and narrow gorges},\ }\href@noop {} {\bibfield  {journal} {\bibinfo  {journal} {Quantum Science \& Technology}\ }\textbf {\bibinfo {volume} {7}},\ \bibinfo {pages} {045015} (\bibinfo {year} {2022})}\BibitemShut {NoStop}%
\bibitem [{\citenamefont {Leng}\ \emph {et~al.}(2023)\citenamefont {Leng}, \citenamefont {Hickman}, \citenamefont {Li},\ and\ \citenamefont {Wu}}]{leng2023quantum}%
  \BibitemOpen
  \bibfield  {author} {\bibinfo {author} {\bibfnamefont {J.}~\bibnamefont {Leng}}, \bibinfo {author} {\bibfnamefont {E.}~\bibnamefont {Hickman}}, \bibinfo {author} {\bibfnamefont {J.}~\bibnamefont {Li}},\ and\ \bibinfo {author} {\bibfnamefont {X.}~\bibnamefont {Wu}},\ }\bibfield  {title} {\bibinfo {title} {{Quantum Hamiltonian descent}},\ }\href@noop {} {\bibfield  {journal} {\bibinfo  {journal} {Bulletin of the American Physical Society}\ }\textbf {\bibinfo {volume} {68}} (\bibinfo {year} {2023})}\BibitemShut {NoStop}%
\bibitem [{\citenamefont {Leng}\ \emph {et~al.}(2025)\citenamefont {Leng}, \citenamefont {Zheng}, \citenamefont {Jia}, \citenamefont {Fan}, \citenamefont {Zhao}, \citenamefont {Peng},\ and\ \citenamefont {Wu}}]{leng2025quantum}%
  \BibitemOpen
  \bibfield  {author} {\bibinfo {author} {\bibfnamefont {J.}~\bibnamefont {Leng}}, \bibinfo {author} {\bibfnamefont {Y.}~\bibnamefont {Zheng}}, \bibinfo {author} {\bibfnamefont {Z.}~\bibnamefont {Jia}}, \bibinfo {author} {\bibfnamefont {L.}~\bibnamefont {Fan}}, \bibinfo {author} {\bibfnamefont {C.}~\bibnamefont {Zhao}}, \bibinfo {author} {\bibfnamefont {Y.}~\bibnamefont {Peng}},\ and\ \bibinfo {author} {\bibfnamefont {X.}~\bibnamefont {Wu}},\ }\bibfield  {title} {\bibinfo {title} {{Quantum Hamiltonian descent for non-smooth optimization}},\ }\href@noop {} {\bibfield  {journal} {\bibinfo  {journal} {arXiv preprint arXiv:2503.15878}\ } (\bibinfo {year} {2025})}\BibitemShut {NoStop}%
\bibitem [{\citenamefont {Chen}\ \emph {et~al.}(2025)\citenamefont {Chen}, \citenamefont {Lu}, \citenamefont {Wang}, \citenamefont {Liu},\ and\ \citenamefont {Li}}]{chen2025quantum}%
  \BibitemOpen
  \bibfield  {author} {\bibinfo {author} {\bibfnamefont {Z.}~\bibnamefont {Chen}}, \bibinfo {author} {\bibfnamefont {Y.}~\bibnamefont {Lu}}, \bibinfo {author} {\bibfnamefont {H.}~\bibnamefont {Wang}}, \bibinfo {author} {\bibfnamefont {Y.}~\bibnamefont {Liu}},\ and\ \bibinfo {author} {\bibfnamefont {T.}~\bibnamefont {Li}},\ }\bibfield  {title} {\bibinfo {title} {Quantum langevin dynamics for optimization},\ }\href@noop {} {\bibfield  {journal} {\bibinfo  {journal} {Communications in Mathematical Physics}\ }\textbf {\bibinfo {volume} {406}},\ \bibinfo {pages} {52} (\bibinfo {year} {2025})}\BibitemShut {NoStop}%
\bibitem [{\citenamefont {Catli}\ \emph {et~al.}(2026)\citenamefont {Catli}, \citenamefont {Simon},\ and\ \citenamefont {Wiebe}}]{catli2026efficient}%
  \BibitemOpen
  \bibfield  {author} {\bibinfo {author} {\bibfnamefont {A.~B.}\ \bibnamefont {Catli}}, \bibinfo {author} {\bibfnamefont {S.}~\bibnamefont {Simon}},\ and\ \bibinfo {author} {\bibfnamefont {N.}~\bibnamefont {Wiebe}},\ }\bibfield  {title} {\bibinfo {title} {Efficient quantum optimization via dynamical simulation},\ }\href@noop {} {\bibfield  {journal} {\bibinfo  {journal} {PRX Quantum}\ }\textbf {\bibinfo {volume} {7}},\ \bibinfo {pages} {033026} (\bibinfo {year} {2026})}\BibitemShut {NoStop}%
\bibitem [{\citenamefont {Gily{\'e}n}\ \emph {et~al.}(2019)\citenamefont {Gily{\'e}n}, \citenamefont {Su}, \citenamefont {Low},\ and\ \citenamefont {Wiebe}}]{gilyen2019quantum}%
  \BibitemOpen
  \bibfield  {author} {\bibinfo {author} {\bibfnamefont {A.}~\bibnamefont {Gily{\'e}n}}, \bibinfo {author} {\bibfnamefont {Y.}~\bibnamefont {Su}}, \bibinfo {author} {\bibfnamefont {G.~H.}\ \bibnamefont {Low}},\ and\ \bibinfo {author} {\bibfnamefont {N.}~\bibnamefont {Wiebe}},\ }\bibfield  {title} {\bibinfo {title} {Quantum singular value transformation and beyond: exponential improvements for quantum matrix arithmetics},\ }\href@noop {} {\bibfield  {journal} {\bibinfo  {journal} {Proceedings of the 51st annual ACM SIGACT symposium on theory of computing}\ ,\ \bibinfo {pages} {193}} (\bibinfo {year} {2019})}\BibitemShut {NoStop}%
\bibitem [{\citenamefont {Sato}\ \emph {et~al.}(2023)\citenamefont {Sato}, \citenamefont {Kondo}, \citenamefont {Koide},\ and\ \citenamefont {Kajita}}]{sato2023quantum}%
  \BibitemOpen
  \bibfield  {author} {\bibinfo {author} {\bibfnamefont {Y.}~\bibnamefont {Sato}}, \bibinfo {author} {\bibfnamefont {R.}~\bibnamefont {Kondo}}, \bibinfo {author} {\bibfnamefont {S.}~\bibnamefont {Koide}},\ and\ \bibinfo {author} {\bibfnamefont {S.}~\bibnamefont {Kajita}},\ }\bibfield  {title} {\bibinfo {title} {Quantum topology optimization of ground structures for near-term devices},\ }\href@noop {} {\bibfield  {journal} {\bibinfo  {journal} {2023 IEEE International Conference on Quantum Computing and Engineering (QCE)}\ }\textbf {\bibinfo {volume} {1}},\ \bibinfo {pages} {168} (\bibinfo {year} {2023})}\BibitemShut {NoStop}%
\bibitem [{\citenamefont {Kim}\ and\ \citenamefont {Wang}(2025)}]{kim2025variational}%
  \BibitemOpen
  \bibfield  {author} {\bibinfo {author} {\bibfnamefont {J.~E.}\ \bibnamefont {Kim}}\ and\ \bibinfo {author} {\bibfnamefont {Y.}~\bibnamefont {Wang}},\ }\bibfield  {title} {\bibinfo {title} {Variational quantum algorithm for constrained topology optimization},\ }\href@noop {} {\bibfield  {journal} {\bibinfo  {journal} {Quantum Science and Technology}\ }\textbf {\bibinfo {volume} {10}},\ \bibinfo {pages} {045025} (\bibinfo {year} {2025})}\BibitemShut {NoStop}%
\bibitem [{\citenamefont {Stein}\ \emph {et~al.}(2023)\citenamefont {Stein}, \citenamefont {M{\"u}ller}, \citenamefont {H{\"o}lscher}, \citenamefont {Chnitidis}, \citenamefont {Jojo}, \citenamefont {Farea}, \citenamefont {{\c{C}}elebi}, \citenamefont {Bucher}, \citenamefont {Wulf}, \citenamefont {Fischer} \emph {et~al.}}]{stein2023exponential}%
  \BibitemOpen
  \bibfield  {author} {\bibinfo {author} {\bibfnamefont {J.}~\bibnamefont {Stein}}, \bibinfo {author} {\bibfnamefont {L.}~\bibnamefont {M{\"u}ller}}, \bibinfo {author} {\bibfnamefont {L.}~\bibnamefont {H{\"o}lscher}}, \bibinfo {author} {\bibfnamefont {G.}~\bibnamefont {Chnitidis}}, \bibinfo {author} {\bibfnamefont {J.}~\bibnamefont {Jojo}}, \bibinfo {author} {\bibfnamefont {A.}~\bibnamefont {Farea}}, \bibinfo {author} {\bibfnamefont {M.~S.}\ \bibnamefont {{\c{C}}elebi}}, \bibinfo {author} {\bibfnamefont {D.}~\bibnamefont {Bucher}}, \bibinfo {author} {\bibfnamefont {J.}~\bibnamefont {Wulf}}, \bibinfo {author} {\bibfnamefont {D.}~\bibnamefont {Fischer}}, \emph {et~al.},\ }\bibfield  {title} {\bibinfo {title} {Exponential quantum speedup for simulation-based optimization applications},\ }\href@noop {} {\bibfield  {journal} {\bibinfo  {journal} {arXiv preprint arXiv:2305.08482}\ } (\bibinfo {year} {2023})}\BibitemShut {NoStop}%
\bibitem [{\citenamefont {H{\"o}lscher}\ \emph {et~al.}(2026)\citenamefont {H{\"o}lscher}, \citenamefont {Karch}, \citenamefont {Samimi},\ and\ \citenamefont {Danzig}}]{holscher2026quantum}%
  \BibitemOpen
  \bibfield  {author} {\bibinfo {author} {\bibfnamefont {L.}~\bibnamefont {H{\"o}lscher}}, \bibinfo {author} {\bibfnamefont {L.}~\bibnamefont {Karch}}, \bibinfo {author} {\bibfnamefont {O.}~\bibnamefont {Samimi}},\ and\ \bibinfo {author} {\bibfnamefont {T.}~\bibnamefont {Danzig}},\ }\bibfield  {title} {\bibinfo {title} {Quantum simulation-based optimization for cooling system design},\ }\href@noop {} {\bibfield  {journal} {\bibinfo  {journal} {Journal of Physics A: Mathematical and Theoretical}\ }\textbf {\bibinfo {volume} {59}},\ \bibinfo {pages} {125301} (\bibinfo {year} {2026})}\BibitemShut {NoStop}%
\bibitem [{\citenamefont {Low}\ and\ \citenamefont {Wiebe}(2018)}]{low2018hamiltonian}%
  \BibitemOpen
  \bibfield  {author} {\bibinfo {author} {\bibfnamefont {G.~H.}\ \bibnamefont {Low}}\ and\ \bibinfo {author} {\bibfnamefont {N.}~\bibnamefont {Wiebe}},\ }\bibfield  {title} {\bibinfo {title} {Hamiltonian simulation in the interaction picture},\ }\href@noop {} {\bibfield  {journal} {\bibinfo  {journal} {arXiv preprint arXiv:1805.00675}\ } (\bibinfo {year} {2018})}\BibitemShut {NoStop}%
\bibitem [{\citenamefont {Friedlander}\ and\ \citenamefont {Schmidt}(2012)}]{friedlander2012hybrid}%
  \BibitemOpen
  \bibfield  {author} {\bibinfo {author} {\bibfnamefont {M.~P.}\ \bibnamefont {Friedlander}}\ and\ \bibinfo {author} {\bibfnamefont {M.}~\bibnamefont {Schmidt}},\ }\bibfield  {title} {\bibinfo {title} {Hybrid deterministic-stochastic methods for data fitting},\ }\href@noop {} {\bibfield  {journal} {\bibinfo  {journal} {SIAM Journal on Scientific Computing}\ }\textbf {\bibinfo {volume} {34}},\ \bibinfo {pages} {A1380} (\bibinfo {year} {2012})}\BibitemShut {NoStop}%
\bibitem [{\citenamefont {Devolder}\ \emph {et~al.}(2013)\citenamefont {Devolder}, \citenamefont {Glineur}, \citenamefont {Nesterov} \emph {et~al.}}]{devolder2013first}%
  \BibitemOpen
  \bibfield  {author} {\bibinfo {author} {\bibfnamefont {O.}~\bibnamefont {Devolder}}, \bibinfo {author} {\bibfnamefont {F.}~\bibnamefont {Glineur}}, \bibinfo {author} {\bibfnamefont {Y.}~\bibnamefont {Nesterov}}, \emph {et~al.},\ }\bibfield  {title} {\bibinfo {title} {First-order methods with inexact oracle: the strongly convex case},\ }\href@noop {} {\bibfield  {journal} {\bibinfo  {journal} {CORE Discussion Papers}\ }\textbf {\bibinfo {volume} {2013016}},\ \bibinfo {pages} {47} (\bibinfo {year} {2013})}\BibitemShut {NoStop}%
\bibitem [{\citenamefont {Sakamoto}\ and\ \citenamefont {Fujii}(2026)}]{sakamoto2026quantum}%
  \BibitemOpen
  \bibfield  {author} {\bibinfo {author} {\bibfnamefont {K.}~\bibnamefont {Sakamoto}}\ and\ \bibinfo {author} {\bibfnamefont {K.}~\bibnamefont {Fujii}},\ }\bibfield  {title} {\bibinfo {title} {On the quantum computational complexity of classical linear dynamics with geometrically local interactions: Dequantization and universality},\ }\href@noop {} {\bibfield  {journal} {\bibinfo  {journal} {Quantum}\ }\textbf {\bibinfo {volume} {10}},\ \bibinfo {pages} {2182} (\bibinfo {year} {2026})}\BibitemShut {NoStop}%
\bibitem [{\citenamefont {Javadi-Abhari}\ \emph {et~al.}(2024)\citenamefont {Javadi-Abhari}, \citenamefont {Treinish}, \citenamefont {Krsulich}, \citenamefont {Wood}, \citenamefont {Lishman}, \citenamefont {Gacon}, \citenamefont {Martiel}, \citenamefont {Nation}, \citenamefont {Bishop}, \citenamefont {Cross} \emph {et~al.}}]{javadi2024quantum}%
  \BibitemOpen
  \bibfield  {author} {\bibinfo {author} {\bibfnamefont {A.}~\bibnamefont {Javadi-Abhari}}, \bibinfo {author} {\bibfnamefont {M.}~\bibnamefont {Treinish}}, \bibinfo {author} {\bibfnamefont {K.}~\bibnamefont {Krsulich}}, \bibinfo {author} {\bibfnamefont {C.~J.}\ \bibnamefont {Wood}}, \bibinfo {author} {\bibfnamefont {J.}~\bibnamefont {Lishman}}, \bibinfo {author} {\bibfnamefont {J.}~\bibnamefont {Gacon}}, \bibinfo {author} {\bibfnamefont {S.}~\bibnamefont {Martiel}}, \bibinfo {author} {\bibfnamefont {P.~D.}\ \bibnamefont {Nation}}, \bibinfo {author} {\bibfnamefont {L.~S.}\ \bibnamefont {Bishop}}, \bibinfo {author} {\bibfnamefont {A.~W.}\ \bibnamefont {Cross}}, \emph {et~al.},\ }\bibfield  {title} {\bibinfo {title} {Quantum computing with qiskit},\ }\href@noop {} {\bibfield  {journal} {\bibinfo  {journal} {arXiv preprint arXiv:2405.08810}\ } (\bibinfo {year} {2024})}\BibitemShut {NoStop}%
\bibitem [{\citenamefont {Hull}\ and\ \citenamefont {Basu}(2016)}]{hull2016options}%
  \BibitemOpen
  \bibfield  {author} {\bibinfo {author} {\bibfnamefont {J.~C.}\ \bibnamefont {Hull}}\ and\ \bibinfo {author} {\bibfnamefont {S.}~\bibnamefont {Basu}},\ }\href@noop {} {\emph {\bibinfo {title} {Options, futures, and other derivatives}}}\ (\bibinfo  {publisher} {Pearson Education India},\ \bibinfo {year} {2016})\BibitemShut {NoStop}%
\bibitem [{\citenamefont {Gonzalez-Conde}\ \emph {et~al.}(2023)\citenamefont {Gonzalez-Conde}, \citenamefont {Rodr{\'\i}guez-Rozas}, \citenamefont {Solano},\ and\ \citenamefont {Sanz}}]{gonzalez2023efficient}%
  \BibitemOpen
  \bibfield  {author} {\bibinfo {author} {\bibfnamefont {J.}~\bibnamefont {Gonzalez-Conde}}, \bibinfo {author} {\bibfnamefont {{\'A}.}~\bibnamefont {Rodr{\'\i}guez-Rozas}}, \bibinfo {author} {\bibfnamefont {E.}~\bibnamefont {Solano}},\ and\ \bibinfo {author} {\bibfnamefont {M.}~\bibnamefont {Sanz}},\ }\bibfield  {title} {\bibinfo {title} {Efficient hamiltonian simulation for solving option price dynamics},\ }\href@noop {} {\bibfield  {journal} {\bibinfo  {journal} {Physical Review Research}\ }\textbf {\bibinfo {volume} {5}},\ \bibinfo {pages} {043220} (\bibinfo {year} {2023})}\BibitemShut {NoStop}%
\bibitem [{\citenamefont {Jin}\ \emph {et~al.}(2026{\natexlab{b}})\citenamefont {Jin}, \citenamefont {Tang}, \citenamefont {Yin},\ and\ \citenamefont {Zhang}}]{jin2026quantum}%
  \BibitemOpen
  \bibfield  {author} {\bibinfo {author} {\bibfnamefont {S.}~\bibnamefont {Jin}}, \bibinfo {author} {\bibfnamefont {Z.}~\bibnamefont {Tang}}, \bibinfo {author} {\bibfnamefont {X.}~\bibnamefont {Yin}},\ and\ \bibinfo {author} {\bibfnamefont {L.}~\bibnamefont {Zhang}},\ }\bibfield  {title} {\bibinfo {title} {Quantum circuits for black-scholes equations via schr{\"o}dingerisation},\ }\href@noop {} {\bibfield  {journal} {\bibinfo  {journal} {Communications in Computational Physics}\ }\textbf {\bibinfo {volume} {39}},\ \bibinfo {pages} {1} (\bibinfo {year} {2026}{\natexlab{b}})}\BibitemShut {NoStop}%
\bibitem [{\citenamefont {Iten}\ \emph {et~al.}(2016)\citenamefont {Iten}, \citenamefont {Colbeck}, \citenamefont {Kukuljan}, \citenamefont {Home},\ and\ \citenamefont {Christandl}}]{iten2016quantum}%
  \BibitemOpen
  \bibfield  {author} {\bibinfo {author} {\bibfnamefont {R.}~\bibnamefont {Iten}}, \bibinfo {author} {\bibfnamefont {R.}~\bibnamefont {Colbeck}}, \bibinfo {author} {\bibfnamefont {I.}~\bibnamefont {Kukuljan}}, \bibinfo {author} {\bibfnamefont {J.}~\bibnamefont {Home}},\ and\ \bibinfo {author} {\bibfnamefont {M.}~\bibnamefont {Christandl}},\ }\bibfield  {title} {\bibinfo {title} {Quantum circuits for isometries},\ }\href@noop {} {\bibfield  {journal} {\bibinfo  {journal} {Physical Review A}\ }\textbf {\bibinfo {volume} {93}},\ \bibinfo {pages} {032318} (\bibinfo {year} {2016})}\BibitemShut {NoStop}%
\bibitem [{\citenamefont {Barenco}\ \emph {et~al.}(1995)\citenamefont {Barenco}, \citenamefont {Bennett}, \citenamefont {Cleve}, \citenamefont {DiVincenzo}, \citenamefont {Margolus}, \citenamefont {Shor}, \citenamefont {Sleator}, \citenamefont {Smolin},\ and\ \citenamefont {Weinfurter}}]{barenco1995elementary}%
  \BibitemOpen
  \bibfield  {author} {\bibinfo {author} {\bibfnamefont {A.}~\bibnamefont {Barenco}}, \bibinfo {author} {\bibfnamefont {C.~H.}\ \bibnamefont {Bennett}}, \bibinfo {author} {\bibfnamefont {R.}~\bibnamefont {Cleve}}, \bibinfo {author} {\bibfnamefont {D.~P.}\ \bibnamefont {DiVincenzo}}, \bibinfo {author} {\bibfnamefont {N.}~\bibnamefont {Margolus}}, \bibinfo {author} {\bibfnamefont {P.}~\bibnamefont {Shor}}, \bibinfo {author} {\bibfnamefont {T.}~\bibnamefont {Sleator}}, \bibinfo {author} {\bibfnamefont {J.~A.}\ \bibnamefont {Smolin}},\ and\ \bibinfo {author} {\bibfnamefont {H.}~\bibnamefont {Weinfurter}},\ }\bibfield  {title} {\bibinfo {title} {Elementary gates for quantum computation},\ }\href@noop {} {\bibfield  {journal} {\bibinfo  {journal} {Physical review A}\ }\textbf {\bibinfo {volume} {52}},\ \bibinfo {pages} {3457} (\bibinfo {year} {1995})}\BibitemShut {NoStop}%
\bibitem [{\citenamefont {H{\"a}ner}\ \emph {et~al.}(2018)\citenamefont {H{\"a}ner}, \citenamefont {Roetteler},\ and\ \citenamefont {Svore}}]{haner2018optimizing}%
  \BibitemOpen
  \bibfield  {author} {\bibinfo {author} {\bibfnamefont {T.}~\bibnamefont {H{\"a}ner}}, \bibinfo {author} {\bibfnamefont {M.}~\bibnamefont {Roetteler}},\ and\ \bibinfo {author} {\bibfnamefont {K.~M.}\ \bibnamefont {Svore}},\ }\bibfield  {title} {\bibinfo {title} {Optimizing quantum circuits for arithmetic},\ }\href@noop {} {\bibfield  {journal} {\bibinfo  {journal} {arXiv preprint arXiv:1805.12445}\ } (\bibinfo {year} {2018})}\BibitemShut {NoStop}%
\bibitem [{\citenamefont {Knill}\ \emph {et~al.}(2007)\citenamefont {Knill}, \citenamefont {Ortiz},\ and\ \citenamefont {Somma}}]{knill2007optimal}%
  \BibitemOpen
  \bibfield  {author} {\bibinfo {author} {\bibfnamefont {E.}~\bibnamefont {Knill}}, \bibinfo {author} {\bibfnamefont {G.}~\bibnamefont {Ortiz}},\ and\ \bibinfo {author} {\bibfnamefont {R.~D.}\ \bibnamefont {Somma}},\ }\bibfield  {title} {\bibinfo {title} {Optimal quantum measurements of expectation values of observables},\ }\href@noop {} {\bibfield  {journal} {\bibinfo  {journal} {Physical Review A—Atomic, Molecular, and Optical Physics}\ }\textbf {\bibinfo {volume} {75}},\ \bibinfo {pages} {012328} (\bibinfo {year} {2007})}\BibitemShut {NoStop}%
\end{thebibliography}%

\end{document}